\documentclass[fleqn,usenatbib]{mnras}
\usepackage{amsmath,graphicx,xspace,natbib,enumitem}
\usepackage[varg]{txfonts}
\usepackage[T1]{fontenc}
\usepackage{ae,aecompl}
\usepackage[normalem]{ulem}
\allowdisplaybreaks
\newcommand{\eg}{{e.g.\/}}
\newcommand{\ie}{{i.e.\/}}
\newcommand{\etc}{{etc.\/}}

\newcommand{\cf}{{c.f.\/}}

\newcommand{\azeus}{\textsf{AZEuS}\xspace}
\newcommand{\zeusddd}{\textsl{ZEUS-3D}\xspace}
\newcommand{\zeus}{\textsl{ZEUS}\xspace}
\newcommand{\pol}{\ensuremath{_{\rm p}}}
\newcommand{\dgr}{\ensuremath{^{\circ}}}

\newcommand{\veps}{\varepsilon}
\newcommand{\rmi}{\ensuremath{_{\rm i}}}

\newcommand{\qandq}{\ensuremath{\quad \text{and}\quad}}

\newlength{\VSpaceBeforeTabBib}
\newlength{\VSpaceBeforeTabFoot}
\newcommand*\tablefootname{Notes}
\newcommand*\tablefootfont{\small}
\newcommand*\tablefootnamefont{\small\bfseries}
\newcommand\tablefoot[1]{\VSpaceBeforeTabBib=1ex%
  \par\vspace{\VSpaceBeforeTabFoot}
  \noindent
  \begin{minipage}{\linewidth}
    {\tablefootnamefont\tablefootname.}~%
    \tablefootfont
    \ignorespaces
    #1%
  \end{minipage}%
}
\newcommand*\tablefootmark[1]{%
  \unskip
  \hbox{\textsuperscript{\normalfont\itshape\ignorespaces#1}}%
  \,%
  \ignorespaces
}
\newcommand\tablefoottext[2]{%
  \hbox{\textsuperscript{\normalfont({\itshape\ignorespaces#1})}}%
  ~%
  \ignorespaces
  #2\ \ignorespaces%
}
\newcommand{\shortauthor}{J.~P. Ramsey \& D.~A. Clarke}
\newcommand{\shorttitle}{Formation and propagation of protostellar jets.}
\title[\shorttitle]{MHD simulations of the formation and propagation of protostellar jets to observational length scales}
\author[\shortauthor]{Jon P.\ Ramsey$^{1,3}$\thanks{jpramsey@virginia.edu (JPR)} and David A.\ Clarke$^{2}$\thanks{dclarke@ap.smu.ca (DAC)}\\
$^{1}$Centre for Star and Planet Formation, Natural History Museum of Denmark and the Niels Bohr Institute,\\ University of Copenhagen, \O ster Voldgade 5-7, 1350 Copenhagen K, Denmark\\
$^{2}$Department of Astronomy \& Physics, Saint Mary's University, Halifax, Nova Scotia B3H 3C3, Canada\\
$^{3}$Department of Astronomy, University of Virginia, Charlottesville, VA 22904, USA
}
\date{Accepted Jan 9 2019}
\pubyear{?}
\defcitealias{rcm12}{RCM12}
\defcitealias{rc11}{RC11}
\defcitealias{bp82}{BP}
\begin{document}
\label{firstpage}
\pagerange{\pageref{firstpage}--\pageref{lastpage}}
\maketitle
\begin{abstract}
We present 2.5-D global, ideal MHD simulations of magnetically and rotationally driven protostellar jets from Keplerian accretion discs, wherein only the initial magnetic field strength at the inner radius of the disc, $B\rmi$, is varied.  Using the AMR-MHD code \azeus, we self-consistently follow the jet evolution into the observational regime ($>10^3\,\mathrm{AU}$) with a spatial dynamic range of $\sim6.5\times10^5$.  The simulations reveal a three-component outflow:  1) A hot, dense, super-fast and highly magnetised `jet core'; 2) a cold, rarefied, trans-fast and highly magnetised `sheath' surrounding the jet core and extending to a tangential discontinuity; and 3) a warm, dense, trans-slow and weakly magnetised shocked ambient medium entrained by the advancing bow shock.  The simulations reveal power-law relationships between $B\rmi$ and the jet advance speed, $v_{\rm jet}$, the average jet rotation speed, $\langle v_\varphi\rangle$, as well as fluxes of mass, momentum, and kinetic energy.  Quantities that do not depend on $B\rmi$ include the plasma-$\beta$ of the transported material which, in all cases, seems to asymptote to order unity.  Jets are launched by a combination of the `magnetic tower' and `bead-on-a-wire' mechanisms, with the former accounting for most of the jet acceleration---even for strong fields---and continuing well beyond the fast magnetosonic point.  At no time does the leading bow shock leave the domain and, as such, these simulations generate large-scale jets that reproduce many of the observed properties of protostellar jets including their characteristic speeds and transported fluxes.
\end{abstract}
% Currently 235 words.
%
\maketitle
\section{Introduction}
\label{sec:intro}
One of the most important epochs in the early evolution of most stars is the short period during which it throws a small fraction of the accreted gas back into the interstellar medium (ISM) as a pair of collimated, bipolar, supersonic jets.  This period, lasting typically $10^4$--$10^5$ years, is to the star's entire lifetime what a few hours is to a human's.  Yet, in this single `afternoon', the protostellar system manages to shed itself of sufficient angular momentum to enable significant accretion from the protoplanetary accretion disc onto the protostar.  Without protostellar jets, stars as we know them would not exist.

Thought to be stars in their own right when first observed by \citet{burnham1890}, protostellar jets only started to be appreciated for what they are by \citet{snelletal80}.  In this seminal work, a detailed bipolar outflow model for L1551 is described that is still basic to the modern view, and done without ever using the word \emph{jet}\footnote{While the term `jet' was first used in an astrophysical context by \citet{baademinkowski54} in describing the optical `protrusion' on M87, it did not enter the protostellar vernacular until \citet{mundtfried83}.}. It is now known that protostellar jets reach lengths of 0.1--5 pc \citep{ballyetal07_ppv} and can transport $\sim\! 10$\% of the accreted mass and $\sim\! 70$\% of the angular momentum \citep{woitasetal05} out of the protostellar system and back into the ISM.  They can appear as straight, ballistic, highly supersonic flows complete with leading bow shocks (\eg\ HH\,34; \citealt{devineetal1997_hh34}), or more like effluent from a smokestack; wide, twisted and with no particular evidence of a supersonic nature (\eg\ HH\,47; \citealt{hartiganetal2005_hh47}).

\citet{reipurth99} and \citet{wuetal04} catalogue some 1,000 Herbig-Haro (HH) objects and molecular outflows from protostellar objects, imaged with atomic (\eg, H$\alpha$, O\textsc{iii}, S\textsc{ii}) and/or molecular (\eg, CO) line emission and from which important kinematical and dynamical quantities are measured.  While observational properties of individual outflows vary widely, ranges of values for parameters most useful for constraining numerical magnetohydrodynamical (MHD) simulations can be established.  These include the advance speed of the jet into the ISM ($v_{\rm jet}$), the speed of entrained material swept up by the bow shock leading the jet ($v_{\rm entr}$), the average rotational speed of jet material about its propagation axis ($\langle v_{\rm rot}\rangle$), fluxes in mass ($\dot M$), linear momentum ($\dot p$), and angular momentum ($\dot L$).  Table \ref{tab:obsparms} displays a summary of what is currently known of these parameters as reported by \citet{hartiganetal94}, \citet{reipurthbally01_araa}, \citet{podioetal06}, \citet{mckeeostriker07}, \citet{rayetal07_ppv}, \citet{coffeyetal08, coffeyetal11} and \citet{franketal14_ppvi}.  While these observations provide an extensive and highly detailed picture of protostellar jets, the quantities listed in Table \ref{tab:obsparms} are largely inferred and indirect, and should not be taken as hard limits.

\begin{table}
\begin{center}
\caption{\label{tab:obsparms} Physical quantities as measured/inferred from observations of protostellar outflows taken from references listed in the text.  The quantities include from top to bottom: $v_{\rm jet}$, advance speed of the jet; $\langle v_{\rm rot}\rangle$, jet rotational speed; $\dot M$, mass flux; $\dot p$, linear momentum flux; $\dot L$, angular momentum flux; and $v_{\rm entr}$, advance speed of entrained outflow.}
\begin{tabular}{r|l}
\hline
$v_{\rm jet}$ & 100 -- 1000 km\,s$^{-1}$\\
$\langle v_{\rm rot}\rangle$ & 5 -- 25 km\,s$^{-1}$\\
$\dot M$ & $10^{-9}$ -- $10^{-5}\,M_{\odot}$\,yr$^{-1}$\\
$\dot p$ & $10^{-6}$ -- $1.4\times10^{-4}\,M_{\odot}$\,yr$^{-1}$\,km\,s$^{-1}$\\
$\dot L$ & $10^{-6}$ -- $10^{-5}$ $M_{\odot}$\,yr$^{-1}$\,AU\,km\,s$^{-1}$\\
$v_{\rm entr}$ & 1 -- 30 km\,s$^{-1}$\\
\hline
\end{tabular}
\end{center}
\end{table}

For example, direct evidence of jet rotation remains somewhat controversial since only recently has observational resolution been sufficient (on the order of 10 AU; \eg\ \citealp{coffeyetal08,bjerkelietal16,leeetal2017_hh212}) to yield reliable radial profiles across the jet.  \citet{woitasetal05} report line-of-sight velocity gradients which they interpret as rotation, though \citet{soker05} suggest this might indicate an interaction of the jet with a warped disc, while \citet{fendt11} suggests MHD shocks in a helical field.  Still, it is widely believed that protostellar jets must rotate if they are to succeed in their presumed task of ridding the protostar of its angular momentum.

\begin{table}
\begin{center}
\caption{\label{tab:numparms} Estimates made from local simulations of the observational parameters listed in Table \ref{tab:obsparms}, taken from \citet{op97a, ustyugovaetal99, andersonetal05, fendt09, staffetal10, sheikhnezamietal12, stepanovsfendt14}.}
\begin{tabular}{r|l}
\hline
$v_{\rm jet}$ & 35 -- 1,300 km\,s$^{-1}$\\
$\langle v_{\rm rot}\rangle$ & $<50$ km\,s$^{-1}$\\
$\dot M$ & $10^{-8}$ -- $10^{-5}\,M_{\odot}$\,yr$^{-1}$\\
$\dot p$ & --- \\
$\dot L$ & $3\times 10^{-8}$ -- $9\times 10^{-6} M_{\odot}$\,yr$^{-1}$\,AU\,km\,s$^{-1}$\\
$v_{\rm entr}$ & ---\\
\hline
\end{tabular}
\end{center}
\end{table}

An important physical quantity missing from Table \ref{tab:obsparms} is the magnetic field strength.  Direct measurements of $\mathbfit{B}$ in a protostellar jet remain elusive, with just two indirect measures reported to date \citep{rayetal97, carrascogonzalezetal10} which, almost by definition, represent extreme cases.  Still, the theoretical evidence for magnetic fields pervading protostellar jets is overwhelming, and it is nearly universally accepted as being a critical ingredient to jet dynamics \citep{hartiganetal07}.  Certainly, strong fields are known to exist within the inner regions of protostellar discs ($\sim\!1$\,kG; \citealp{donatietal05}), and it is difficult to imagine how this is not transported outward by the jet.

To a large extent, the base of a jet can be characterised by the presence of strong gravitational and magnetic fields, and rapid rotation.  In such an environment, \citet{bp82}, based on the `bead-on-a-wire' model first suggested by \citet{henriksenrayburn71}, showed that the formation of a super-fast jet is virtually \emph{inevitable}.  In their model, a Keplerian disc is threaded with `frozen in' vertical magnetic flux.  As the disc rotates, magnetic field lines are twisted and, once the angle at which they emerge from the disc falls below the critical value of $60\dgr$, the centrifugal force overwhelms gravity and drives material outward as `beads sliding along a rotating wire'.

This model, also known as `magneto-centrifugal driving', has precipitated a plethora of numerical simulations to investigate its consequences.  Since a 2-D axisymmetric, ideal Keplerian disc is unstable to the magneto-rotational instability (MRI, \citealp{balbushawley92} and references therein; only in 3-D is the instability saturated, \citealp{stoneetal96}), most simulations of magneto-centrifugally launched jets (\eg\ \citealp{uchidashibata85, ustyugovaetal95, ustyugovaetal99, meieretal97, op97a, op97b, op99, klb99, krasnopolskyetal03, fendtcemeljic02, vitorinoetal02, vonrekowskietal03, ocp03, andersonetal05, andersonetal06, porthfendt10, stuteetal14, tesileanuetal14, staffetal10, staffetal15}) treat the accretion disc as a boundary condition, allowing the jet dynamics to be studied independently of the disc.

There are a number of 2-D studies which do include the disc as part of the simulations, even if in a somewhat idealised fashion (\eg\ \citealp{cassekeppens02, cassekeppens04, zannietal07, tzeferacosetal09, murphyetal10, sheikhnezamietal12, fendtsheikhnezami13, stepanovsfendt14, stepanovsfendt16, surianoetal2017, zhustone2017, bai2017, surianoetal2018}).  These simulations typically use a magnetic resistivity to prevent excessive disc turbulence, and are more realistic by including the disc evolution self-consistently.  However, they are much more expensive computationally because of the significantly shorter physical time scales in the disc and it is because of this we have chosen here to treat the disc as a boundary condition.

Because most of the magneto-centrifugal `action' occurs near the inner radius of the disc, simulations must be performed at a resolution of 0.01 AU or less in order to resolve the important physics there. Thus, even the most ambitious of the works listed above have followed the jet to just 100 AU (\citealt{andersonetal05}), and more recently to 150 AU \citep{stepanovsfendt14,stepanovsfendt16}, above the disc.  Accordingly, we refer to these simulations collectively as `local' simulations.

There also exists a class of `global' simulations which follow the gravitational collapse of isolated, magnetised molecular cloud cores including the formation of the protostar and accretion disc (\eg\ \citealp{seifriedetal11, seifriedetal12, tomidaetal13, tomidaetal15, massonetal2016, kolligankuiper2018}).  While these simulations include, by design, observational length scales, due to the extreme computational costs involved, they cannot include the sub-0.01 AU scales necessary to suitably resolve the physics of the jet launching mechanism for any substantial length of time.

Notably, the large scale difference between local simulations and observed jets ($10^3$--$10^6$ AU; \eg\ \citealt{devineetal1997_hh34,asoetal15}) makes direct comparisons impossible, and to make any comparison at all one must make severe assumptions on how local variables relate to global properties of the jet.  As an example, most local simulations continue their calculations long after the leading bow shock or Alfv\'en wave has left the grid.  To say nothing of the change to the (thermo)dynamics of the jet that the sudden loss of a confining bow shock must cause, a proxy for $v_{\rm jet}$ must be used.  Typically, this is the speed at the Alfv\'en point ($v_{\rm A}$) or, if it is still in the domain, the fast magnetosonic point ($v_{\rm f}$).  In the only simulation performed to date where the jet launching conditions are controlled and the leading bow shock remains within the computational domain \citep{rc11}, we find that the jet continues to accelerate well beyond the fast point, and thus $v_{\rm A}$ and $v_{\rm f}$ are poor proxies for the final $v_{\rm jet}$.

Table \ref{tab:numparms} summarises estimates of the parameters in Table \ref{tab:obsparms} made from the local simulations cited in the caption. Because of the assumptions and extrapolations inherent in these estimates, we offer them only as an `order-of-magnitude' check with the observations.  Notably, we are not aware of any estimates of $\dot p$ that can be gleaned from local simulations.

This work is a continuation of \citet{rc11}.  Here, we present eight 2.5-D axisymmetric \emph{global} simulations in which the jet is followed from its launching point with $0.00625$ AU resolution to a length of up to 4,000 AU, well into the observational regime.  Even still, this represents only about 1\% of the age and length of the largest jets from class 0/I young stellar objects which, as we will see, puts some limitations on what can be inferred.

As `immature' as our simulations may be, they are still global in nature (both resolving the region where the jet is launched, and following the jet to observational scales), and imply a dynamic range in length scale of $\sim 6.5\times 10^5$ for our most highly resolved simulation.  A single-grid (4,$096\times 256$ AU) 2-D MHD simulation with a resolution of $0.00625$ AU would require $>100$ billion zones over 100 million time steps to complete.  Thus, to perform these simulations, we have used the adaptive mesh refinement (AMR) MHD code, \azeus\ \citep{rcm12}.  Each simulation differs from the others only in the strength of the magnetic field at the inner radius of the disc, $B\rmi$, which is used to scale an initially force-free, global `hour-glass' magnetic field distribution.  The central gravitating mass ($0.5\,{\rm M}_\odot$), as well as the parameters governing the Keplerian disc (initially in gravito-centrifugal balance) and the coronal atmosphere (initially in hydrostatic balance) are the same for all simulations.  The purpose of this study is to determine if this is sufficient to produce a jet with the right observational characteristics, and further, what role if any $B\rmi$ has on determining observational and physical properties of the jet.  Neither of these fundamental questions can be answered by local simulations.

Finally, a comment on the choice of axisymmetry is in order.  Even with AMR and distributing the calculations over 16--24 CPU cores, some of the simulations discussed herein required more than six months to complete, and a fully 3-D treatment was simply impractical.  Buoyed by the knowledge that many stellar jets appear axisymmetric \citep[e.g.\ HH\,34;][]{devineetal1997_hh34}, a 2-D axisymmetric approach was adopted at the outset of this project.

Still, the cost in realism is undeniable.  Even in systems with a high degree of \emph{apparent} axisymmetry, the fluid is subject to all modes of Kelvin-Helmholtz (K-H) instabilities \citep[e.g.][]{hardeeclarke95} on both the large- and small-scale, which take their toll as the jet propagates.  On the large scale, \citet{clarke1993} showed that the otherwise perfectly stable nose-cone found in a 2-D axisymmetric magnetically-confined jet was periodically sloughed off to the side in 3-D, resulting in a blunter, more slowly propagating jet.  On the smaller scale, \citep{clarke1996} found that, in simulations of a propagating jet with a weak magnetic field, the 3-D jet once again propagated more slowly and formed a blunter bow shock.  In this case, the numerous modes of K-H instabilities in 3-D---all but one unavailable in 2-D axisymmetry---effectively converts directed kinetic energy of outflow to turbulent and ultimately thermal energy in the expanding cocoon.  Such considerations should therefore be borne in mind as discussion of the present simulations unfolds.

In Sect.\ \ref{sec:steadystate}, we review some of the relevant steady state theory which applies to portions of our numerical solutions.  In Sect.\ \ref{sec:numerics}, we describe briefly the numerical methodology and how the simulations are initialised.  Sections \ref{sec:simulations} and \ref{sec:analysis} comprise the bulk of the paper in which the simulations are described and analysed in detail.  Finally, conclusions are drawn in Sect.\ \ref{sec:conclusions}.

\section{Steady state analysis}
\label{sec:steadystate}
Analogous to Bernoulli's constant for hydrodynamics, in a steady state ($\partial_t=0$)\footnote{Throughout this paper, we use the `abbreviated Leibniz notation' for derivatives.  Thus, $\partial_t\equiv\partial/\partial t$.}, axisymmetric ($\partial_\varphi=0$), ideal MHD fluid, there are four conserved quantities along a given magnetic field line, which themselves are contours of the flux function\footnote{$\psi=rA_\varphi$, where $A_\varphi$ is the toroidal component of the vector potential.}, $\psi$.  In Gaussian cgs units, these are \citep{weberdavis67,mestel68,pudritznorman1983,pp92,spruit96}:
\begin{align}
  \label{eq:massload}
  \eta(\psi) &= \frac{\rho v\pol}{B\pol} = M_{\rm A}\sqrt{\frac{\rho}{4\pi}};\\
  \label{eq:angmom}
  l(\psi) &= r \left(v_\varphi - \frac{a_\varphi}{M_{\rm A}}\right);\\
  \label{eq:rotspeed}
  \Omega(\psi) &= \frac{1}{r}\left(v_\varphi - M_{\rm A}a_\varphi\right);\\
  \label{eq:energy}
  \varepsilon(\psi) &= \frac{v\pol^2}{2} - \frac{v_\varphi^2}{2} + M_{\rm A}v_\varphi a_\varphi + \frac{c_{\rm s}^2}{\gamma - 1} + \phi,
\end{align}
where $\eta$ is the mass load, $l$ and $\Omega$ are, respectively, the specific angular momentum and angular speed of the field line (including terms describing the magnetic torque)\footnote{$\Omega(\psi)$ is sometimes referred to as the iso-rotation parameter \citep[e.g.][]{fendtmemola2001,porthetal2011}.}, and $\varepsilon$ is the specific energy of the fluid.  In axisymmetry, $\psi$ describes surfaces of constant magnetic flux, and eqs. (\ref{eq:massload})--(\ref{eq:energy}) are therefore also constant on these flux surfaces.  Henceforth, we refer to these as the `Weber-Davis (WD) constants'. Note that Eq.\ (\ref{eq:energy}) is essentially Bernoulli's constant generalised for MHD.  Other variables include the density, $\rho$, the poloidal velocity, $v\pol$, the toroidal velocity, $v_{\varphi}$, the poloidal magnetic field, $B\pol$, the Alfv\'en Mach number, $M_{\rm A}=v\pol/a\pol$, the poloidal Alfv\'en speed, $a\pol=B\pol/\!\sqrt{4\pi\rho}$, the toroidal Alfv\'en speed, $a_\varphi=B_\varphi/\!\sqrt{4\pi\rho}$, the toroidal magnetic field, $B_\varphi$, the adiabatic sound speed, $c_{\rm s}=\!\sqrt{\gamma p/\rho}$, the ratio of specific heats, $\gamma$, and the gravitational potential of the protostar, $\phi=-GM_*/R$. Here, $R=\!\sqrt{z^2+r^2}$, is the spherical polar radial coordinate, $z$, $r$ and $\varphi$ are the cylindrical coordinates\footnote{Note the difference here between $\phi$, the gravitational potential, and $\varphi$, the cylindrical coordinate.}.  Equations (\ref{eq:massload})--(\ref{eq:energy}) are a straight-forward extension from Eqs.\ (12, 13, 19, and 31) in \citet{spruit96}, where we set $f^\prime=\Omega$.

As an example, by definition, the mass flux (${\cal F}_\rho=\rho v\pol \delta A$) is conserved along a streamtube of cross section $\delta A$ and the magnetic flux (${\cal F}_B=B\pol\delta A$) is conserved along a magnetic flux tube.  In the steady state when $v\pol\parallel B\pol$, streamlines are everywhere parallel to magnetic flux lines, and $\eta(\psi)={\cal F}_\rho/{\cal F}_B$ is constant along a field line.  Arguments establishing the constancy of $l$, $\Omega$, and $\varepsilon$ along field lines in the steady state follow similar lines.

\begin{figure}
  \begin{center}
    \includegraphics[width=0.80\columnwidth,clip=true]{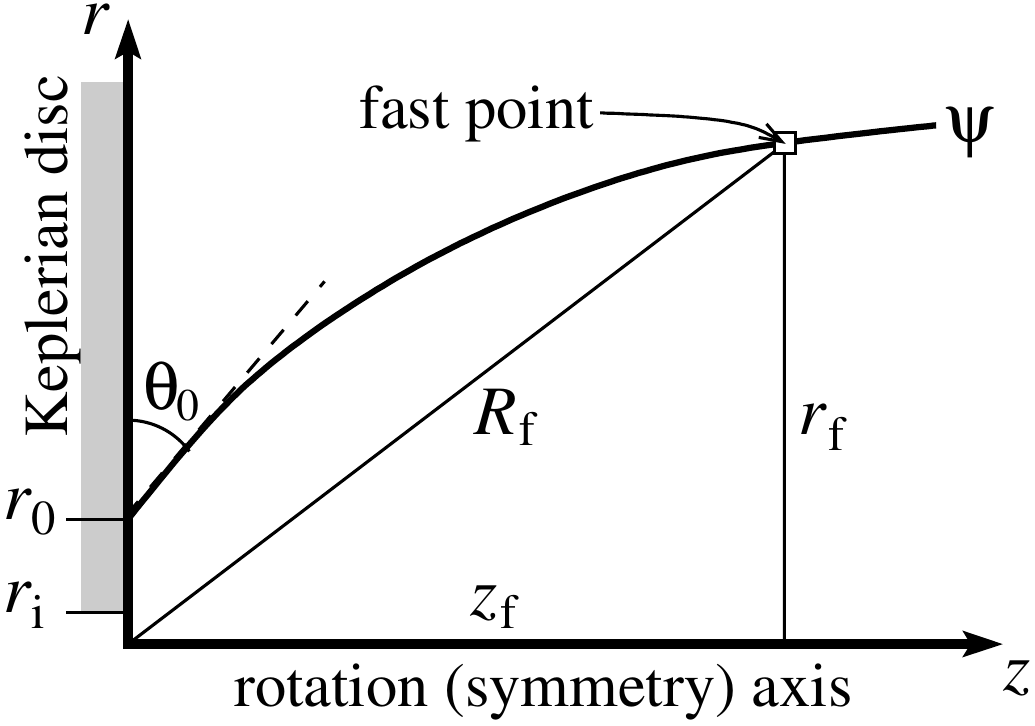}
  \end{center}
  \caption{\label{fig:anchor} A magnetic field line, $\psi$, anchored in the accretion disc at distance $r_0$ from the rotation axis emerging from the disc at an angle $\theta_0$.  Other quantities are defined in the text.}
\end{figure}

Following \citet{spruit96}, if $r_0(\psi)$ is the radial coordinate of the disc where a particular field line, $\psi$, is anchored (Fig.\ \ref{fig:anchor}), one can evaluate $\Omega(\psi)$ and ${\varepsilon}(\psi)$ at the anchor point.  At $(z,r)=(0,r_0)$, $v\pol\sim0\,\Rightarrow\,M_{\rm A}\sim0$, $v_\varphi=v_{\rm K,0}=\sqrt{GM_*/r_0}$ (assuming the disc is Keplerian), $\phi(r_0)=-v_{\rm K,0}^2$, and Eqs.\ (\ref{eq:rotspeed}) and (\ref{eq:energy}) reduce to:
\begin{equation}
\label{eq:omegaE}
\Omega(\psi)=\frac{v_{\rm K,0}}{r_0} \qquad {\rm and}\qquad \varepsilon(\psi)=-\frac{3v_{\rm K,0}^2}{2},
\end{equation}
assuming a cold fluid ($\beta\ll 1; c_{\rm s}\sim0$).  Conversely, $\eta$ and $l$ are most conveniently evaluated at the Alfv\'en point ($M_{\rm A}=1$) where:
\begin{equation}
  \eta(\psi)=\sqrt{\frac{\rho_{\rm A}}{4\pi}} \qquad {\rm and} \qquad l(\psi)=r_{\rm A}^2\Omega(\psi) = v_{\rm K,0}\frac{r_{\rm A}^2}{r_0}.
  \label{eq:etaL}
\end{equation}

Local simulations treat the leading Alfv\'en torsional wave and bow shock as transients and, irrespective of the dynamical consequences, allow them to leave the computational domain.  Thereafter, most local simulations reach some sort of steady state from which various comparisons with analytical theory are made. For example, being in a near-steady state, $\eta(\psi)$ from Eq.\ (\ref{eq:massload}) is expected to be constant, and thus many investigations use $\eta(\psi)$ as a parameter to specify the nature of the outflow (\eg\ \citealt{op99,andersonetal05}); if $\rho v\pol$ and $B\pol$ are changed in proportion to each other in a steady state jet, then the character of the outflow should remain unaltered.

Local simulations reaching a steady state typically show that the jet speed saturates at or just beyond the fast point, assuming that this point remains inside the grid.  Those that report on asymptotic jet speeds find $v_{\rm p,max}\sim \eta^\alpha$, with the majority finding $\alpha<0$ (\ie\ flow speed increases as the field strength increases or the poloidal momentum at the disc decreases; \eg\ \citealt{andersonetal05,zannietal07,porthfendt10}), as is expected from steady state theory \citep{spruit96}\footnote{A notable exception is \citet{op99}, who find the opposite trend.}. Indeed, it should come as no surprise that local simulations confirm various aspects of steady state theory, since the assumptions of no transients is common to both.  As soon as $v\pol\parallel B\pol$ is realised over much of the computational domain, the conclusions of steady state theory become inescapable.

Since the leading bow shock and Alfv\'en torsion wave never leave the grid in our simulations, none reach a global steady state.  However, regions near the disc of some simulations (more so for stronger $B\rmi$) do reach (locally) a quasi-steady state (as confirmed by the constancy of $\eta$, $l$, $\Omega$, and $\varepsilon$ along field lines), and we exploit this observation in some of the analysis.  Under the assumption that portions of the jet are in steady state, we can determine how the flow speed at the fast point, for example, varies with $B\rmi$, and then attempt to relate this to $v_{\rm jet}$.

To this end, from Eqs.\ (\ref{eq:rotspeed}), (\ref{eq:energy}), and (\ref{eq:omegaE}), we can write:
\begin{equation}
\label{eq:junk1}
2\varepsilon - 2\phi + r^2\Omega^2=v_{\rm K,0}^2 \left(-3 + 2\frac{r_0}{R} + \frac{r^2}{r_0^2}\right) = v_p^2 + M_{\rm A}^2a_\varphi^2 = \frac{v\pol ^2a^2}{a\pol^2},
\end{equation}
where $a^2=a_\varphi^2+a\pol^2=B^2/(4\pi\rho)$ is the fast speed squared when $c_{\rm s}^2=0$ (cold flow).  Thus, at the fast point where $(z,r)=(z_{\rm f},r_{\rm f})$, $a\pol=a_{\rm p,f}=B_{\rm p,f}/\!\sqrt{4\pi\rho_{\rm f}}$, and $v\pol=a= v_{\rm p,f}$, Eq.\ (\ref{eq:junk1}) becomes:
\begin{equation}
\label{eq:fastpoint}
v_{\rm p,f}=\sqrt{v_{\rm K,0} B_{\rm p,f}}\left[\frac1{4\pi\rho_{\rm f}}\left(\frac{r_{\rm f}^2}{r_0^2}+\frac{2r_0}{R_{\rm f}}-3\right)\right]^{1/4},
\end{equation}
where $r_{\rm f}$ and $R_{\rm f}$ are, respectively, the cylindrical and spherical polar radial coordinates to the fast point, as shown in Fig.\ \ref{fig:anchor}.  We have verified this formula directly from our simulations, and find agreement to better than 3\% along the field line anchored at $r_0=1$\,AU with measures from other field lines in steady state giving similar results (\eg, Table \ref{tab:fastpoint} on page \pageref{tab:fastpoint}).

If we choose the same field line foot print, $r_0$, for each simulation, $v_{\rm K,0}$ becomes a constant and, in as much as the quantity in square brackets to the 1/$4^{\rm th}$ power in Eq.\ (\ref{eq:fastpoint}) depends weakly on $B\rmi$, we might expect:
\begin{equation}
\label{eq:vpfofBi}
v_{\rm p,f}\sim\sqrt{B_{\rm p,f}}\sim B\rmi^{1/2},
\end{equation}
since the magnetic field profile scales with $B\rmi$.  This can be contrasted with the asymptotic flow speed predicted for steady state flow and a purely radial magnetic field (\eg\ \citealt{spruit96}; Eq.\ 74):
\begin{equation}
\label{eq:vpmax}
  v_{\rm p,max} = \left( \frac{\Omega^2(\psi) r_0^2B_{\rm p}(r_0)}{4\pi \eta(\psi)} \right)^{1/3} \sim B\rmi^{2/3},
\end{equation}
since $\eta(\psi)\sim B_{\rm p}^{-1}$.  We return to these predicted dependencies on $B\rmi$ in Sect.\ \ref{sub:steadystate}.

\section{Numerical considerations}
\label{sec:numerics}
\subsection{\azeus}
\label{sub:azeus}
The simulations presented herein are performed with the adaptive mesh refinement (AMR) MHD code, \azeus (Adaptive Zone Eulerian Scheme; \citealt{rcm12}; \url{http://people.virginia.edu/~jpr8yu/azeus}), based on Version 3.6 of \zeusddd (\citealp {clarke96, clarke10}; \url{http://www.ica.smu.ca/zeus3d}).  The \zeus family of codes is among the best tested, documented, and most widely used astrophysical MHD codes available.  Our version allows one to choose to solve the internal energy or total energy equations, the latter being conservative in energy to machine round-off error. As the simulations presented here are only mildly super-magnetosonic ($M_{\rm f}<8$; Table \ref{tab:simsum2}), the internal energy equation does an adequate job of conserving energy, while guaranteeing a positive-definite pressure, which is of greater importance here than strict energy conservation.

Like \zeusddd, \azeus solves the ideal equations of MHD on a fully staggered mesh (zone-centred scalars, face-centred vector components) in an operator split fashion (source terms computed separately from fluxes), using directional splitting for compressive terms (scalar transport, pressure gradient, transport of the $i^{\rm th}$ component of momentum in the $i$-direction), and planar splitting for transverse terms (magnetic induction, transverse Lorentz forces, transport of the $i^{\rm th}$ component of momentum in the $j$-direction, $i\neq j$).  \azeus is upwinded in the entropy and Alfv\'en waves and relies on a modest amount of \citet{vonneumannrichtmyer50} artificial viscosity to stabilise compressive (fast and slow magnetosonic) waves.  Interpolations are performed using the second order, monotonised scheme of \citet{vanleer77} and, for transverse terms, interpolations are performed implicitly in each plane using the Consistent Method of Characteristics (CMoC; \citealp{clarke96}).

As for the AMR module, we have adapted the block-based method of \citet{bc89} and \citet{belletal94} for the staggered mesh of \azeus.  Significant effort was spent minimising errors caused by waves passing across grid boundaries, which is of particular importance to this work.  This includes the development and implementation of third-order interpolation schemes in which mass, momentum, and energy are conserved to machine round-off error.  Prolongation of magnetic field is done using a method based on \citet{lili04}, ensuring the validity of the solenoidal condition to machine round-off error regardless of how various 2-D meshes abut, overlap, and overlay each other.  Indeed, we find it critical for the solenoidal condition to be valid to machine round-off error even within the boundaries. The interested reader is referred to \citet{rcm12} for details.

All simulations are initialised with nine static, nested grids (including the base grid) with a refinement ratio $\nu=2$.  Table \ref{tab:grids} gives the extent (in AU) of each of the 2-D grids ($z_{\rm max}$ and $r_{\rm max}$) excluding the boundary regions, along with their resolution, $\Delta$, in each of the $z$- and $r$-directions.  Thus, level 1---the coarsest `base' grid---is resolved with $2,\!548 \times 160$ zones (including 2 boundary zones at each edge), while each of levels 2--9 are resolved with $640 \times 160$ zones.

\begin{table}
  \begin{center}
    \caption{\label{tab:grids} Initial static grids used in all simulations (refinement ratio $\nu=2$).  All zones are square, and thus $\Delta z = \Delta r=\Delta$. The near powers-of-two for the grid dimensions is a consequence of requiring the number of zones---including boundary zones---in each dimension of grids 1--8 to be a multiple of the number of OpenMP threads used (typically 16).}
    \begin{tabular}{crrl}
      \hline\hline
      Level & $z_{\max}$ (AU) & $r_{\max}$ (AU) & $\Delta$ (AU)\\
      \hline
      1 & 4070.4~~~~ & 249.6~~~~~ & 1.6 \\
      2 &  508.8~~~~ & 124.8~~~~~ & 0.8 \\
      3 &  254.4~~~~ &  62.4~~~~~ & 0.4 \\
      4 &  127.2~~~~ &  31.2~~~~~ & 0.2 \\
      5 &   63.6~~~~ &  15.6~~~~~ & 0.1 \\
      6 &   31.8~~~~ &   7.8~~~~~ & 0.05 \\
      7 &   15.9~~~~ &   3.9~~~~~ & 0.025 \\
      8 &    7.95~~~~ &   1.95~~~~~ & 0.0125 \\
      9 &    3.975~~~~ &  0.975~~~~~ & 0.00625 \\
      \hline
    \end{tabular}
  \end{center}
\end{table}

In addition, smaller grids are added and removed dynamically based on how well the radial gradient of $B_\varphi$ is resolved near the symmetry axis. By definition, in axisymmetry both $v_\varphi$ and $B_\varphi$ should be zero on axis.  In our simulations, we find that while $v_\varphi$ obliges, $B_\varphi$ does not always. Specifically, as a jet propagates, a hot, low-velocity `spine' of strong helical field develops along the symmetry axis.  With insufficient resolution, the decline of $B_\varphi$ from its maximum value off-axis to zero on-axis is buried within a single zone, creating an `inverted profile' for $B_\varphi$, one whose magnitude \emph{declines} away from the symmetry axis.  This generates an axial current density, $J_z\propto\partial_r(rB_\varphi)$ of \emph{opposite} sign to $B_\varphi$ (physically, in this situation, $J_z$ and $B_\varphi$ should have the \emph{same} sign) which, in turn, exerts a Lorentz force, $F_r\propto -J_zB_\varphi$, directed radially \emph{outward} (instead of inward).  Left unchecked, these unphysical forces occasionally trigger rather dramatic `numerical explosions', sending vast bubbles of hot, rarefied gas expanding into the solution.  As interesting as these events are to watch, they are completely numerical in origin and destroy the integrity of the simulation.

We have therefore imposed a `Lorentz criterion' in which a level of refinement is added in any region near the axis where the gradient in $B_\varphi$ is insufficiently resolved.  Specifically, we require the unitless gradient:
\begin{equation*}
  \partial_r B_\varphi \frac{\Delta}{B} > \frac{1}{N},
\end{equation*}
where $\Delta$ is the zone size, $B=\sqrt{B\pol^2+B_\varphi^2}$ is the local magnetic field strength, and $N=6$ is the minimum number of zones we require to resolve the radial profile of $B_\varphi$.  To avoid `mesh trashing' \citep{khokhlov98}, the threshold for removing a grid is $2N$.  In practise, we must also guard against `frivolous' inverted $B_\varphi$ profiles.  Frequently, noisy and dynamically inactive values of $B_\varphi$ can create inverted profiles near the symmetry axis and such occurrences should not trigger the insertion of a new grid.  In these simulations, we do not go beyond refinement level 9.

\subsection{Initial conditions}
\label{sub:initial}
Young protostellar discs can extend for hundreds of AU, but have inner radii, $r\rmi$, of 3--5 stellar radii, $R_*$ \citep{calvetetal00_ppiv}. For a typical T Tauri star ($M_*=0.5\,M_{\odot}$), $R_*=2.5R_{\odot}$.  Thus, we adopt $r\rmi = 0.05$ AU and use this as our length scale.  Our finest static grid (level 9; Table \ref{tab:grids}) resolves $r\rmi$ with eight zones which, based on test simulations of different resolutions, is sufficient for numerical convergence with respect to the physics of the jet launching mechanism.

\subsubsection{The atmosphere}
\label{subsub:atmos}
The atmosphere is initialised in hydrostatic equilibrium (HSE),
\begin{equation}
  \label{eq:hseeqn}
  \nabla p+\rho\nabla\phi=0,
\end{equation}
where $\phi$ is the gravitational potential of $M_*$.  Since the second term is not a perfect gradient, differencing it directly on a staggered-mesh commits sufficient truncation error to render the atmosphere numerically unstable.  \citet{op97a,op99}, and continuing with \cite{staffetal10,staffetal15}, address this problem by assuming a strict polytropic equation of state, $p=\kappa\rho^\gamma$ where $\kappa$ is constant throughout the grid even across shocks.  While this has the advantage of allowing the $\rho\nabla\phi$ term to be written as a perfect gradient which eliminates the numerical truncation error and stabilises the atmosphere, it also replaces energy with entropy as the primary conserved variable.  This has the unintended consequences of forbidding the formation of contact/tangential discontinuities and generating isentropic shocks, which we find adversely affects the global solution.  Thus, in all of our work, we have retained the adiabatic equation of state ($p\propto\rho^\gamma$, but where the proportionality constant remains a function of entropy) to allow the correct entropy jump across shocks and the spontaneous formation of contact discontinuities with their required discontinuities in entropy.

This leaves, however, the numerical instability of the HSE atmosphere unsettled.  We address this problem by replacing $\nabla\phi$ in Eq.\ (\ref{eq:hseeqn}) with the corresponding poloidal gravitational acceleration vector,
\begin{equation}
\label{eq:hseaccel}
  \mathbfit{g}=-\frac{1}{\rho_{\rm h}}\nabla{p_{\rm h}},
\end{equation}
where $\rho_{\rm h}$ and $p_{\rm h}$ are the hydrostatic density and pressure:
\begin{equation}
  \label{eq:hserhop}
  \rho_{\rm h}=\rho\rmi\left(\frac{r\rmi}{\sqrt{r^2+z^2}}\right)^{\frac{1}{\gamma-1}}\qandq~~p_{\rm h}=p\rmi\left(\frac{\rho_{\rm h}}{\rho\rmi}\right)^{\gamma}.
\end{equation}
Here, $\rho\rmi$ and $p\rmi$ are the initial density and pressure\footnote{The factor of $1/\gamma$ appearing in Eq.\ (3) of \cite{rc11} is in error.} at $r=r\rmi$, and $p\propto\rho^\gamma$ ($\gamma=5/3$) is assumed throughout the atmosphere at $t=0$.  In this way, differencing Eq.\ (\ref{eq:hseeqn}) maintains HSE to machine round-off error \emph{indefinitely}.  While it is true that $\mathbfit{g}$ determined from Eq.\ (\ref{eq:hseaccel}) is not numerically irrotational (no scalar function $\phi$ exists such that $\mathbfit{g}=-\nabla\phi$ to machine round-off error), this turns out to be an unnecessary requirement on $\mathbfit g$.

Still, Eq.\ (\ref{eq:hseaccel}) alone is insufficient to guarantee the numerical integrity of the atmosphere. Regardless of resolution, the singular nature of Eqs.\ (\ref{eq:hserhop}) generates sufficient truncation errors at the origin to produce a steady, outwardly directed pressure gradient that drives a supersonic, narrow jet along the symmetry axis, destroying the integrity of the solution.  This numerical effect is fixed by replacing the point mass at the origin with a uniform sphere of the same mass and a radius $R_{\rm sph}$, thus modifying the first of Eqs.\ (\ref{eq:hserhop}) to:
\begin{equation}
\label{eq:hsesmooth}
\rho_{\rm h}= \rho\rmi\begin{cases} \left(\dfrac{r\rmi}{\sqrt{r^2+z^2}}\right)^{\frac{1}{\gamma-1}},&r^2+z^2\geq{R_{\rm sph}}^2;\\[12pt]
\left(\dfrac{r\rmi}{R_{\rm sph}}\,\dfrac{3{R_{\rm sph}}^2-r^2-z^2}{2{R_{\rm sph}}^2}\right)^{\frac{1}{\gamma-1}},&r^2+z^2<{R_{\rm sph}}^2.\end{cases}
\end{equation}
If $R_{\rm sph}$ is sufficiently resolved (\eg\ four zones), the numerical jet is eliminated.  The resulting `rounded potential' is superior to a `softened potential' since the former has no measurable effects beyond $R_{\rm sph}$. Here, we use $R_{\rm sph}=r\rmi$.

As a final comment on the initialisation of pressure, unlike, e.g., \citet{op97a,op97b,op99}, we find no need to attribute a portion of the thermal pressure to Alfv\'enic turbulent pressure.  This was used as a mechanism to reduce what \citeauthor{op97a} felt was an unrealistically high temperature in their outflow.  In local simulations, as soon as the confining bow shock leaves the grid, expansion of jet material is free and thus isothermal and the temperature of the fluid does not fall.  In our case, the jet expansion is always confined and the gas continuously does $P\,dV$ work to expand thereby reducing the temperature in the inner portions of the jet.  Indeed, without additional physics such as radiative heating, the temperature within the inner portions of our jets may be unrealistically \emph{low}, and certainly we have no need to assign a portion of the thermal pressure to `Alfv\'enic turbulence'.

Similar to \citet{op97a}, the atmosphere is initialised with a force-free `hour-glass' magnetic field distribution.  The toroidal component of the magnetic field, $B_\varphi$, is initially zero whereas the poloidal components are set by specifying the toroidal component of the vector potential:
\begin{equation}
  \label{eq:aphi}
A_\varphi=A_0\frac{\sqrt{r^2+(z+z_{\rm d})^2}-(z+z_{\rm d})}{r},
\end{equation}
where $z_{\rm d}$ is the `disc thickness' (which we set to $r\rmi$ for convenience) and where, on \azeus' staggered mesh, $A_\varphi$ is edge-centred.  The $r$- and $z$-components of the magnetic field are then given by:
\begin{equation}
\label{eq:binit}
B_z=\frac{1}{r}\partial_r(rA_\varphi)\qandq B_r=-\partial_zA_\varphi,
\end{equation}
which, when differenced, locates the poloidal magnetic field components at the face-centres and ensures $\nabla\cdot\mathbfit{B}=0$ to machine round-off error.  Thus, at $(z,r)=(0,r\rmi)$ where $B\pol=B\rmi$,
\begin{equation*}
A_0=\frac{B\rmi r\rmi}{\sqrt{2-\sqrt{2}}},
\end{equation*}
where, in terms of the initial plasma beta at $(0,r\rmi)$, $\beta\rmi$, used to characterise the simulations (\eg\ Table \ref{tab:simsum}),
\begin{equation}
\label{eq:betai}
  B\rmi=\sqrt{\frac{8\pi p\rmi}{\beta\rmi}}.
\end{equation}

We have adopted an `hour-glass' initial magnetic field distribution, in part, because of its simplicity and the availability of a closed analytical form, but we do acknowledge that the true magnetic field distribution in protostellar systems is generally unknown, and indeed will vary from system to system due to environmental differences.  Other authors have studied the effect of different field distributions on outflow launching, and find that it significantly affects the collimation of the outflow \citep[e.g.][]{pro06,fendt2006}.  While this aspect is worth exploring, for the current effort, we choose instead to focus only on varying the initial magnetic field strength.

Finally, to ensure the declining density and magnetic field profiles do not fall below observational limits, we add floor values $\rho_{\rm floor}\sim10^{-6}\rho\rmi$ and $B_{z,{\rm floor}}\sim10^{-5}B\rmi$ (\cf\ \citealt{bergintafalla07}, \citealt{vallee03}) to Eqs.\ (\ref{eq:hsesmooth}) and (\ref{eq:binit}).  Thus, the atmosphere attains its asymptotic values by $z\sim500\,$AU.  By imposing HSE and the adiabatic gas law at $t=0$, a floor value on $\rho$ also imposes effective floor values on $\mathbfit{g}$, $p$, $T$, \etc

\begin{figure}
  \begin{center}
    \includegraphics[width=0.90\columnwidth,clip=true]{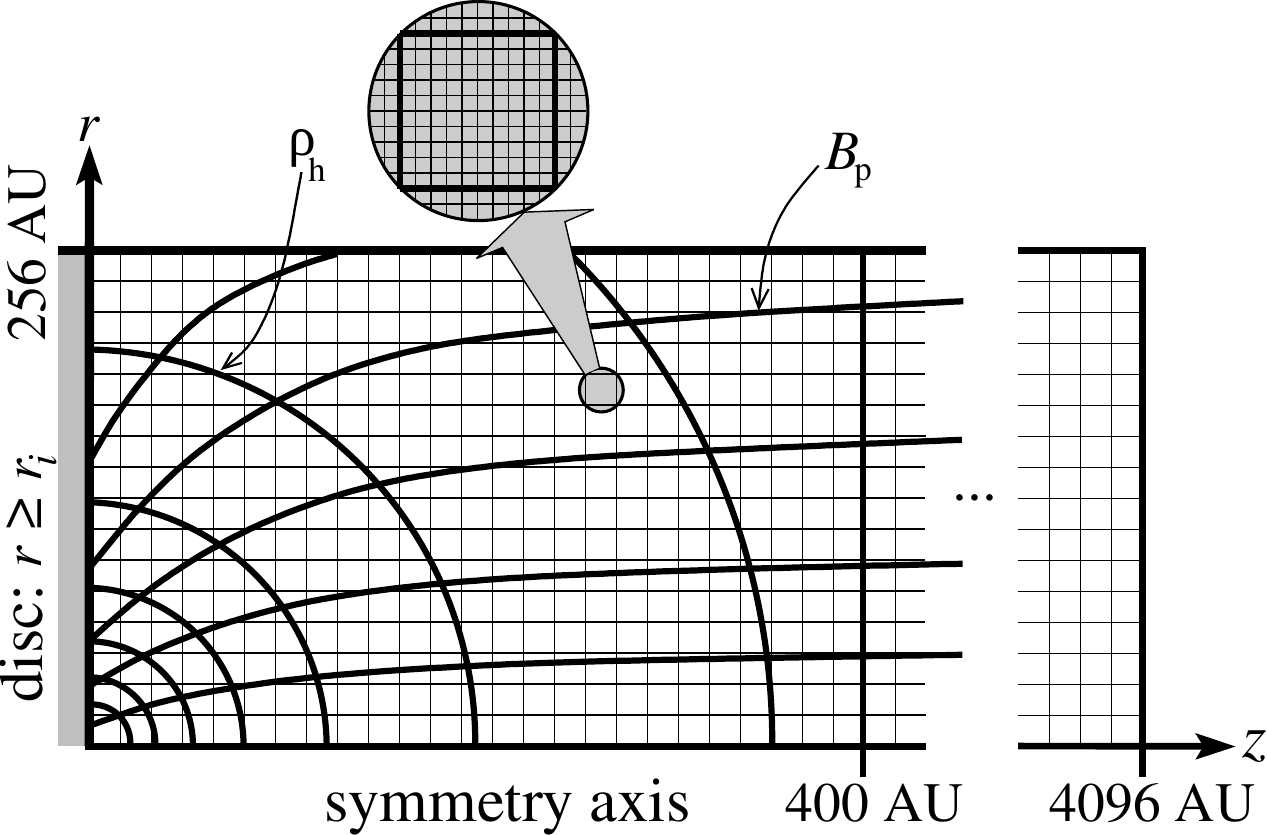}
  \end{center}
  \caption{\label{fig:init} A schematic representation of the base (coarsest) grid (level 1), showing representative contours of the initial density, $\rho_{\rm h}$, field lines of the initial force-free `hour-glass' poloidal magnetic field, $B\pol$, and the grid.  As indicated by the inset, the actual grid is ten times finer than shown.}
\end{figure}

A schematic of the base (coarsest) grid is shown in Fig.\ \ref{fig:init}, where contours of $\rho_{\rm h}$, the `hour-glass' $B\pol$, and every tenth grid line are plotted.

\subsubsection{The accretion disc, and other boundary conditions}
\label{subsub:bcs}
The accretion disc, maintained in $z \leq 0$ as a boundary condition, is initially assumed to be in gravito-centrifugal balance (\ie\ Keplerian), and to have a force-free magnetic field. Thus, for $z\leq 0$ and $r\geq r\rmi$, $v_\varphi=v_{\rm K}=\sqrt{GM_*/r}$.   We also assume a gentle `evaporation speed', $v_z=\zeta v_{\rm K}$ with $\zeta=10^{-3}$ to transfer mass from the disc surface to the atmosphere, preventing any outflow from being `starved' of material.  The disc and atmosphere are initially in pressure balance with a density contrast $\eta=\rho_{\rm disc}/\rho_{\rm atm}=100$, while $\mathbfit{B}$ is initialised from Eqs.\ (\ref{eq:binit}).

Following \citet{klb99}, $\rho$, $p$, and $v_z$ are held to their initial conditions, while $v_r$ and $v_\varphi$ are allowed to `evolve' in time according to:
\begin{equation*}
v_r=v_zB_r/B_z;\qquad v_\varphi=v_{\rm K}+v_zB_\varphi/B_z.
\end{equation*}
Magnetic boundary conditions are maintained by imposing conditions on the edge-centred induced electric field, $\mathbfit{E}=-\mathbfit{v}\times\mathbfit{B}$:

\vspace{-9pt}

\begin{equation}
\label{eq:discE}
\left.\begin{aligned}
&&E_z(-z) & = E_z(z);\\
E_r(0)&=-v_{\rm K}B_z(0); & E_r(-z)&=2E_r(0)-E_r(z);\\
E_\varphi(0)&=0; & E_\varphi(-z)&=-E_\varphi(z),
\end{aligned}
\quad\right\}
\end{equation}
where $E_z(0)$ is allowed to `float'.  Since $v_z$ is sub-slow, these conditions are formally over-determined and, in principle, $p$ should be allowed to evolve as well.  Testing this conjecture, we find that, since $\nabla p$ is $\lesssim 1\%$ of the net Lorentz force at the disc surface, a floating $p$ has only the slightest quantitative effects in the computational domain, yet rather severe consequences within the boundary.  Owing to the incomplete dynamics, unphysically high temperatures develop inside the `disc', forcing unnecessarily small time steps on the rest of the simulation.  Thus, we fix $p$ to its initial value (matching the initial atmospheric profile) as a numerical convenience throughout the simulations.

Ideally, one would perform a full characteristic analysis at the boundary, setting amplitudes of the outwardly directed waves to zero and using the inwardly directed waves to determine properly upwinded boundary values (\eg\ App.\ A in \citealt {delzannaetal01}).  Such a capacity has not yet been implemented in \azeus.

Within the inner radius of the accretion disc ($z\leq0$ and $r<r\rmi$), we apply reflecting and conducting boundary conditions ($\mathbfit{J}=\frac1{4\pi}\nabla\times\mathbfit{B}\neq0$).  Thus, $\rho$, $p$, and $\mathbfit{v}$ are reflected across $z=0$, and magnetic boundary conditions are set according to $E_z(-z)=-E_z(z)$, $E_r(-z)=E_r(z)$, and $E_\varphi(-z)=E_\varphi(z)$. At $z=0$, $E_r$ and $E_\varphi$ are evolved using the full MHD equations.

Finally, we use reflecting boundary conditions along the $r=0$ symmetry axis with inversion of $v_\varphi$ and $B_\varphi$, and outflow conditions along the outermost $z$- and $r$-boundaries, neither of which are ever crossed by anything significant to the simulations.

\begin{table*}
  \begin{center}
  \caption{\label{tab:simsum} Summary of simulations. $t_{\rm end}$ and $z_{\rm end}$ are the time and jet length at simulation end, respectively.}
  \begin{tabular}{c|cccccccc}
    \hline\hline
    Simulation & A & B & C & D & E & F & G & H \\
    \hline
    $\beta\rmi$ & 0.1 & 0.4 & 1.0 & 2.5 & 10 & 40 & 160 & 640 \\
    $B\rmi$ (G) & 200 & 100 & 63.2 & 40 & 20 & 10 & 5 & 2.5 \\
    $t_{\rm end}$ (yr) & 47 & 64 & 77 & 88 & 121 & 153 & 153 & 153 \\
    $z_{\rm end}$ (AU) & 4070 & 4070 & 4070 & 4070 & 4070 & 3800 & 2770 & 2380 \\
    \hline
  \end{tabular}
  \end{center}
\end{table*}

\subsection{Scaling relations}
\label{sub:scaling}
All simulations are performed in units where $\rho\rmi=r\rmi=c_{\rm s,i}=1$, and where $c_{\rm s,i}$ is the sound speed at $r=r\rmi$.  Physical units can be restored as follows. First, from Eq.\ (\ref{eq:hseeqn}) and the adiabatic gas law, one can show that:
\begin{equation}
\label{eq:csvk}
c_{\rm s}^2=\gamma\frac{p}{\rho}=\left(\gamma-1\right)\frac{GM_{\rm *}}{\sqrt{r^2+z^2}}=\left(\gamma-1\right)v_{\rm K}^2.
\end{equation}
Then, from Eqs.\ (\ref{eq:betai}), (\ref{eq:csvk}), and the ideal gas law ($p=\rho{kT}/\langle{m}\rangle$, where $\langle{m}\rangle$ is $\sim$half a proton mass), the following scaling relations to convert from unitless to physical quantities may be derived:
\begin{align}
\label{eq:pscale}
p\rmi &= \left(160~{\rm dyne\,cm}^{-2}\right)\left(\frac{\beta\rmi}{40}\right)\left(\frac{B\rmi}{10\,{\rm G}}\right)^2;\\
\label{eq:rhoscale}
\frac{\rho\rmi}{\langle{m}\rangle}&=\left(5.4\times10^{12}\,{\rm cm}^{-3}\right)\!\left(\frac{\beta\rmi}{40}\right)\!\left(\frac{B\rmi}{10\,{\rm G}}\right)^2\!\!\left(\frac{r\rmi}{0.05\,{\rm AU}}\right)\!\left(\frac{0.5M_{\odot}}{M_{\rm *}}\right);\\
\label{eq:tempscale}
T\rmi&=\left(2.2\times10^5~{\rm K}\right)\,\left(\frac{0.05\,{\rm AU}}{r\rmi}\right)\left(\frac{M_{\rm *}}{0.5\,M_{\odot}}\right);\\
\label{eq:csvkscale}
c_{\rm s,i}&=\left(77~{\rm km\,s}^{-1}\right)~\left(\frac{0.05\,{\rm AU}}{r\rmi}\right)^{1/2}\left(\frac{M_{\rm *}}{0.5\,M_{\odot}}\right)^{1/2};\\
\label{eq:tscale}
\tau\rmi&=\frac{r\rmi}{c_{\rm s,i}}=\left(9.7\times10^4~{\rm s}\right)~\left(\frac{r\rmi}{0.05\,{\rm AU}}\right)^{3/2}\left(\frac{0.5\,M_{\odot}}{M_{\rm *}}\right)^{1/2},
\end{align}
where $\gamma=5/3$ and a nominal magnetic field strength of 10 G at $(z,r)=(0,r\rmi)$ have been used.  Evidently, $T\rmi$ is the temperature at $(z,r)=(0,r\rmi)$ and $\tau\rmi$ is the time scale which, for the chosen parameters, is slightly more than a day.  As a representative example, simulation E required $\sim\! 4.2\times 10^7$ time steps on the finest grid ($\sim\! 164,\!000$ on the coarsest grid) over a span of $121$ yr to reach the end of the computational domain (4070 AU). The time step in these simulations is typically controlled by the Alfv\'en speed within several $r\rmi$ of the disc, close to the symmetry axis.

\section{Description of the simulations}
\label{sec:simulations}
Table \ref{tab:simsum} lists the values of $\beta\rmi$ and $B\rmi$ for the eight simulations, A--H, as well as the problem time and jet length at simulation end, assuming the scaling parameters in Sect.\ \ref{sub:scaling}. The simulations were stopped when the tip of the leading bow shock reached the end of the coarsest grid (4,070 AU), or after $t = 50,\!000\,t_i \simeq 153$ yr, whichever came first. For reference, after 153 yr, the inner and outer edges of the disc (at $r\rmi$ and $4,\!992\,r\rmi$) have undergone $\sim$9,750 and $\sim$0.03 Keplerian orbits, respectively. Time-lapse animations of the simulations described below can be found at \url{http://people.virginia.edu/\~jpr8yu/azeus/proto\_jets.html}.

\subsection{Overview}
\label{sub:overview}

The simulations listed in Table \ref{tab:simsum} can be divided into three categories based on the strength of the initial magnetic field: strong (A--D); moderate (E and F); and weak (G and H).  Discussion in this subsection on the origins of the jet and its principle morphological features applies mostly to the strong and moderate-field cases, and less so to the weak-field cases.  While simulations G and H do generate sustained outflow, they are much more turbulent with far fewer distinctive features than in the stronger field runs.

With this in mind, when any of the simulations begin, a torsion Alfv\'en wave is launched into the initially stationary atmosphere at $r\geq r\rmi$ by the rotating disc.  The wave propagates outward at the local Alfv\'en speed, $a\pol$, leaving in its wake atmospheric material rotating in the same sense as the disc, and a toroidal magnetic field, $B_\varphi$, is twisted out of the initial $B\pol$ in a direction opposite to the rotation.  Note that $v_\varphi$ and $B_\varphi$ remain nearly zero within the inner disc radius ($r<r\rmi$) for most of the simulations (with the notable exceptions of the weak-field cases).  The torsion wave is a transient feature, borne from the unrealistic initial conditions in which the stationary atmospheric magnetic field threads the rotating disc.  That said, it is quickly overcome and absorbed by the leading bow shock of a super-fast jet launched almost immediately from the disc surface, and the torsion wave thus plays a negligible role in the overall appearance of the simulations.

With the passage of the torsion wave, the magnetic field distribution is no longer force-free, and the radial Lorentz force\footnote{$\mathbfit J\times\mathbfit B$ is actually a force \emph{density} which needs to be integrated over a volume to get an actual force.} becomes:
\begin{equation}
\label{eq:radial_lorentz}
F_r=J_\varphi B_z-J_zB_\varphi\sim-\frac{B_\varphi}r\partial_r(rB_\varphi),
\end{equation}
where $J_\varphi$ (as determined from Eqs.\ \ref{eq:binit}) remains approximately zero, at least early in the simulations and for the stronger field cases.

\begin{figure*}
  \begin{center}
    \includegraphics[width=1.0\textwidth,clip=true]{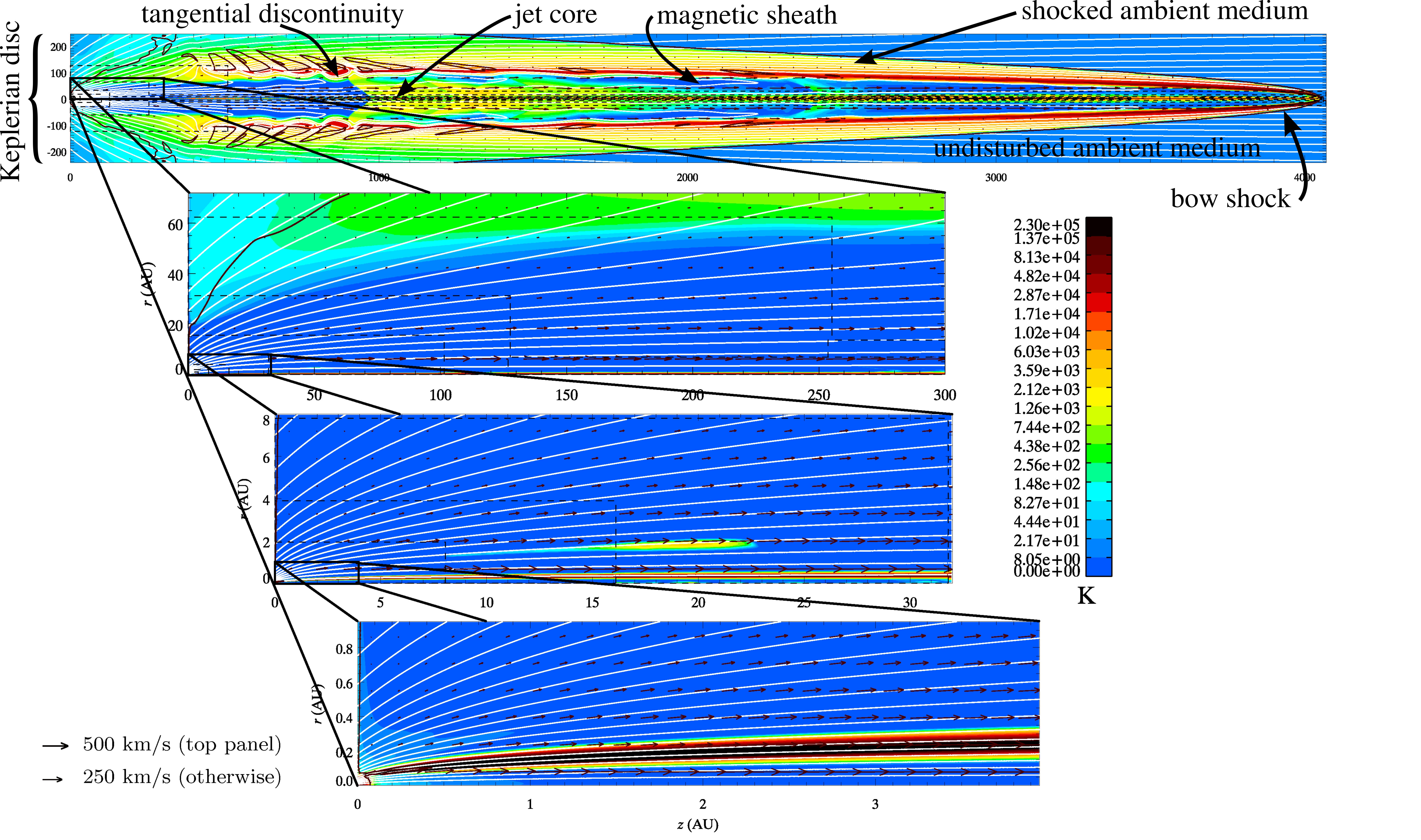}
  \end{center}
  \caption{\label{fig:zoom0.1} Nested images from levels (from top to bottom) 1, 2, 6, and 9 (Table \ref{tab:grids}) of simulation A ($\beta\rmi=0.1$) at $t\sim47$ yr.  Colours indicate temperature, white contours magnetic field lines, black contours the slow magnetosonic surface, and arrows the poloidal velocity.  The annotations in the top panel denote the components of the outflow: the super-fast dense jet core, the super-fast under-dense sheath, the tangential discontinuity (TD), the super-slow shocked ambient medium and the undisturbed ambient medium.  Dashed lines indicate AMR grid boundaries.}
\end{figure*}

\begin{figure*}
  \begin{center}
    \includegraphics[width=0.85\textwidth]{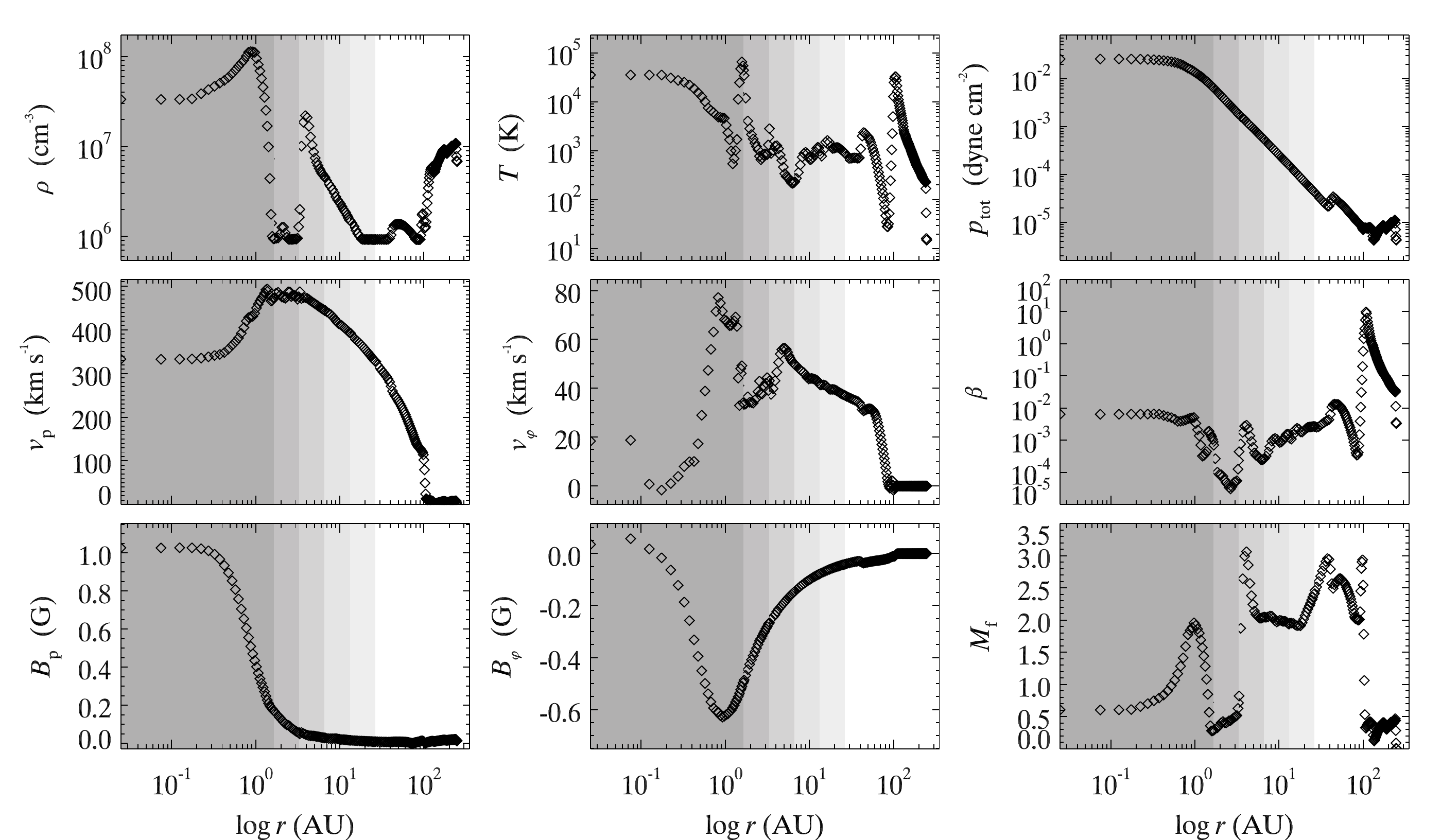}
  \end{center}
  \caption{\label{fig:rslice0.1} A radial slice from simulation A at $z=1,\!200$\,AU and $t=47$\,yr.  Plotted on the left from top to bottom are: density; poloidal velocity; and poloidal magnetic field.  Plotted in the middle from top to bottom are: temperature; toroidal velocity; and toroidal magnetic field. Plotted on the right from top to bottom are: total (thermal + magnetic) pressure; plasma-$\beta$; and fast magnetosonic Mach number.  Shading from medium gray to white indicate grid levels 6--1, respectively.}
\end{figure*}

The radial profile of $rB_\varphi$ is necessarily zero on axis, remains (essentially) zero inside $0<r<r\rmi$ where the torsion wave does not pass, deviates strongly from zero beyond $r=r\rmi$ reaching a global minimum at $r=r_{\rm m}$ ($\sim 1\,{\rm AU}=20r\rmi$), then for the most part returns monotonically and asymptotically to zero as $r\rightarrow\infty$ (\eg, bottom middle panel of Fig.\ \ref{fig:rslice0.1}).  This profile is directly related to the three distinct regions of the jet that develop in the simulations, as depicted in the top panel of Fig.\ \ref{fig:zoom0.1} and described below.

First, inside $0<r<r\rmi$, where no material is driven onto the grid from the $z=0$ boundary and where the axial field remains nearly force-free, a narrow, cold and relatively quiescent magnetic `spine' develops along the symmetry axis. Its integrity is maintained throughout the simulations in all but the weakest-field cases.  In 3-D, however, it is unlikely this feature could survive given the higher mode instabilities that tend to disrupt the axisymmetry of the jet (\eg\ \citealt{hardeeclarke95}).

Second, within $r\rmi<r<r_{\rm m}$, $B_\varphi$ and $\partial_r(rB_\varphi)$ have the same sign and $F_r$ in Eq.\ (\ref{eq:radial_lorentz}) is directed inward, compressing material toward the axis.  This sets up what becomes a dense, hot, strongly magnetised, super-fast `jet core' threaded by the spine.

Third, for $r>r_{\rm m}$, $B_\varphi$ and $\partial_r(rB_\varphi)$ have opposite sign, and $F_r$ in Eq.\ (\ref{eq:radial_lorentz}) is directed outward.  This drives material away from the axis and opens up a wide cavity terminating radially at the tangential discontinuity (TD), located at $r=r_{\rm TD}$ which, in Fig.\ \ref{fig:rslice0.1}), is located at $r\sim 100$ AU.  This cavity is cold, highly magnetised and, at least initially, relatively evacuated, thus allowing material driven from the disc easy passage.  This outflow forms an exceedingly cold, strongly magnetised, trans-fast, and dense (but less dense than the core), `sheath' that surrounds the jet core.

The torsion wave sets material spinning, and material near the disc surface with its frozen-in $B\pol$ experiences a centrifugal acceleration outward.  If the angle between the disc and a magnetic field line, $\theta_0$ (Fig.\ \ref{fig:anchor}), falls below a critical angle $\theta_{\rm c}=60\dgr$, the magnetocentrifugal `bead-on-a-wire' mechanism (BWM; \citealt{henriksenrayburn71,bp82}) is triggered and accelerates material away from the disc.  From Eqs.\ (\ref{eq:aphi}) and (\ref{eq:binit}), one can show that initially at the disc surface, $\theta_0=\tan^{-1}B_z/B_r<\theta_{\rm c}$ for $r\geq\sqrt{3}r\rmi$, and it is only in the region $r\rmi<r<\sqrt{3}r\rmi$ where $\theta_0$ must be reduced before outflow can begin.  As discussed in Sect.\ \ref{sub:knotgen}, this turns out to be key in the persistent generation of `knots' (plasmoids) observed in one of our simulations.

The BWM requires a `rigid' $\mathbfit B\pol$, which can occur only if $\beta\pol=8\pi p/B\pol^2\lesssim 1$.  If $\beta\pol\gg 1$, $B\pol$ cannot provide the magnetic tension necessary to act as the `wire' to guide the `beads' of plasma, and a different mechanism to drive the jet must be invoked.  As discussed further in Sect.\ \ref{sub:driving}, even a weak $\mathbfit B\pol$ in a persistently rotating environment will generate a dynamically important $B_\varphi$ ($\beta_\varphi=8\pi p/B^2_\varphi\sim 1$), and this can provide sufficient magnetic pressure to accelerate an outflow.  This is sometimes referred to as the `magnetic tower mechanism' (MTM; \citealt{lyndenbell96}).

Whether driven magneto-centrifugally and/or via a sustained magnetic pressure, outflow is strongest near $r\rmi$ where the rotation is the most rapid and the magnetic field strongest.  While the driving force gets progressively weaker with increasing $r$, the density in the disc and sheath also fall, and material can still be accelerated to super-fast speeds.  Beyond $r_{\rm TD}$ (outer limit of the sheath), the weak Lorentz force cannot accelerate the much denser ambient material effectively.  In this way, a fairly distinct boundary is visible between fast, outwardly moving material---the actual `jet' consisting of a core and sheath---and the more slowly moving, shocked and much denser atmosphere behind the jet bow shock.

Thus, as annotated in the top panel of Fig.\ \ref{fig:zoom0.1}, the general outflow established in our simulations consists of three outwardly moving components.  First, the super-fast, dense, hot, strongly and helically-magnetised narrow jet core---what we suggest corresponds to the main portion of the observable jet---is threaded by a narrow, quiescent spine of strong axial magnetic field.  Second, the jet core is surrounded by a somewhat slower-moving trans-fast, under-dense, cold, highly magnetised wide sheath, which may, in part, be observable as a `cavity' of emission surrounding the jet.  As noted below, this region is subject to strong reflection shocks triggered by the Kelvin-Helmholtz unstable TD, that can heat sheath material significantly.  As a result, some of the sheath may contribute to the observable outflow.  We note in passing that the development of an under-dense sheath could account for the great distances to which jets can propagate stably in 3-D (\citealt{hardeeclarkerosen97}). Third, between $r_{\rm TD}$ and the bow shock, the sheath is surrounded by a trans-slow, hot, weakly-magnetised, shocked ambient medium, whose density is less than that of the jet core, but greater than that of the undisturbed ambient medium. This may correspond to the `second wind', first described by \citet{stockeetal88} and currently interpreted as `molecular winds' (\eg, \citealt{franketal14_ppvi}), whose forward motion is a result of entrainment by the leading bow shock driven by the advancing jet rather than magnetic stresses near the disc.

\subsection{Strong-field simulations: A--D}
\label{sub:strong-field}

Differences among the four simulations in the strong-field category are relatively minor, and we use as an exemplar the final epoch of simulation A ($\beta\rmi=0.1$, $B\rmi=200\,$G) shown in Fig.\ \ref{fig:zoom0.1}, in which colour contours of temperature are superposed with magnetic field lines (white contours) and the slow surface (where $v\pol=a_{\rm p,s}$, the poloidal slow speed; black contours).  Here, the magnetic field is strong enough to enforce $\mathbfit v\pol\parallel \mathbfit B\pol$ virtually everywhere inside the TD, resulting in what appears to be a largely self-similar, steady state solution.  Indeed, the bottom panel in Fig.\ \ref{fig:zoom0.1} could be taken from any time within the final $\sim$60\% of the run and the relative simplicity of this run makes it prototypical of the description given in Sect.\ \ref{sub:overview}.

The bottom panel of Fig.\ \ref{fig:zoom0.1} bears a resemblance to the early, local simulations of \citet{uchidashibata85}.  Near the disc, outflow is driven almost exclusively by the BWM, with the MTM gradually contributing additional thrust as the poloidal field becomes wound up even before the Alfv\'en point (Sect.\ \ref{sub:driving}).  Outflow is robust, long-lived, steady, and there is no reason to believe it would ever cease so long as the disc continues to provide mass.  Magneto-centrifugal wind launching theory \citep[\eg][]{bp82} predicts that acceleration of the flow ceases beyond the Alfv\'en point, while some local simulations (e.g.\ \citealt{pro06}) show acceleration of the outflow continues until the fast point, with steady flow thereafter.  However, we find that because of the MTM, a gentle acceleration persists well beyond the fast point, accelerating the advance speed of the jet to $\sim420\,{\rm km\,s}^{-1}$ for simulation A ($v_{\rm jet}$ in Table \ref{tab:simsum2}), rendering the jet essentially ballistic (since $v_{\rm p,max}\sim v_{\rm jet}$).  Further, the sound speed in the asymptotic ambient medium is,
\begin{equation*}
c_{\rm s,\infty}=c_{\rm s,i}\sqrt{\frac{T_\infty}{T\rmi}}=c_{\rm s,i}\left(\frac{\rho_\infty}{\rho\rmi}\right)^{\frac{\gamma-1}2}=c_{\rm s,i}(10^{-6})^{1/3}\sim 0.77\,{\rm km\,s}^{-1},
\end{equation*}
using the asymptotic density in Sect.\ \ref{subsub:atmos}.  Thus, the external sonic Mach number of the jet in simulation A is $M_{\rm ext}\sim540$.  Of course, the jet core is much hotter than the ambient medium (by a factor of $\sim 10^4$ and as high as $10^6$ K), and the internal sonic Mach number is therefore considerably less; $M_{\rm int}\sim 22$.  The internal fast magnetosonic Mach number, that which actually governs the `supersonic character' of the jet, is a rather modest 2.1.  See Table \ref{tab:simsum2} for comparative values for the other simulations.

Hugging the axis, but not particularly apparent in even the lower panel of Fig.\ \ref{fig:zoom0.1}, is the relatively quiescent, cold spine with negligible $B_\varphi$, strong $B_{\rm p}$ and a radius which remains nearly constant ($\sim\! 2r\rmi$) (but not resolved by levels $<$ 6).  This spine threads the dense, hot, highly magnetised, super-fast jet core which, as seen in the bottom panel of Fig.\ \ref{fig:zoom0.1}, reaches a radius of $\sim0.3\,$AU (outer extent of the hot region) at $z=4\,$AU from the origin.  Fig.\ \ref{fig:rslice0.1} shows a radial slice of numerous variables across the computational domain at $z=1,\!200\,$AU.  In these slices, the jet core is demarcated by the drop in density from $10^8$ to $10^6$ cm$^{-3}$ in the top left panel, and the global minimum of $B_\varphi$ in the bottom middle panel, and thus has reached a radius of about 1\,AU.  At $z\sim$ 3,500 AU, the jet core reaches its maximum radius of a few AU.

Surrounding the jet core is the magnetic sheath, and the boundary between the sheath and the hotter, denser shocked ambient medium is the tangential discontinuity, best visualised in the top panel of Fig.\ \ref{fig:zoom0.1}.  The distinguishing feature of a TD (as opposed to an ordinary contact discontinuity) is the lack of a normal component of the magnetic field; $\mathbfit B$ is everywhere parallel to the feature we have identified in Fig.\ \ref{fig:zoom0.1} as the TD.  The TD is also apparent in Fig.\ \ref{fig:rslice0.1} at $r\sim100\,$AU as a sudden drop in $v\pol$ (left centre panel), a jump in temperature (middle top panel), the end of a gradual drop in $v_\varphi$ (middle centre panel), a sharp rise in the plasma-$\beta$ (right centre panel), and a drop from super to sub-fast-magnetosonic speed (right bottom panel) all while the total pressure (thermal + magnetic) remains more or less continuous (top right panel).

\setlength{\tabcolsep}{0.20em}
\begin{table*}
  \begin{center}
  \caption{\label{tab:simsum2} Summary of measured quantities from the simulations. Mass-weighted average quantities include: $\langle v_\varphi\rangle$, the jet rotation speed (all speeds in km\,s$^{-1}$); $\langle v_z\rangle$, the mass-weighted axial speed along the last 250 AU of the jet; $v_{\rm jet}$, the advance speed of the jet tip into the ambient medium; $v_{\rm entr}$, the mass-weighted average poloidal velocity of the ambient medium entrained by the bow shock; $M_{\rm ext}=v_{\rm jet}/c_{\rm s,ext}$, the sonic Mach number of the advance speed relative to asymptotic external ambient medium (where $c_{\rm s}=0.77$\,km\,s$^{-1}$ for $T_{\rm asym}=22\,$K); $M_{\rm int}=\langle v_z\rangle/\langle c_{\rm s,int}\rangle$, the internal sonic Mach number, where $\langle c_{\rm s,int}\rangle$ is the mass-weighted average sound speed over the last 500 AU of the jet, and $M_{\rm f,int}=\langle v_z\rangle/\langle a_{\rm f,int}\rangle$, the internal fast magnetosonic Mach number, where $\langle a_{\rm f,int}\rangle$ is the mass-weighted average fast speed over the last 500 AU of the jet.  Data from along the magnetic field line anchored at $r_0=1\,$AU in the disc include: $v_{\rm p,A}$, $v_{\rm p,f}$, $v_{\rm p,max}$, the poloidal speeds at, respectively, the Alfv\'en, fast, and asymptotic points; $(z_{\rm A},r_{\rm A})$, $(z_{\rm f},r_{\rm f})$, the coordinates (all distances are in AU) of the Alfv\'en and fast points respectively; and $s_\times$, the location along the field line where $2 v_\varphi = r\omega_0$ (Sect.\ \ref{sub:driving}).  Fluxes measured inside the TD at $z=1,\!000$ AU include: $\dot M$, mass flux ($M_\odot$\,yr$^{-1}$); $\dot P$, linear momentum flux ($M_\odot$\,yr$^{-1}$km\,s$^{-1}$); $\dot L$, angular momentum flux ($M_\odot$\,yr$^{-1}$AU\,km\,s$^{-1}$); and $\dot K$, kinetic energy flux (erg\,s$^{-1}$).  Quantities which follow a power law in $B\rmi$ include an estimate of the power-law index, $\alpha$, in the last column.   Uncertainties for all quantities are given by the standard-deviation of the data time-averaged over a period of $\sim$1 yr (or $>400$ time steps on the coarsest level), or 1 in the last of three significant digits, whichever is the greater.}
  \begin{tabular}{ccccccccccc}
    \hline\hline
&&&&&&&&&&\\[-9pt]
    & A & B & C & D & E & F & G & H &~&  $\alpha$ \\
    $\beta\rmi$ & 0.1 & 0.4 & 1.0 & 2.5 & 10 & 40 & 160 & 640 && \\
    $B\rmi$ (G) & 200 & 100 & 63.2 & 40 & 20 & 10 & 5 & 2.5 & \\
    \hline
&&&&&&&&&&\\[-9pt]
    \multicolumn{11}{l}{mass-weighted averages:}\\
    $\langle v_\varphi\rangle$ & $21.0\pm0.3$ & $12.6\pm0.4$ & $9.1\pm0.5$ & $6.6\pm0.6$ & $4.3\pm0.5$ & $2.8\pm0.6$ & $2.2\pm0.5$ & $1.8\pm0.4$ &&$0.67\pm0.03$ \\
    $\langle v_z\rangle$  & $419.\pm1.$ & $316.\pm1.$ & $267.\pm1.$ & $234.\pm1.$ & $143.\pm1.$ & $113.\pm1.$ & $82.0\pm0.4$ & $64.0\pm0.1$ && $0.43\pm0.01$ \\
    $v_{\rm jet}$ & $417.\pm1.$ & $309.\pm1.$ & $252.\pm1.$ & $230.\pm1.$ & $166.\pm1.$ & $116.\pm1.$ & $85.6\pm0.1$ & $74.2\pm0.1$ && $0.42\pm0.01$ \\
    $v_{\rm entr}$ & $11.5\pm0.1$ & $7.84\pm0.1$ & $5.93\pm0.02$ & $3.9\pm0.2$ & $2.5\pm0.4$ & $2.01\pm0.03$ & $1.46\pm0.01$ & $1.10\pm0.06$ && $0.56\pm0.01$ \\
    $M_{\rm ext}$ & $536.\pm1.$ & $397.\pm1.$ & $324.\pm1.$ & $296.\pm1.$ & $214.\pm1.$ & $149.\pm1.$ & $110.\pm1.$ & $99.\pm1.$ && $0.40\pm0.02$ \\
    $M_{\rm int}$ & $22.2\pm0.2$ & $26.4\pm0.4$ & $27.7\pm0.4$ & $20.9\pm0.3$ & $23.\pm4.$ & $9.5\pm0.1$ & $11.9\pm0.2$ & $9.2\pm0.2$ && --- \\
    $M_{\rm f,int}$ & $2.00\pm0.01$ & $2.52\pm0.01$ & $2.93\pm0.01$ & $3.99\pm0.02$ & $4.3\pm0.1$ & $6.02\pm0.01$ & $7.46\pm0.01$ & $6.29\pm0.02$ && --- \\
    \hline
&&&&&&&&&&\\[-9pt]
    \multicolumn{11}{l}{$r_0=1\,$AU field line:}\\
    $v_{\rm p,A}$ & $161.\pm1.$ & $101.\pm1.$ & $73.0\pm0.1$ & $51.2\pm0.2$ & $29.6\pm0.1$ & $14.9\pm0.1$ & $7.5\pm0.2$\tablefootmark{$\dagger$} & n/a\tablefootmark{$\ddagger$} && $0.76\pm0.02$ \\
    $v_{\rm p,f}$ & $204.\pm1.$ & $129.\pm1.$ & $94.6\pm0.1$ & $69.5\pm0.5$ & $43.9\pm0.1$ & $27.9\pm0.2$ & $18.5\pm0.2$ & n/a && $0.667\pm0.001$ \\
    $v_{\rm p,max}$ & $297.\pm1.$ & $219.\pm1.$ & $166.\pm1.$ & $129.\pm1.$ & $87.1\pm0.1$ & $60.2\pm0.1$ & $33.2\pm0.4$ & n/a && $0.55\pm0.01$ \\
    $r_{\rm A}$ & $11.6\pm0.1$ & $7.24\pm0.01$  & $5.46\pm0.01$ & $4.15\pm0.01$  & $2.84\pm0.01$ & $2.04\pm0.01$ & $1.66\pm0.04$ & n/a && $0.62\pm0.02$ \\
    $r_{\rm f}$ & $18.9\pm0.1$ & $12.0\pm0.1$ & $9.16\pm0.09$ & $7.0\pm0.1$ & $5.06\pm0.01$ & $4.18\pm0.01$ & $4.4\pm0.3$ & n/a && $0.54\pm0.02$ \\
    $z_{\rm A}$ & $106.\pm1.$ & $33.1\pm0.1$ & $17.0\pm0.1$ & $8.78\pm0.01$ & $3.26\pm0.01$ & $1.22\pm0.01$ & $0.51\pm0.04$ & n/a && $1.47\pm0.02$ \\
    $z_{\rm f}$ & $315.\pm2.$ & $141.\pm1.$ & $68.\pm2.$ & $34.\pm2.$ & $13.3\pm0.1$ & $6.79\pm0.04$ & $4.9\pm0.5$ & n/a && $1.41\pm0.04$ \\
    $s_\times$ & $19.3\pm0.1$ & $8.56\pm0.02$ & $5.10\pm0.01$ & $3.12\pm0.01$ & $1.68\pm0.01$ & $0.982\pm0.009$ & $0.64\pm0.01$ & n/a && $1.10\pm0.04$ \\
    \hline
&&&&&&&&&&\\[-9pt]
    \multicolumn{11}{l}{jet fluxes at $z=1,\!000$ AU:}\\
    $\dot M$ $(\times 10^{-6})$ & $2.0\pm0.1$ & $1.5\pm0.1$ & $1.36\pm0.07$ & $1.16\pm0.04$ & $1.01\pm0.06$ & $0.82\pm0.05$ & $0.51\pm0.04$ & $0.039\pm0.006$ && $0.32\pm0.03$\\
    $\dot p$ $(\times 10^{-4})$ & $4.9\pm0.6$ & $2.8\pm0.3$ & $2.0\pm0.1$ & $1.37\pm0.06$ & $0.83\pm0.06$ & $0.46\pm0.03$ & $0.22\pm0.02$ & $0.017\pm0.003$ && $0.82\pm0.03$\\
    $\dot L$ $(\times 10^{-4})$ & $14.\pm2.$ & $5.\pm1.$ & $3.6\pm0.4$ & $1.8\pm0.1$ & $0.97\pm0.04$ & $0.36\pm0.03$ & $0.14\pm0.01$ & $0.0024\pm0.0005$ && $1.24\pm0.05$\\
    $\dot K$ $(\times 10^{33})$ & $96.\pm18.$ & $37.\pm7.$ & $20.\pm2.$ & $11.2\pm0.7$ & $4.6\pm0.4$ & $1.7\pm0.1$ & $0.60\pm0.06$ & $0.053\pm0.009$ && $1.37\pm0.02$\\
    \hline
  \end{tabular}
  \end{center}
  \tablefoot{
  \tablefoottext{$\dagger$}{It is debatable whether the 1 AU field line can be considered in steady state in simulation G.}
  \tablefoottext{$\ddagger$}{The 1 AU field line in simulation H shows no evidence of being in steady state.}
  }
\end{table*}

Dynamically, the most important of these characteristics is the drop in $v\pol$, which means the TD is a shear layer and subject to the Kelvin-Helmholtz instability, manifest in the top panel of Fig.\ \ref{fig:zoom0.1} as gentle undulations along its length.  Between $500\,{\rm AU}\lesssim z\lesssim 900\,{\rm AU}$, these undulations have a wavelength of about $100\,$AU and gradually grow in amplitude.  By $z\sim900\,$AU, the severity of the undulation triggers a fairly strong reflection (criss-cross) shock in the sheath that thermalises enough of the kinetic energy to warm the sheath significantly (from several to a few hundred kelvins), and redirect flow parallel to the jet axis.  This ceases the expansion of the jet sheath, whose outer radius at this epoch of simulation A is about 100 AU.  After $z\sim900$\,AU, the K-H undulations continue with lesser amplitude and a wavelength of 250--300\,AU, and trigger a series of gentler reflection shocks and rarefaction fans similar to those described for a hydrodynamical jet by \citet{normanetal82}. These are all visible as discontinuities and gradations in temperature in Fig.\ \ref{fig:zoom0.1}. At $z\sim2,\!400$\,AU, one final, relatively strong reflection shock is triggered, after which the flow remains rather laminar and featureless, bearing a strong resemblance to the `nose-cone' described by \citet{clarkenormanburns86} for a propagating jet dominated by a toroidal magnetic field.  We note in passing that the strong poloidal magnetic field, which provides some stability against the $m=0$ `pinch mode' apparent in these axisymmetric simulations, would also provide some stability against higher mode instabilities in 3-D (\eg\ \citealt{hardeeclarkerosen97}), and possibly enough to preserve the nearly axisymmetric appearance of jets such as HH\,34 \citep[e.g.][]{devineetal1997_hh34}.

The shocked ambient medium (sometimes referred to as the `second wind') lies between the TD and the bow shock (top panel of Fig.\ \ref{fig:zoom0.1}) and is characterised as a warm, trans-slow, dense, relatively weakly magnetised medium.  It owes its forward motion entirely to entrainment by the leading bow shock, and has virtually zero rotation and toroidal field.  The sub-slow `islands' (closed black contours in the top panel of Fig.\ \ref{fig:zoom0.1}) indicate the forward motion is trans-slow ($\sim\! 10\,{\rm km\,s}^{-1}$). Where the flow is sub-slow, streamlines diverge (slightly) from the jet axis, and where the flow is super-slow, streamlines converge.  These transitions are precisely coupled to the K-H undulations described above.  The temperature ranges from $\sim30,\!000\,$K just above the TD, to a few hundred K just inside the bow shock (top centre panel of Fig.\ \ref{fig:rslice0.1}), and thus permits the survival of molecules such as CO, which are frequently used to observe entrained outflow material \citep[e.g.][]{zhangetal2016_hh4647}.  Finally, the plasma-$\beta$ jumps 4.5 orders of magnitude across the TD (thermal pressure suffers a sudden increase while magnetic pressure undergoes a commensurate decrease to maintain a near-continuous total pressure), and the shocked ambient medium is dominated by thermal pressure ($\beta\sim10$) just above the TD. The plasma-$\beta$ then drops continuously between $r_{\rm TD}$ and the bow shock where $\beta<0.1$, and the flow is once again magnetically dominated, but not to the extent (by a factor of a few) observed in the jet core and sheath (right centre panel of Fig.\ \ref{fig:rslice0.1}).

Last, what remains of the undisturbed ambient medium is visible above the bow shock in the top panel of Fig.\ \ref{fig:zoom0.1}, where the asymptotic levels for $\rho$ and $B_z$ (Sect.\ \ref{subsub:atmos}) are reached (to within 10\%) by $z\sim2$,300 AU.

\begin{figure}
  \begin{center}
    \includegraphics[width=0.95\columnwidth]{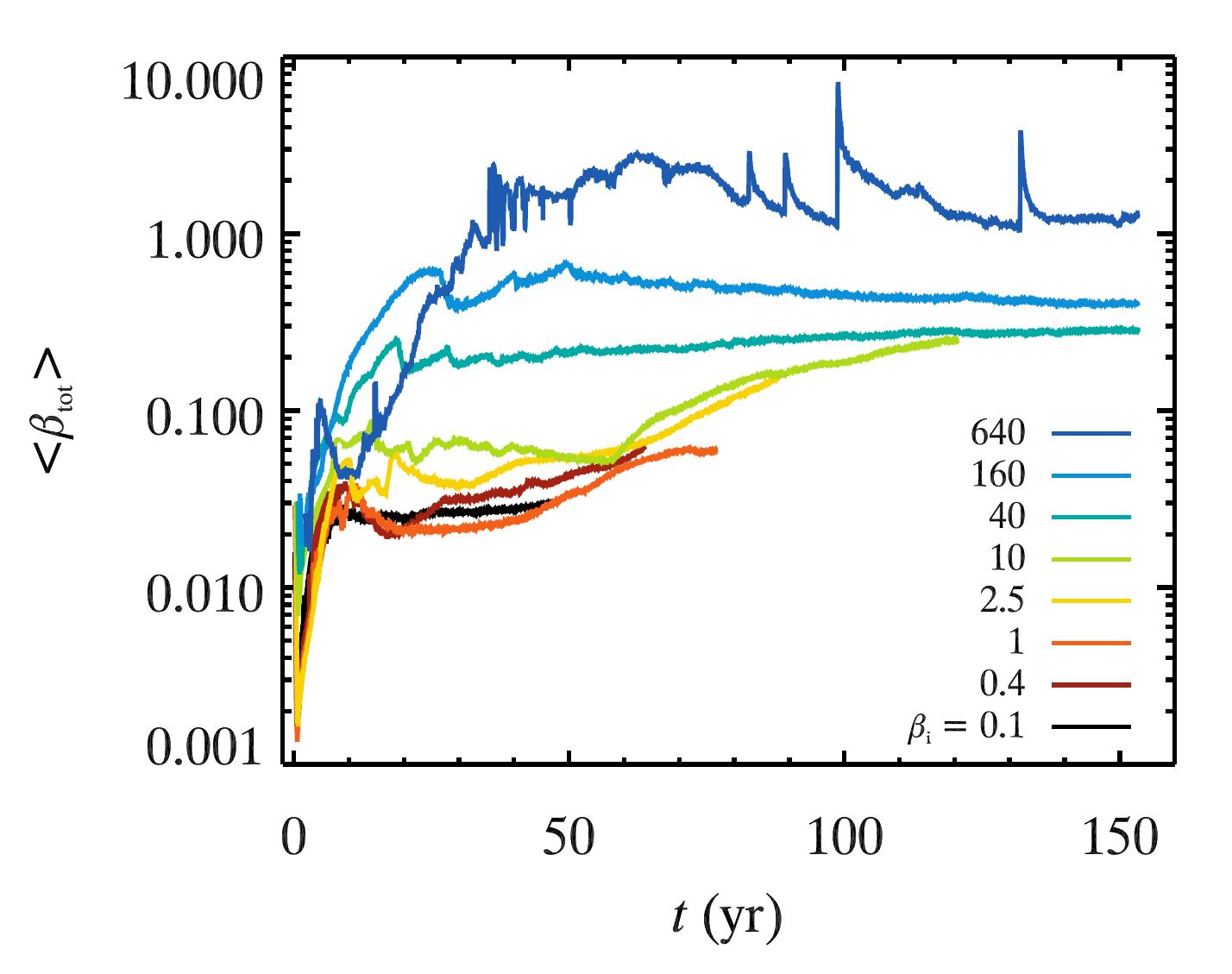}
  \end{center}
  \caption{\label{fig:avgbeta} The volume-averaged plasma-$\beta$ (defined in the text) as a function of time for simulations A (black) through H (dark blue).  Despite three orders of magnitude difference in $\beta\rmi$ for simulations A--G, all loci seem to converge to $\langle \beta_{\rm tot}\rangle\rightarrow0.2$--0.4, characteristic of a magnetically dominated outflow.  Even simulation H seems to converge toward equipartition ($\langle \beta_{\rm tot}\rangle\rightarrow1$).}
\end{figure}

While for convenience, the simulations are identified by the single value $\beta\rmi$, this does not represent the average magnetic field strength in the resulting outflow.  Figure \ref{fig:avgbeta} shows the evolution of the average plasma beta, defined as,
\begin{equation*}
\langle\beta_{\rm tot}\rangle = \frac{8\pi \langle p \rangle}{ \langle B^{2}_\varphi+B^{2}_{\rm p}\rangle},
\end{equation*}
where quantities in angle brackets on the right hand side are volume averages over the outflow identified as regions where $M>5$ and $v_\varphi>10^{-3}$.   The $M=5$ contour was chosen as it was found to hug tightly inside the TD for all jets thus eliminating the shocked ambient medium and less organised outflow near the accretion disc from the average. The limit on rotational speed eliminated material from the jet spine.  Mass-weighted averages were also computed (not shown), giving qualitatively similar loci as in the figure, with values for $\langle\beta_{\rm tot}\rangle$ consistently about 0.2 lower.

What is striking about Fig.\ \ref{fig:avgbeta} is that all loci (with the exception of simulation H) seem to be converging on $\langle\beta_{\rm tot}\rangle\rightarrow0.2$--0.4, characteristic of a magnetically dominated outflow.  Even simulation H, which of all the simulations had the most difficult time organising itself into an outflow, seems to be converging toward magnetic equipartition ($\langle\beta_{\rm tot}\rangle\rightarrow1$).  From these observations, we speculate that \emph{however weak or strong the initial magnetic field in the corona may be that launches the outflow, the outflow itself ends up being magnetically dominated, with an average plasma beta asymptoting to $\lesssim 1$.}  If true, an immediate consequence is that the magnetic properties of an observed jet may not be useful in determining what the magnetic environment may be near the protostar.

\begin{figure*}
  \begin{center}
    \includegraphics[width=1.07\textwidth,clip=true]{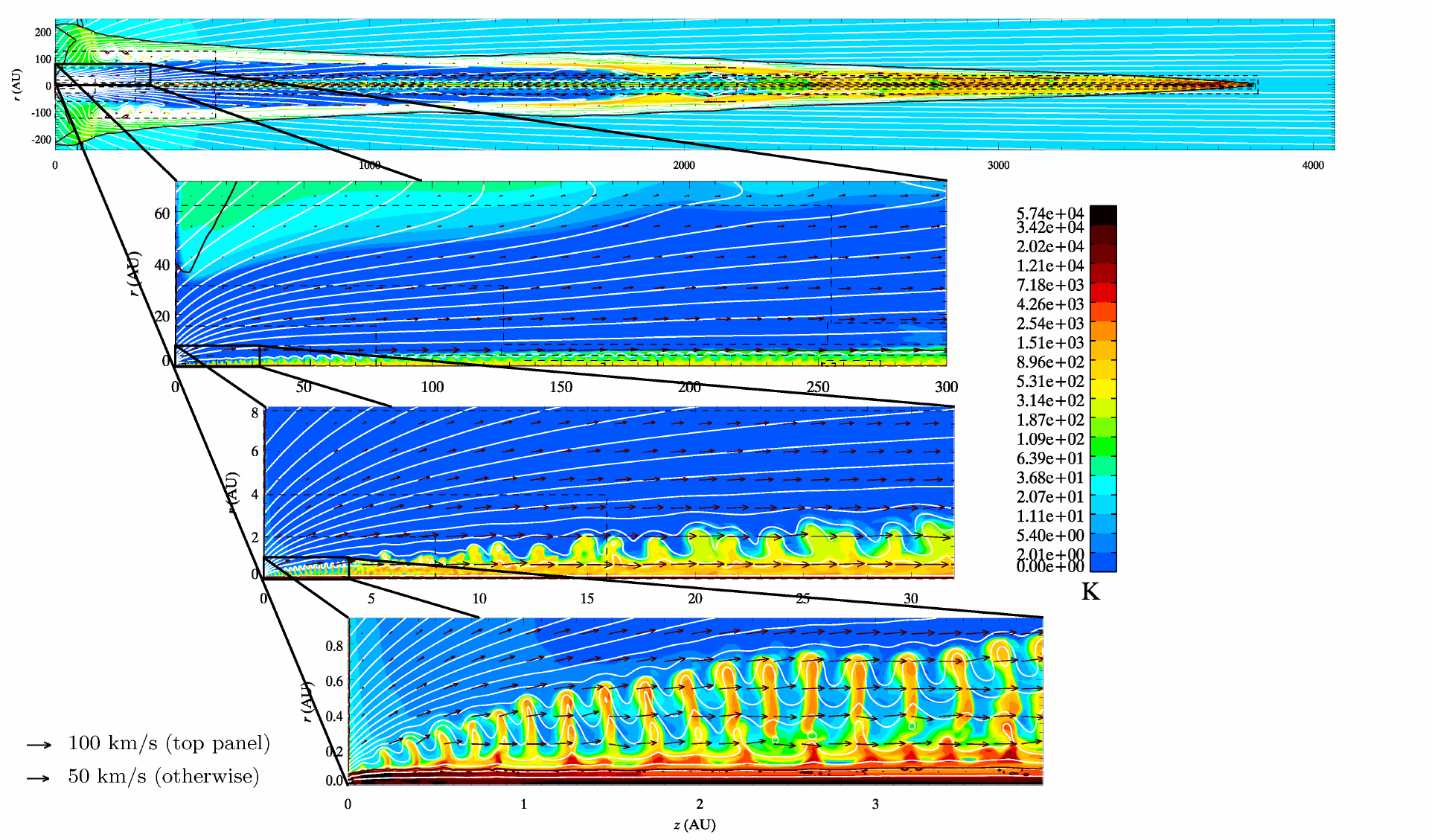}
  \end{center}
  \caption{\label{fig:zoom40} Similar to Fig.\ \ref{fig:zoom0.1}, but for simulation F ($\beta\rmi=40$) at $t = 153$ yr.}
\end{figure*}

The nature of the magnetic field within the jet also evolves with time.  At first, most magnetic field in the outflow is poloidal, reflecting the initial conditions.  As the jet advances and rotates, $B\pol$ is twisted to produce a significant $B_\varphi$ which, by the end of the simulation, accounts for $>90\%$ of the transported magnetic energy density even for simulation A with the strongest initial poloidal field.  This is why we find in our simulations that the MTM is the dominant acceleration mechanism even in simulation A (Sect.\ \ref{sub:driving}), and why our jets continue to accelerate well beyond the fast point.

The higher-$\beta\rmi$ runs (B--D) are qualitatively identical to simulation A, but with significant quantitative differences.  As $\beta\rmi$ increases, jet speeds decrease as do the sonic Mach numbers, although the fast magnetosonic Mach number actually increases monotonically from $\sim2.1$ for simulation A to $\sim4.0$ for simulation D.  These trends continue into the moderate and weak field runs (Table \ref{tab:simsum2}).  The temperature, density, and strength of reflection shocks within the outflow also decrease with increasing $\beta\rmi$, while the time taken for a strong outflow to be organised increases.  This trend is apparent by simulation F (where outflow doesn't really begin until $t\sim 0.2\,$yr), and continues on through simulation H which, in many respects, shows signs of being a `frustrated jet' (Sect.\ \ref{sub:weak-field}).  As part of this trend, the inner-most regions of the jet become less steady for higher $\beta\rmi$ to the point where periodic knots start to form.  These are sporadic for simulation E but, by simulation F, the knots are steady, long-lived, and dominate the inner-jet structure (Sect.\ \ref{sub:medium-field} and \ref{sub:knotgen}).

Finally, near the bottom of the third panel in Fig.\ \ref{fig:zoom0.1} is a prominent `streak' originating from the level 8 grid boundary at $(z,r)\sim(8.0,1.3)$, and stretching to $(z,r)\sim(23,2)$ in the middle of the level 6 grid.  These transient features, more prominent in the stronger field simulations than the weaker ones, are entirely numerical in origin, and triggered where the Alfv\'en surface intersects a grid-boundary.  At such points, and then only rarely, a truncation error occurs in the momentum interpolation which results in a sudden pinch and subsequent local spike in internal energy that is advected downwind with the flow, resulting in the streak seen.  We are not entirely certain why these streaks occur when they do, though they are reminiscent of the `magnetic field explosions' suffered in earlier versions of \zeus (\citealp{clarke96}).  We also know that if one interpolates on velocity in the grid boundaries rather than momentum density (and violate conservation of momentum between grids), these features disappear.

Despite their ominous appearance (in the streak in Fig \ref{fig:zoom0.1}, $T\sim $ a couple thousand kelvins within a cold sheath of a few kelvins), the streaks are not apparent in the other MHD variables, and this region is so magnetically dominated that such a thermal pressure anomaly has negligible dynamical consequences.  We also note that the streaks are entirely absent in the weaker field runs (F--H) and, at this stage, we regard them as cosmetic.  Still, understanding their origin remains an area of current investigation.

\subsection{Medium-field simulations (E, F)}
\label{sub:medium-field}
As the initial magnetic field strength ($B\rmi$) is weakened, all measures of outflow speed decrease (Table \ref{tab:simsum2}) and the outflow itself becomes less steady.  While jets on the observational scale look largely the same (\eg, compare top panels of Figs.\ \ref{fig:zoom0.1} and \ref{fig:zoom40}), the smaller scale structures in the bottom panels are strikingly different.

\begin{figure*}
  \begin{center}
    \includegraphics[width=0.85\textwidth]{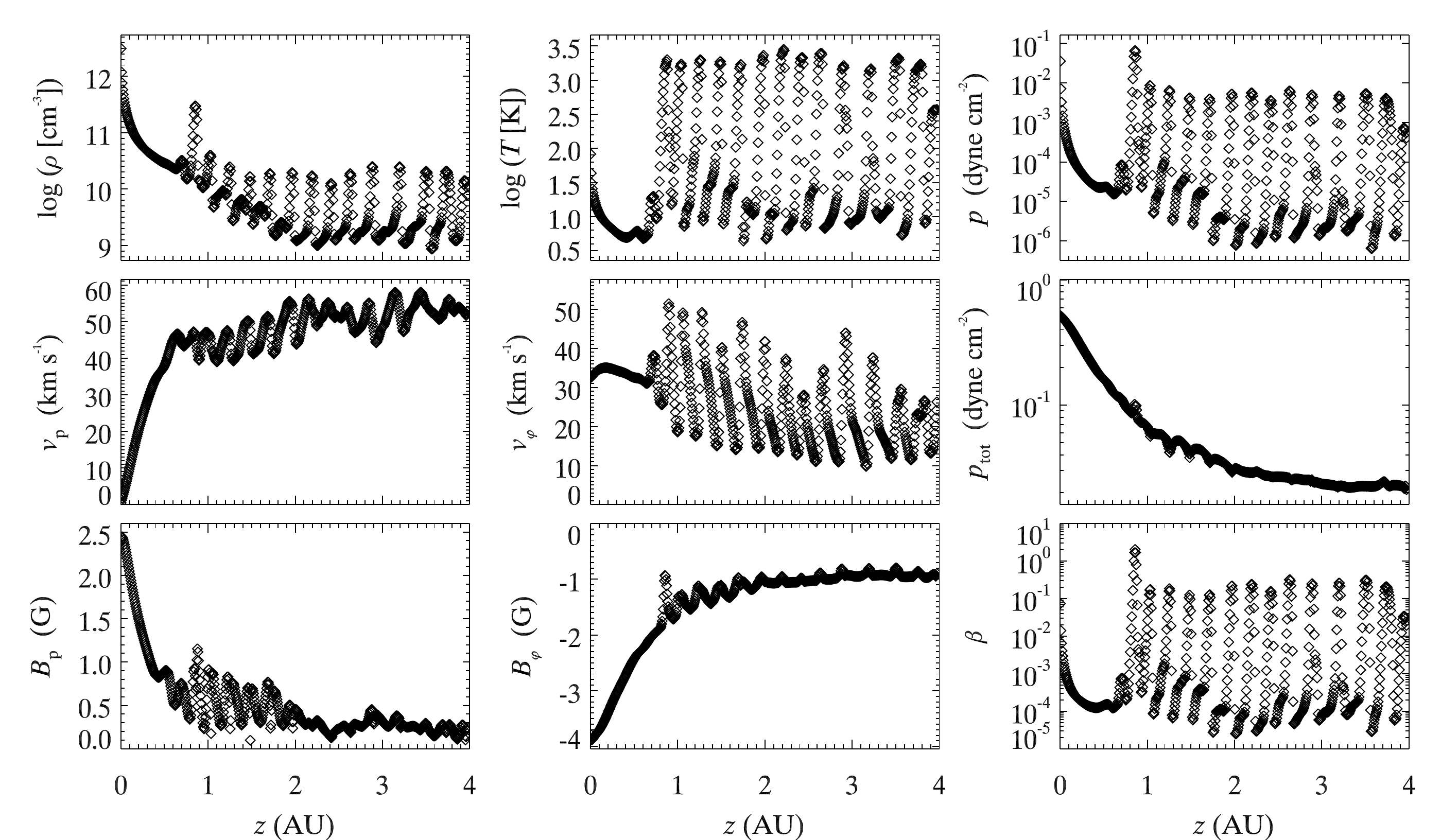}
  \end{center}
  \caption{\label{fig:zslice40} An axial slice of level 9 from simulation F at $r=0.4$\, AU and $t=153$\,yr.  Plotted on the left from top to bottom are: density; poloidal velocity; and poloidal magnetic field. Plotted in the middle from top to bottom are: temperature;  toroidal velocity; and toroidal magnetic field. Plotted on the right from top to bottom are: thermal pressure; total (thermal + magnetic) pressure; and plasma-$\beta$.}
\end{figure*}

Figure \ref{fig:zoom40} shows simulation F at $t = 153$ yr, having reached a length of $\sim3,\!800$\,AU\footnote{A similar simulation taken to 100 yr was described in \citet{rc11}.}. At the jet tip, $v_{\rm jet}\sim116$\,km\,s$^{-1}$, $M_{\rm ext}\sim150$, $M_{\rm int}\sim10$, and $M_{\rm f,int}\sim6$ (Table \ref{tab:simsum2}).  Roughly 99\% of the magnetic energy density within the magnetic sheath is in $B_\varphi$ from which we conclude the MTM is the dominant driver along most of the jet. Indeed, for simulation F (and even more so for weaker $B\rmi$), the Alfv\'en surface lies close to the disk surface ($\lesssim30$ AU) for material interior to the TD, and the BWM is ineffective at any significant height above the disc.

We have chosen simulation F as the exemplar for the medium-field runs because of the propensity and regularity of knots, seen in the lower panel of Fig.\ \ref{fig:zoom40}.  Knots, which are virtually absent in simulations A--D, are present in simulation E, but not to the degree seen in simulation F.  While the production of knots in simulation F is often `steady', they do not represent a steady state; none of the `constants' in Eqs.\ \ref{eq:massload}--\ref{eq:energy} are constant along field lines passing through them.  In this simulation and where they are present in simulation E, only the portion of the sheath devoid of knots and away from the TD is in a quasi-steady state, at least as measured by the constancy of the WD constants (Eqs.\ \ref{eq:massload}--\ref{eq:energy}; Figure \ref{fig:steadystate}).

The knots are launched from $(0,r\rmi)<(z,r)<(r\rmi,2r\rmi)$ where gas is both dense and hot.  Knots are generated nearly from the beginning and, after a few short periods of intermittency and variability, become steady after $t\sim9$ yr with a period ${\cal T}_{\rm knot}\sim0.026$ yr and a wavelength $\lambda_{\rm knot}\sim0.25$ AU.

Once launched, the knots follow a nearly axial trajectory through the sheath, never venturing further from the axis than a few AU.  As can be seen in the lower panels of Fig.\ \ref{fig:zoom40}, they are best described as hot `towers' (tori in 3-D) of plasma following the local magnetic field lines; truly `beads (tori) on a wire'. As they move downstream, they lengthen, merge and, at $z\sim 10$ AU above the disc, coalesce into the continuous hot jet core of radius $\sim2\,$AU, which continues to expand gradually to several AU towards the head of the jet. Never drifting more than a few AU from the axis and losing their identity long before reaching observational scales, we rule these features out as precursors of HH objects.

Along their length, the knots exhibit one, two, and sometimes three extrema in $T$, likely a result of uneven magnetic confinement.  The reader is encouraged to examine animations of this simulation (available on-line\footnote{\url{http://people.virginia.edu/\~jpr8yu/azeus/proto\_jets.html}}), where the propagation of the knots is seen to be extremely dynamic.

Figure \ref{fig:zslice40} is an axial slice of several variables at $r=0.4$ AU ($8r\rmi$) at the final epoch of simulation F.  This plot, however, could have been taken at any time $t>10$ yr, so regular are the knots.  Relative to the cold sheath material, the knots are 10--20 times denser, $\sim\!10^3$ times hotter, and thus have $10^4$ times the thermal pressure of their immediate surroundings!  Were it not for the fact that $\beta$ is still $<1$ within these `plasmoids', they would explode into the ambient gas upon creation.  As it is, the knots are magnetically confined (the total pressure in Fig.\ \ref{fig:zslice40} is nearly continuous) by strong poloidal flux loops that effectively contain them as they propagate intact $\sim50$ knot radii along the jet axis. Other differences in the knot material compared to their immediate surroundings include: $v\pol$ is 10--20\% higher, $v_\varphi$ is $\sim50\%$ lower, $B\pol$ is $\sim50$\% lower, and $|B_\varphi|$ is $\sim20$\% lower.

The fact that the knots move slightly faster than the surrounding sheath material is interesting.  In this simulation, the difference of a few km/s is still super-slow, and thus the knots excite leading slow shocks---much like the jet itself excites a bow shock in the ambient medium---explaining their discontinuous leading edge (Fig.\ \ref{fig:zslice40}) and minimal diffusion.

We resume discussion on the knots in Sect.\ \ref{sub:knotgen}, where the physics of knot generation is addressed.

\subsection{Weak-field simulations (G, H)}
\label{sub:weak-field}

\begin{figure*}
  \begin{center}
    \includegraphics[width=1.07\textwidth,clip=true]{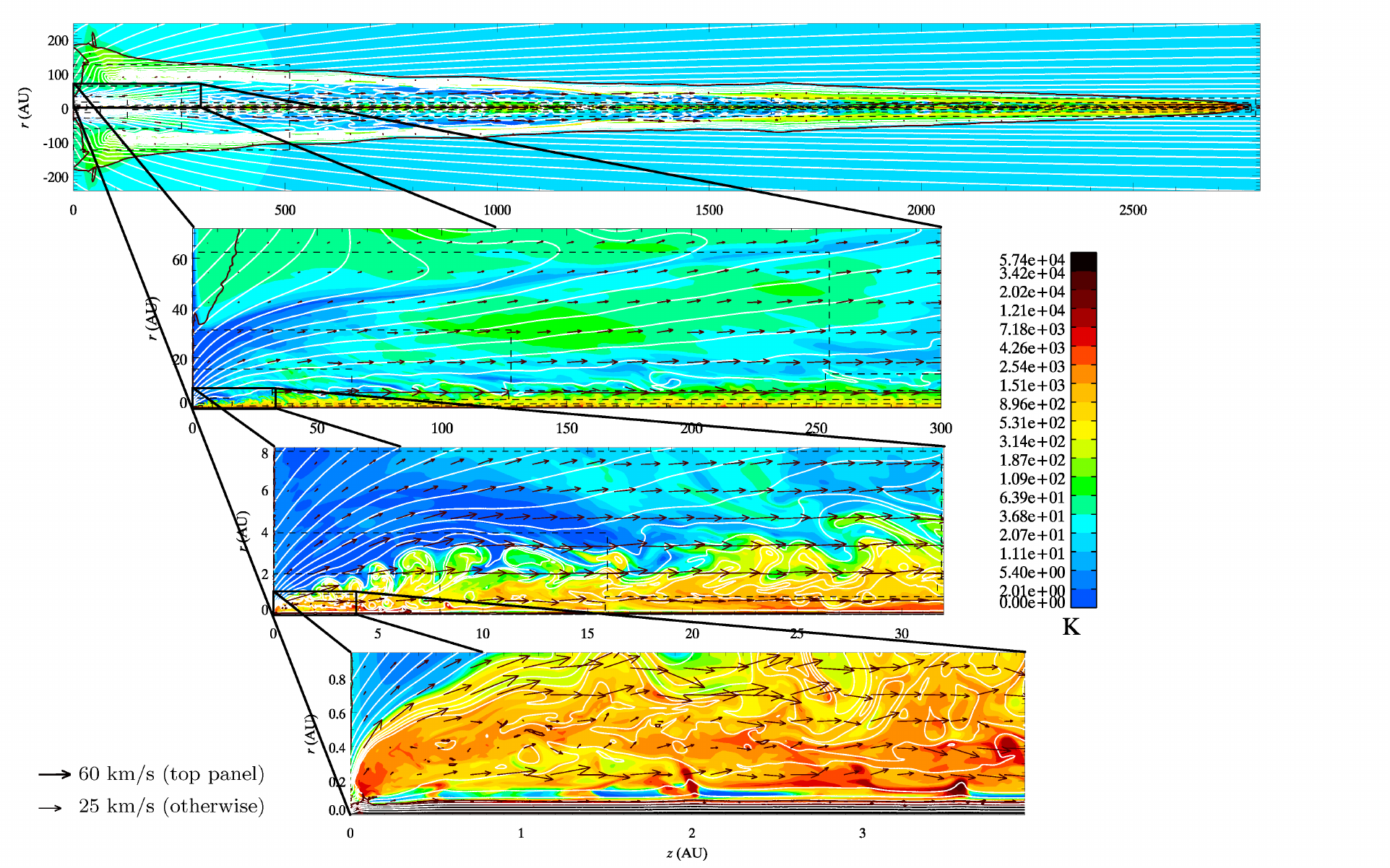}
  \end{center}
  \caption{\label{fig:zoom160} Similar to Fig.\ \ref{fig:zoom0.1}, but for simulation G ($\beta\rmi=160$) at $t = 153$ yr.}
\end{figure*}

Figure \ref{fig:zoom160} shows simulation G at $t = 153$ yr having reached a length of $\sim2,\!800$\,AU.  Evidently, $\beta\rmi$ has passed a critical value as the qualitative appearance of the inner jet is dramatically different even from simulation F with $\beta\rmi$ just a factor of four higher.  The jet speeds continue to diminish with $B\rmi$ ($v_{\rm jet}\sim86$\,km\,s$^{-1}$, $M_{\rm ext}\sim110$, $M_{\rm int}\sim12$) while the fast magnetosonic Mach number increases ($M_{\rm f,int}\sim7.5$; Table \ref{tab:simsum2}).  However, the most striking difference is the nearly complete replacement of organised knots with `turbulent' outflow and displacement of the initial poloidal field lines within the jet core and sheath (vestiges of the knots---the `bases' identified in Sect.\ \ref{sub:medium-field}---are, however, still apparent near the jet axis).  This trend is even more evident in simulation H (not shown), where the turbulent nature of the outflow extends right to the TD, all but eliminating it as a discernible feature. The spine---with its nearly straight axial field---remains prominent and hot next to the axis, and while much of the outflow is still launched from within $r\rmi<r<2r\rmi$, the organisation and steadiness observed in the strong and medium-field simulations is not apparent in the weak-field cases.

Still, on the observational scale (top panel of Fig.\ \ref{fig:zoom160}), the jet looks much the same as the stronger-field simulations.  \emph{Regardless of how weak the magnetic field is at the base of the jet, the gravitational and rotational effects organise to amplify the magnetic field enough to generate a large-scale, supersonic outflow transporting significant magnetic field energy}.  This is true even for simulation H with $\beta\rmi=640$, what ought to be considered an essentially hydrodynamical environment.

As will be shown in Sect.\ \ref{sub:driving}, the weak $B\pol$ is continuously wound up into $B_\varphi$ until such time as $a_\varphi\sim c_{\rm s}$.  At this point, the outwardly-directed gradient in toroidal magnetic energy density is comparable to thermal pressure gradients and gravitational forces, and material can be launched and accelerated outward.  This is true even for the weakest initial magnetic field (simulation H) which, with very little poloidal field to contribute to jet confinement, gives rise to a turbulent jet that entirely fills the magnetic sheath.

As seen in Fig.\ \ref{fig:avgbeta}, once the outflow becomes organised in simulation G ($t\sim1$ yr), $\langle\beta_{\rm tot}\rangle$ quickly falls below $\beta\rmi$ to a surprisingly low value of $\sim0.02$, then rises steadily to $\sim0.6$ after which it declines asymptotically to $\sim0.4$.  Even in the weaker-field simulation H, when outflow first begins, its value of $\langle\beta_{\rm tot}\rangle$ is well under unity, and then rises---rather sporadically---toward unity as the run progresses.

The spikes in $\langle\beta_{\rm tot}\rangle$ seen for simulation H attest to the marginality with which steady outflow is established in this very weak magnetic environment.  Speculating that $\beta\rmi\sim$ several hundred may represent an `end-of-the-line' for a successful jet launch, we ran another simulation with $\beta\rmi=2,\!560$ to see if the MTM would prevail in such a weak magnetic environment.  Sure enough, even in this extreme limit, a jet is launched, albeit even more turbulent than simulation H and requiring even more time to organise itself into an outflow.

Of course, 2-D axisymmetry provides an artificially favour\-able geometry for the MTM in which a weak poloidal field can do nothing but wind up under the relentless rotation of the fluid.  Furthermore, axisymmetry is immune to all but the $m=0$ mode of the Kelvin-Helmholtz instability, which not only allows the toroidal field to develop unimpeded, but provides a perfectly rigid axis along which flow can be directed.

In 3-D, the physics is not so accommodating. With virtually no poloidal field to stabilise the flow, mixing of the shocked ambient medium with the sheath occurs along the K-H unstable TD while the turbulent flow encourages mixing of the jet core with the sheath.  With less contrast in density and temperature between the core and sheath, the core is at risk of higher order MHD instabilities that may cause it to break up into tangled filaments after propagating only several jet radii \citep{hardeeclarkerosen97}.  Further, an advancing, toroidally confined, magnetic column is unstable to the kink instability (\eg\ \citealt{jackson75_book}), exacerbating its ability to propagate to observable length scales.  We speculate, then, that in 3-D and in cases where $\beta\rmi\gtrsim$ a few hundred, a `frustrated jet' could result in which outflow never manages to organise itself to advance very far from the disc.  Other than indicating when 2-D turbulence fills the jet sheath ($160<\beta\rmi<640$), our present simulations are unable to determine what this critical value for $\beta\rmi$ may be.

\section{Analysis}
\label{sec:analysis}
\subsection{Morphology}
\label{sub:morphology}
\begin{figure*}
  \begin{center}
    \includegraphics[width=0.95\textwidth,clip=true]{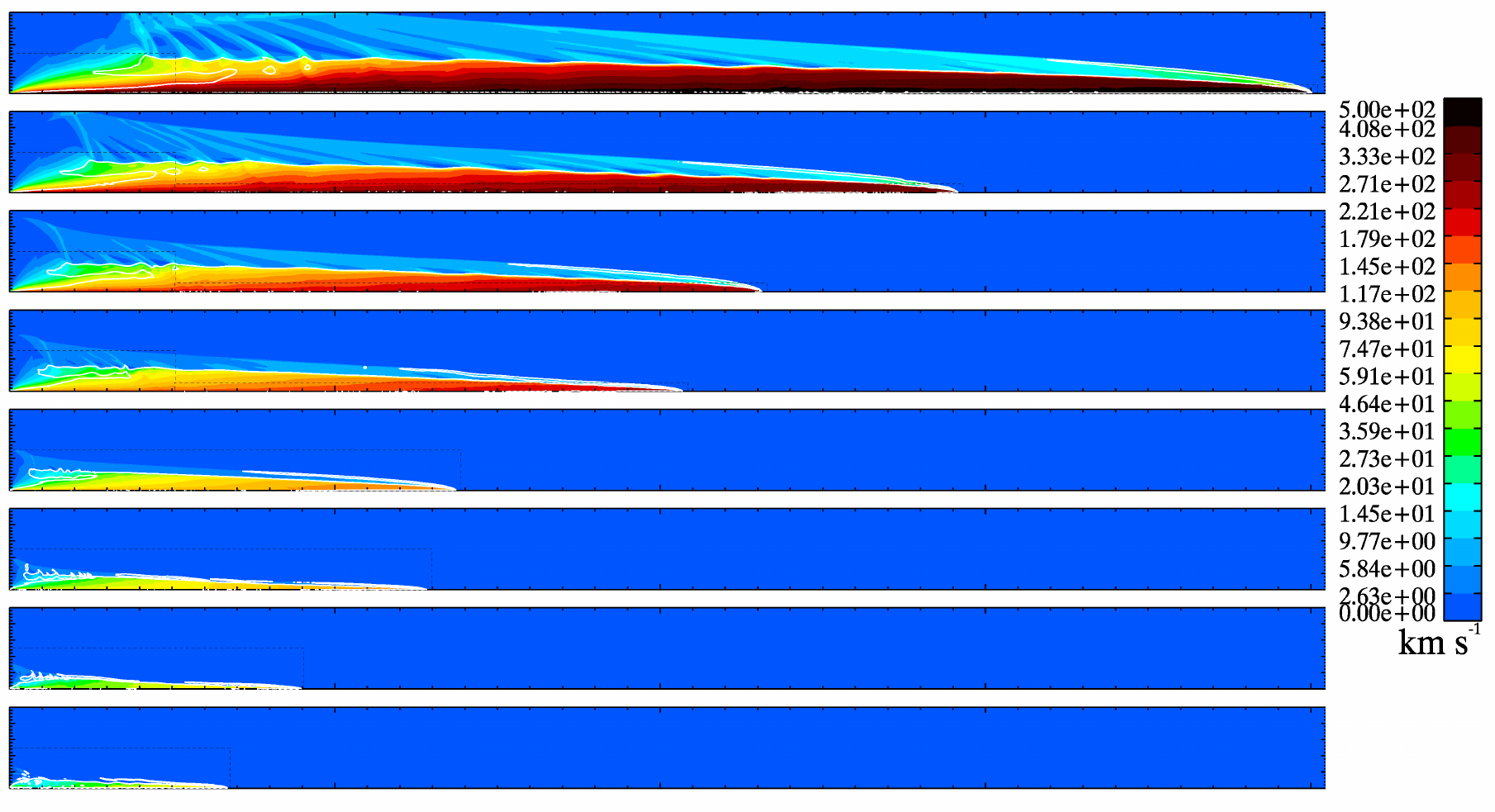}
  \end{center}
  \caption{\label{fig:chrontim_v} Colour contours of the poloidal velocity, $v\pol$, from simulations A--H (top to bottom) at $t\sim47$\,yr with the fast magnetosonic surface ($M_{\rm f} = 1$) shown in white contours. The lower left corner of each panel is at the origin, $(z,r) = (0,0)$, and the upper right corner at $\sim\!(4070, 250)$ AU.}
\end{figure*}

To the best of our knowledge, these are the first jet simulations to link the physics of magnetically driven outflows from Keplerian discs, as first reported by \citet{uchidashibata85}, and the physics of magnetically collimated supersonic outflows, as first reported by \citet{clarkenormanburns86} (hereafter, CNB).

Regardless of $\beta\rmi$, all simulations form a magnetically dominated outflow ($\langle\beta_{\rm tot}\rangle\rightarrow0.2$--0.4, with the exception of simulation H) which come to resemble CNB jets (who, coincidentally, used $\beta_\varphi=0.2$) at observational scales.  Thus, they all possess a hot, supersonic core with an advance speed of 80--420\,km\,s$^{-1}$ and a number of oblique shocks triggered along their length.  The core terminates with a `jet shock' (occasionally a Mach stem), with most of the post shock jet material collected in front of the jet shock forming a `nose-cone'.

As discussed in CNB, the nose-cone consists of trans-fast material---hotter than in the jet core---which, without the confining magnetic field, would form the `back-flowing cocoon' reported by \citet{normanetal82}.  Since the nose-cone is more pointed and denser than the hydrodynamical cocoon created in the absence of $B_\varphi$, a magnetically confined jet pushes into the ambient medium more ballistically than a hydrodynamical jet, consistent with the present simulations at the largest scale.

Surrounding the jet core and as reported by CNB, a largely evacuated `cocoon' filled with cold, highly magnetised rarefied material with a strong toroidal magnetic field confines and stabilises the jet core.  In the present simulations, we refer to this feature as the `sheath', prominent in all simulations except H, the most weakly magnetised jet in our sample.

Surrounding the cocoon (sheath) and the leading nose-cone is a bow shock excited in the quiescent ambient medium by the passage of the supersonic jet.  The shocked ambient material is accelerated forward ($\lesssim$\! 10 km\,s$^{-1}$) forming what we identify as the `second wind', consisting of material not launched from the disc but entrained \emph{in situ} by the bow shock from the ambient atmosphere.  As such, it is warm (300--30,000\,K), weakly magnetised, has virtually zero toroidal velocity and field, with a magnetic field strength directly related to the magnetic conditions in the primordial atmosphere and thus proportional to $B\rmi$.

Because the ambient atmosphere is so cold, the Mach number of the jet relative to the external atmosphere is very high, even for simulation H where $M_{\rm ext}\sim 100$.  Thus, the bow shock is very narrow and the radius of the jet depends very weakly on $B\rmi$ making the jet radius a poor indicator of the magnetic conditions near the disc surface.  Indeed, the morphology of the \emph{observable} jet as a whole---being so similar for all values of $B\rmi$---cannot be used as an indicator of this most elusive of physical quantities.

It is worthwhile noting that our choice of magnetic field distribution (Sect.\ \ref{subsub:atmos}) plays a role in determining the outflow morphology. A steeper radial field distribution ($B(z=0,r) \propto r^{-1}$ in our case) is known to result in less collimated jets \citep{pro06,fendt2006}, and could lead to jets with larger opening angles and larger radii.  We also recognise that, in 3-D, with the availability of additional K-H modes, some of the jet kinetic energy would be converted to turbulent or thermal energy, and we subsequently expect the jet bow shock to not only be blunter and wider, but also propagate more slowly. Finally, on very large scales ($\sim$0.1\,pc; \citealt{franketal1999}), jets are expected to be weakly ionised and ambipolar diffusion will be important, resulting in less efficient magnetic confinement and an additional source of heating \citep[e.g.][]{pintoetal2008,panoglouetal2012}. Like the magnetic field distribution, any of these effects could produce a jet that is less `knife'-like than in the current simulations.

\begin{figure}
  \begin{center}
    \includegraphics[width=1.0\columnwidth,clip=true]{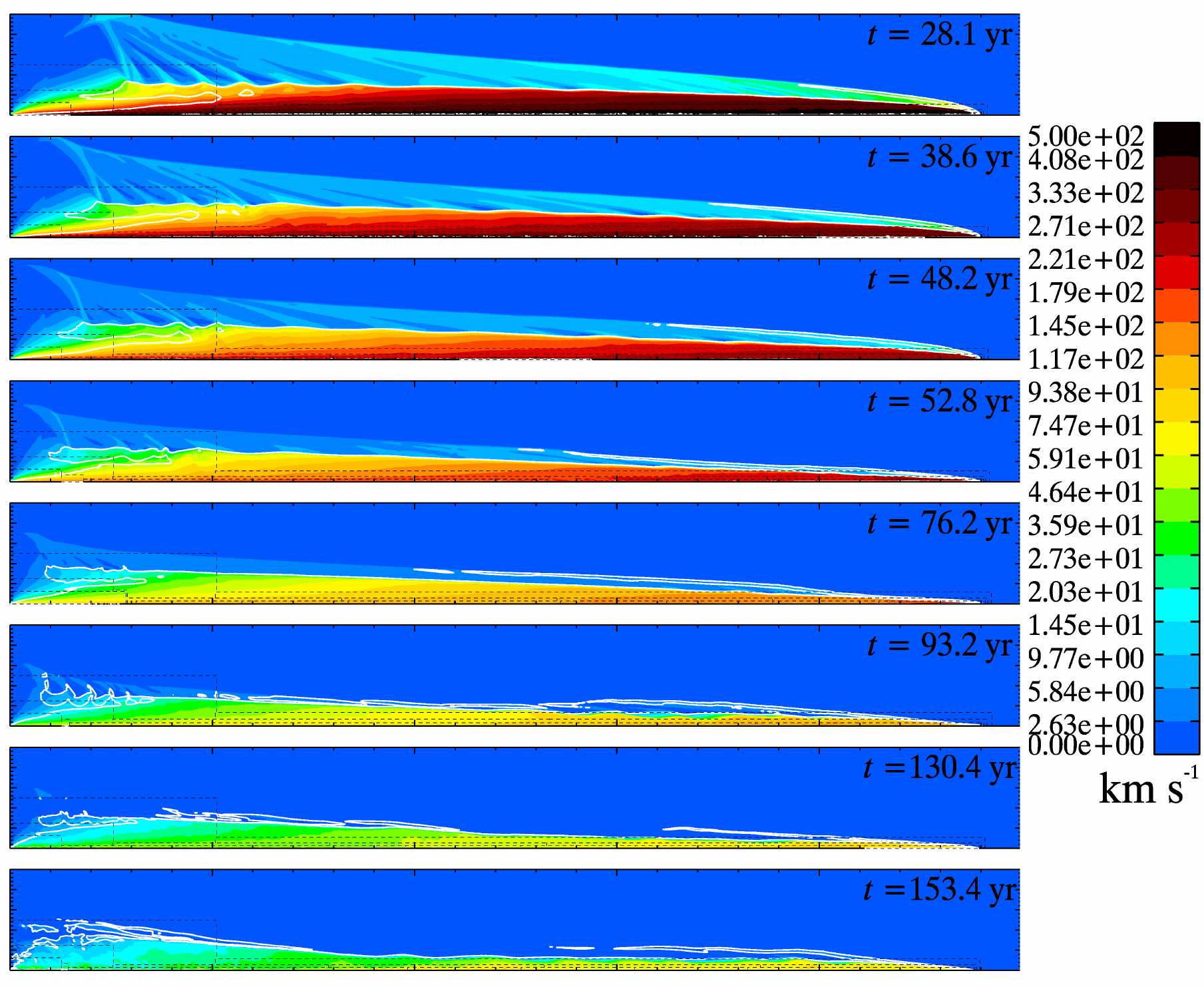}
  \end{center}
  \caption{\label{fig:dyntim_v} Similar to Figure \ref{fig:chrontim_v}, but at the same `dynamical' time, $l=2,\!400$\,AU, with the chronological ages indicated. The radial extent of each panel is $\sim\!250$ AU.}
\end{figure}

\subsection{Qualitative trends}
\label{sub:trends}
\begin{figure*}
  \begin{center}
    \includegraphics[width=0.32\textwidth]{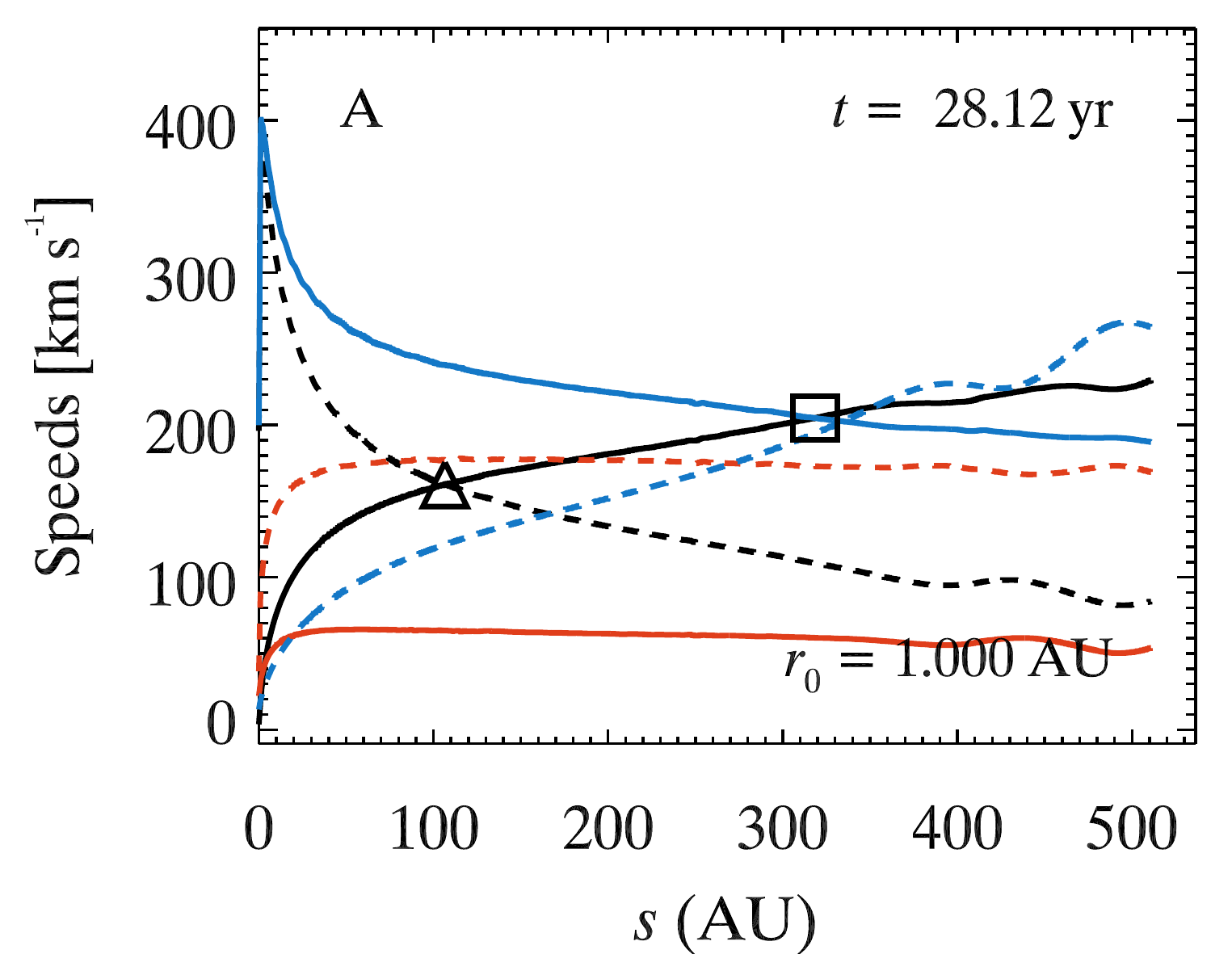}\includegraphics[width=0.32\textwidth]{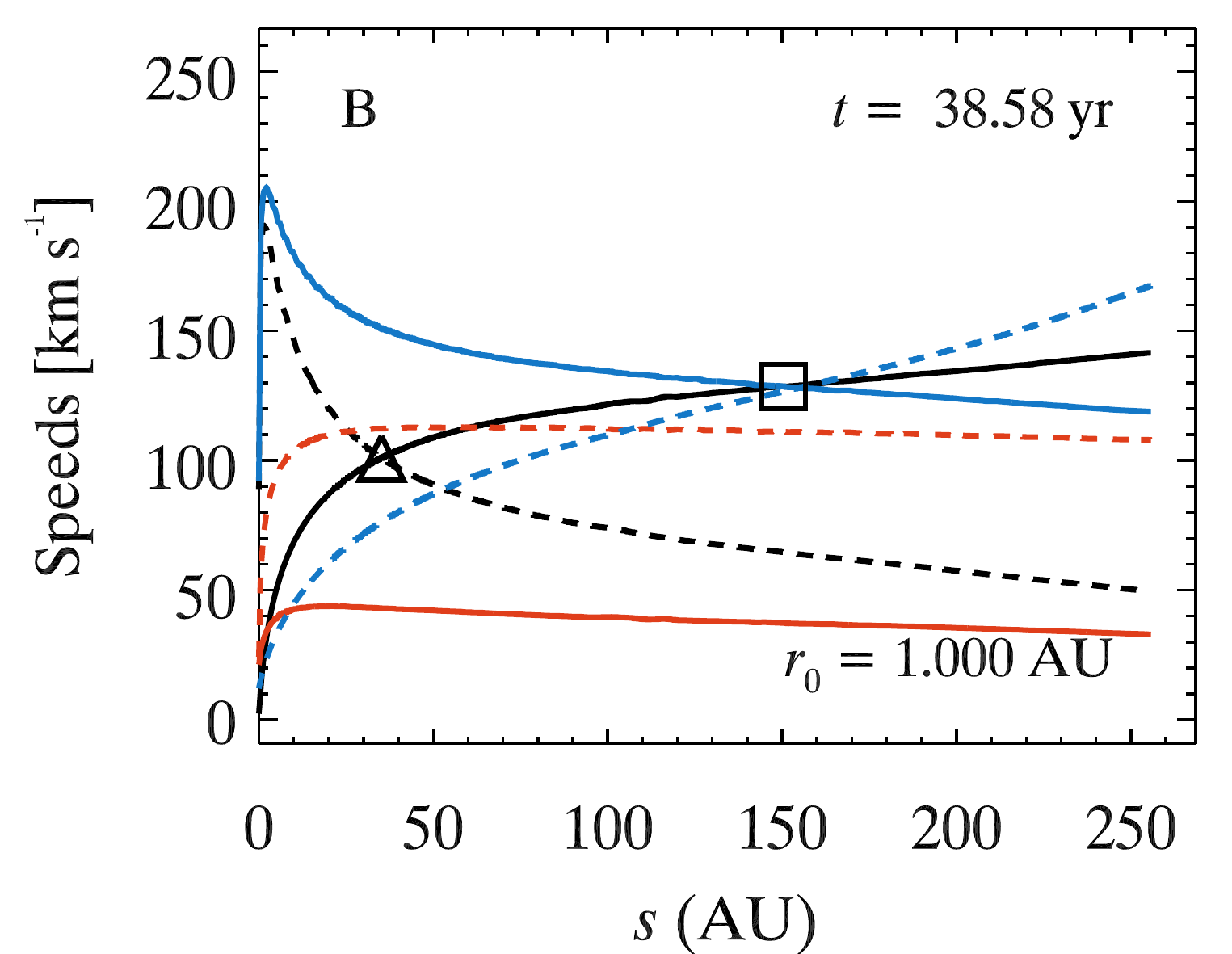}\includegraphics[width=0.32\textwidth]{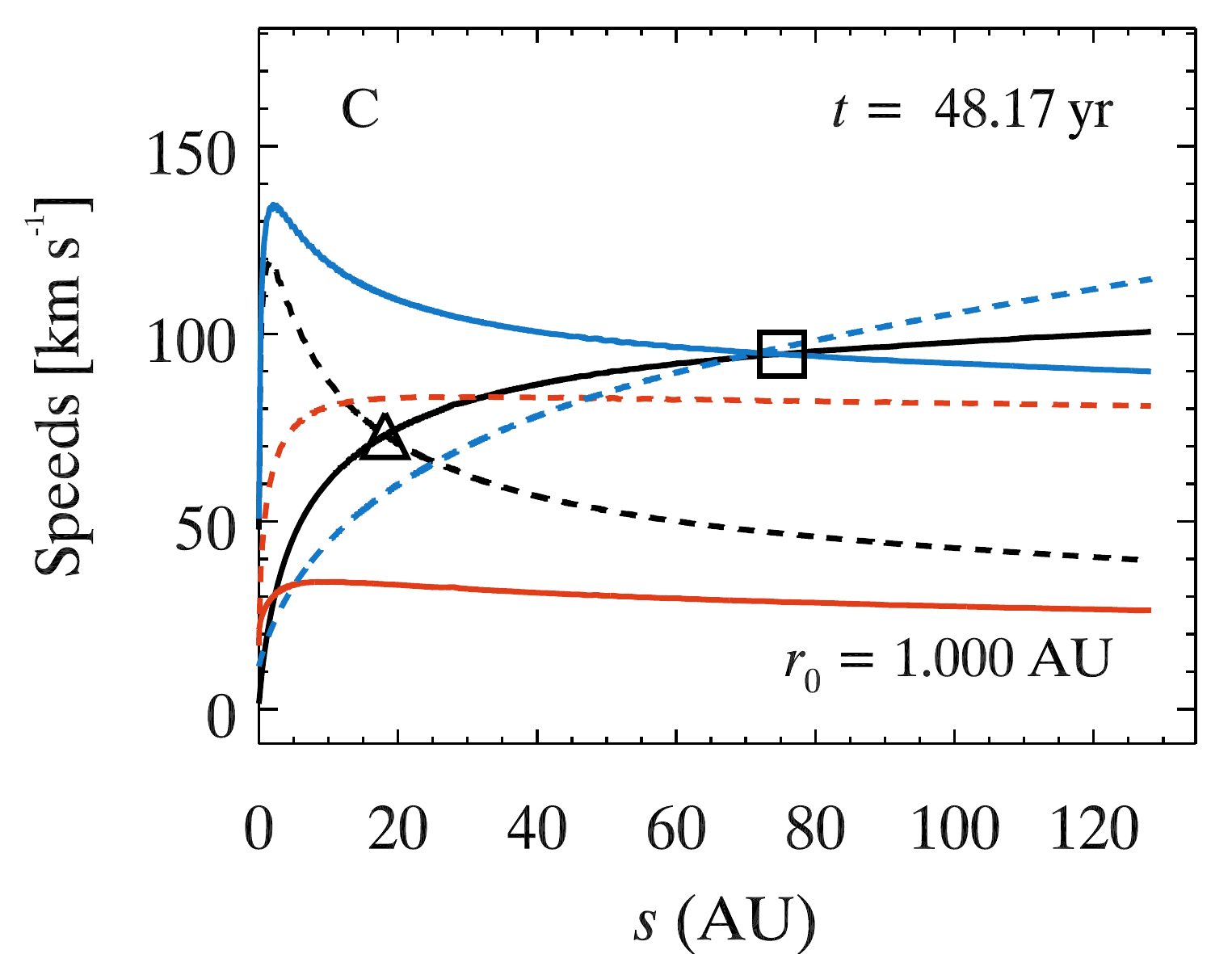}
    \includegraphics[width=0.32\textwidth]{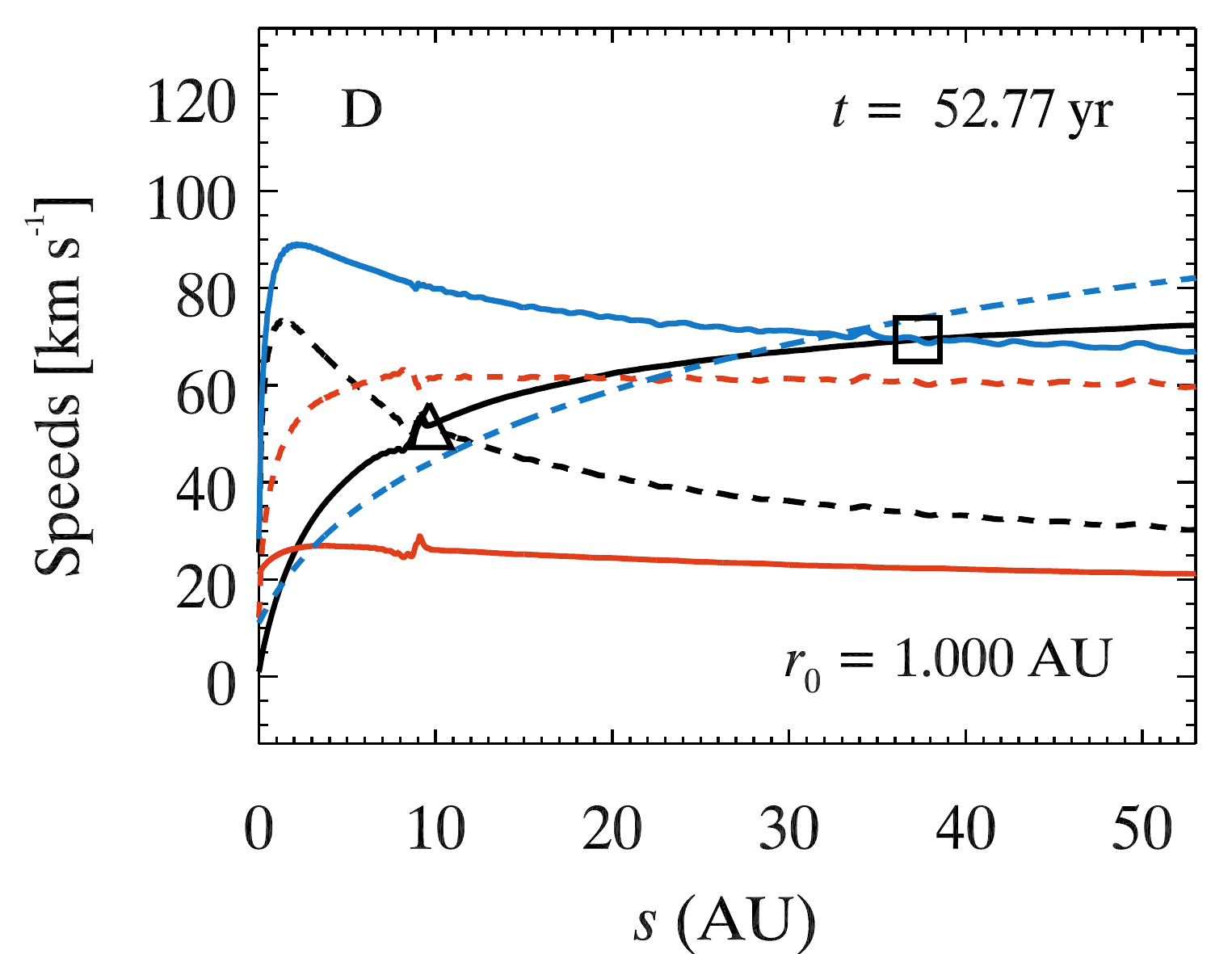}\includegraphics[width=0.32\textwidth]{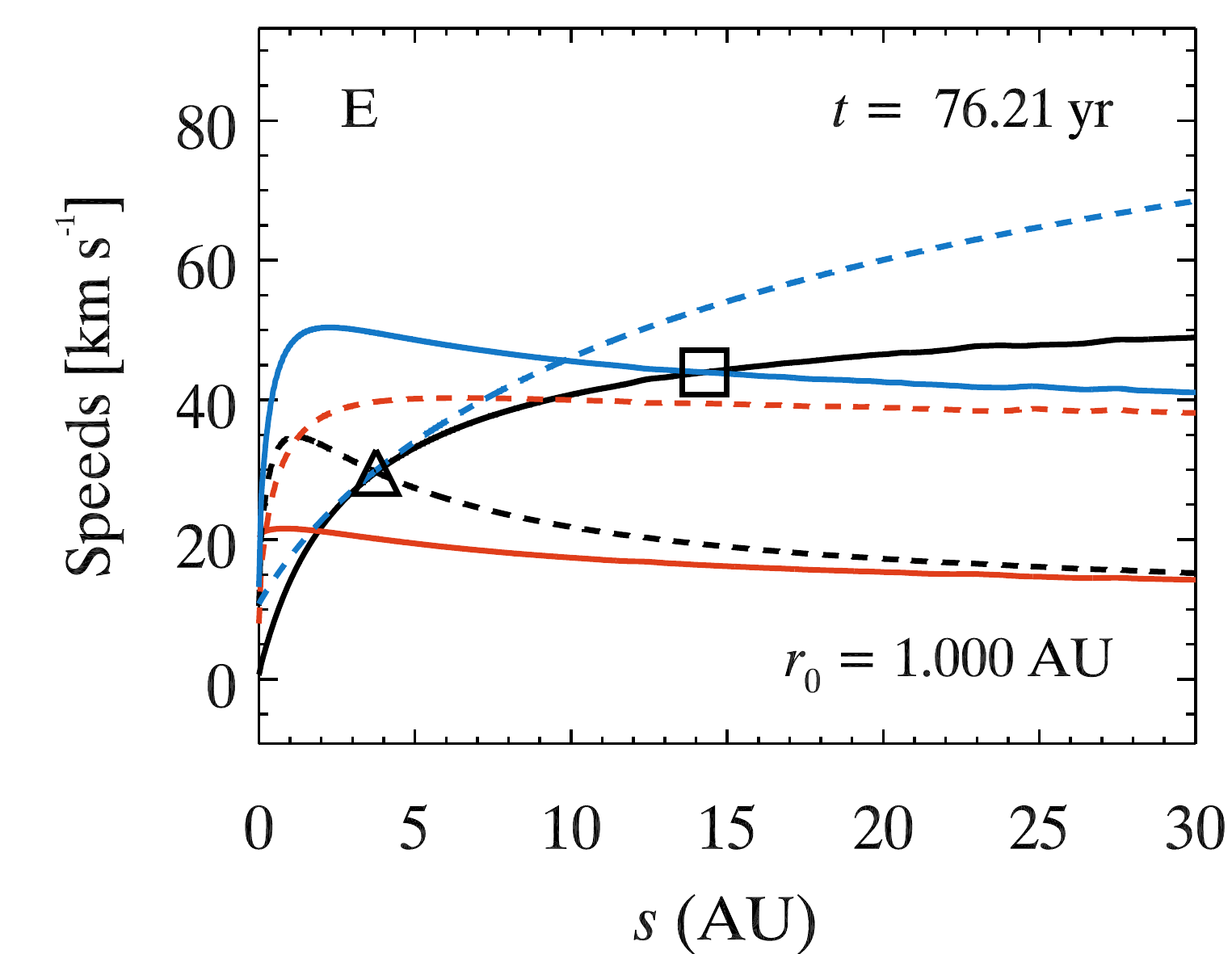}\includegraphics[width=0.32\textwidth]{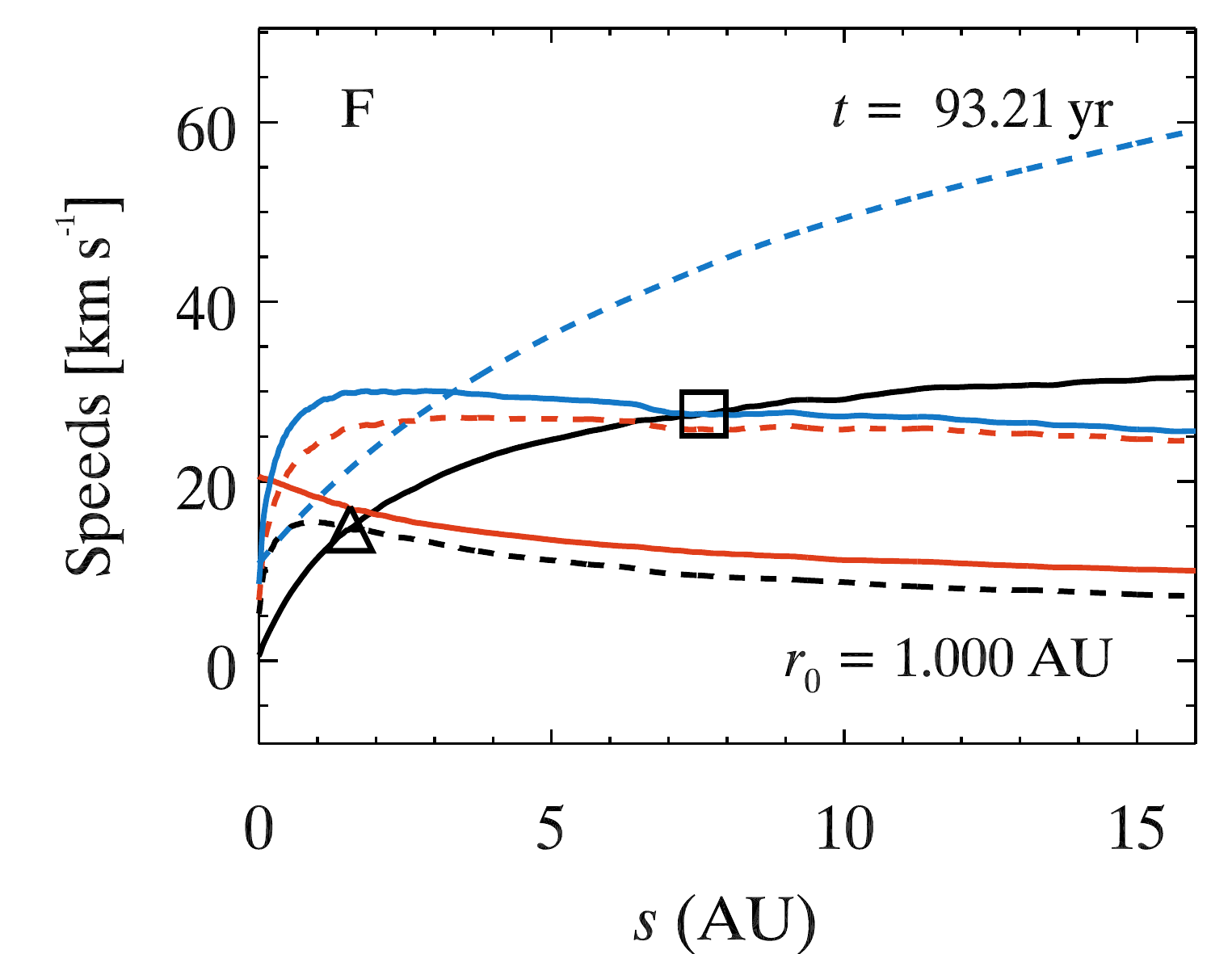}
    \includegraphics[width=0.32\textwidth]{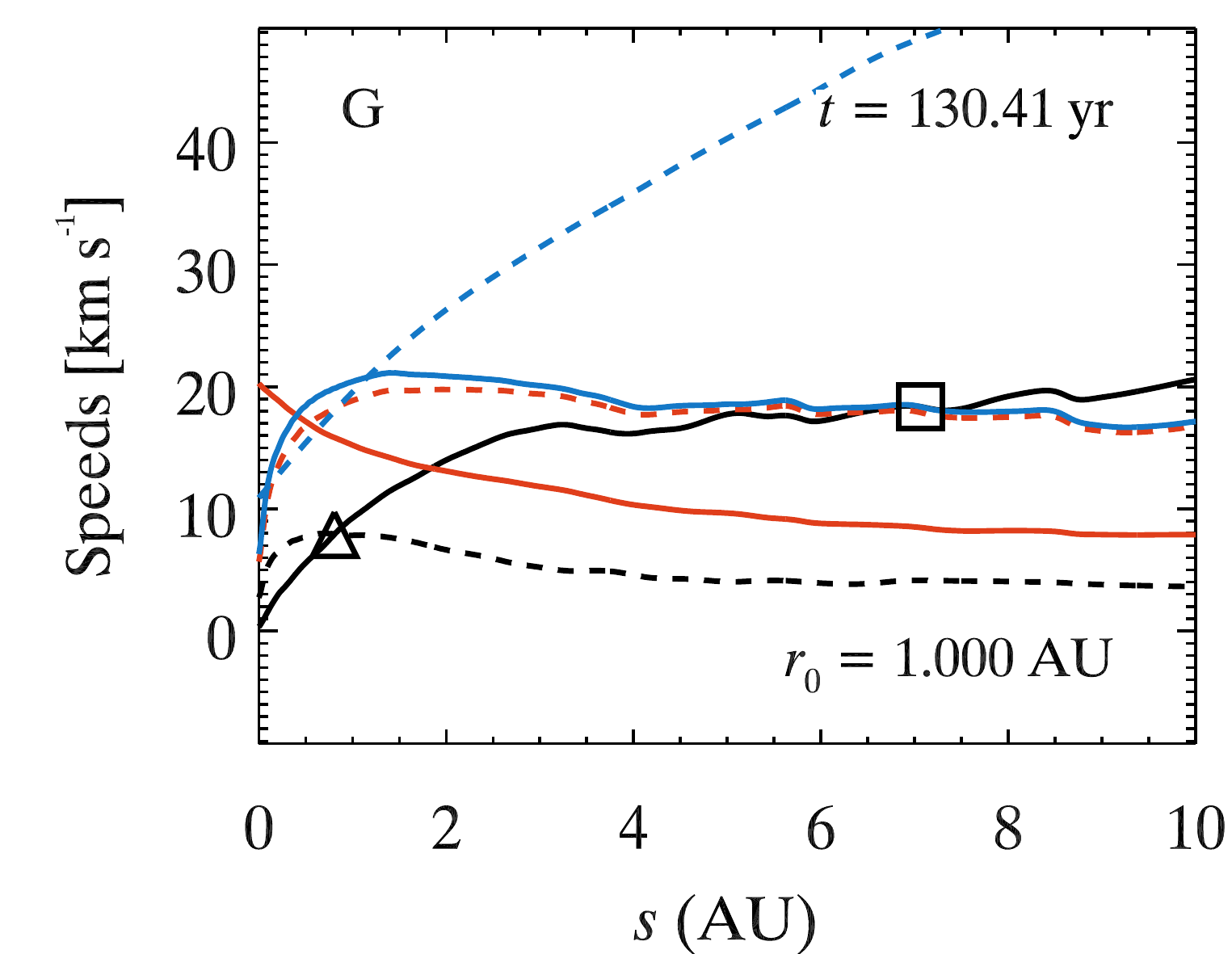}\includegraphics[width=0.32\textwidth]{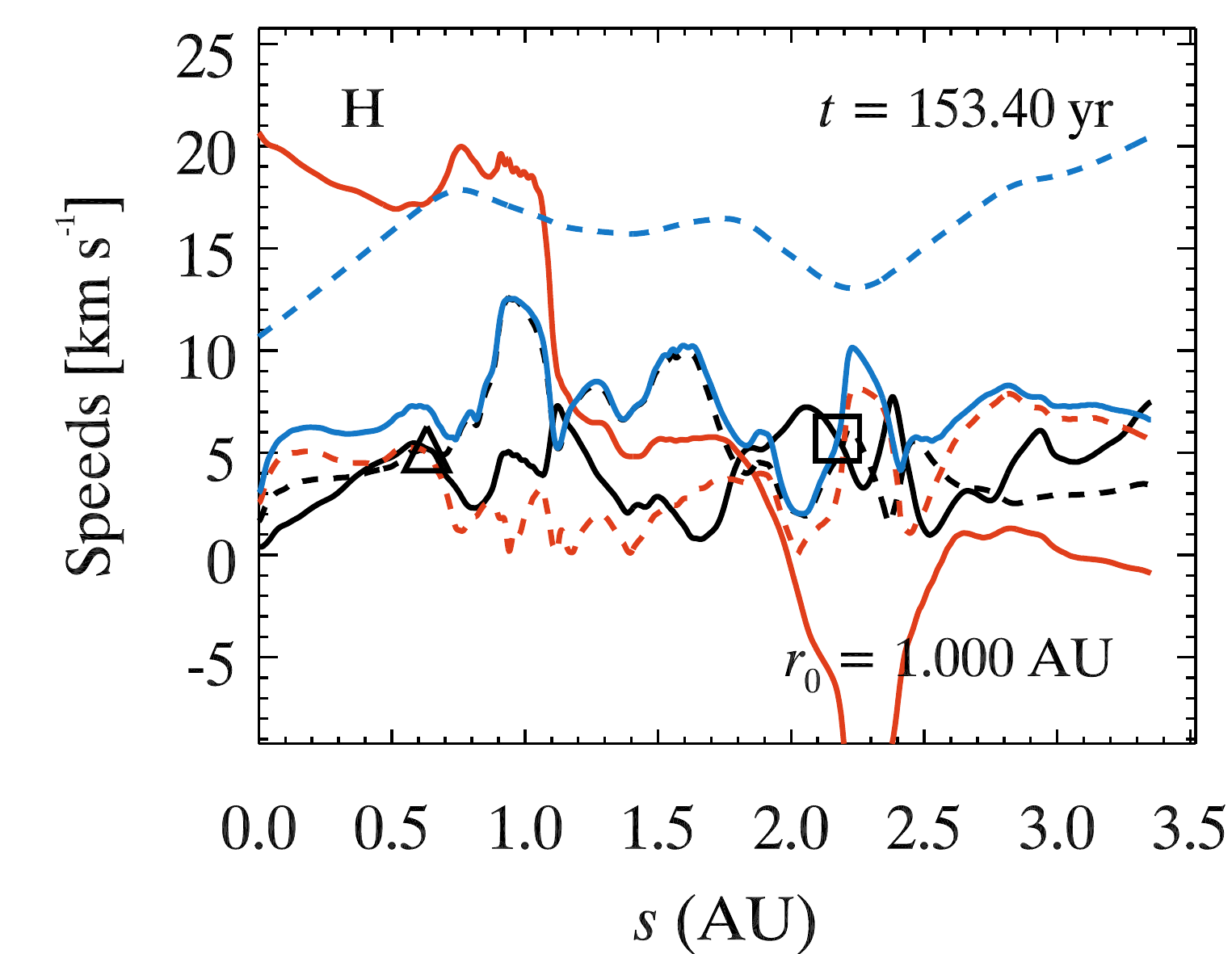}\includegraphics[width=0.32\textwidth]{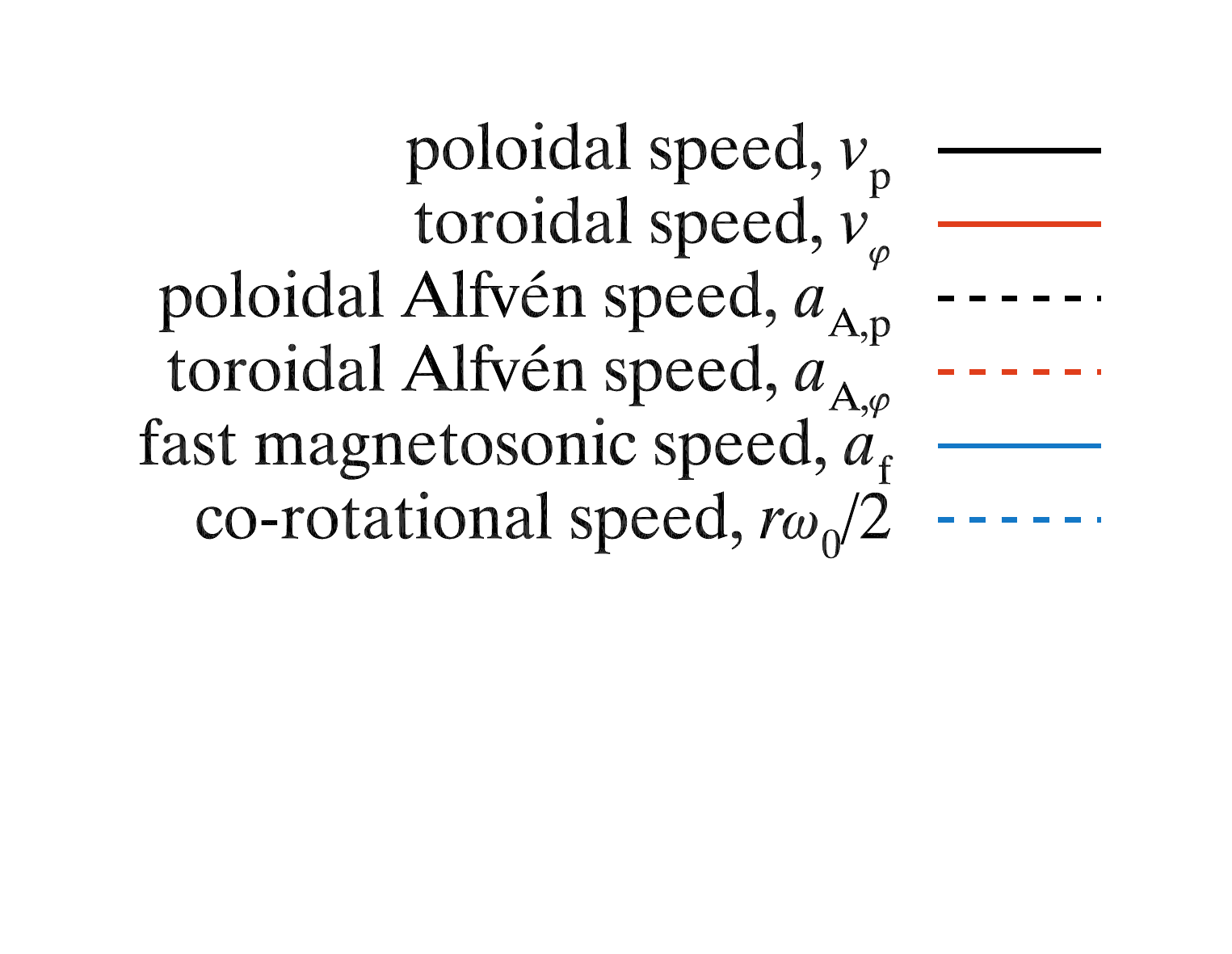}
  \end{center}
  \caption{\label{fig:velocitiesAH} Various speeds as a function of position, $s$, along the field line anchored at $r_0=1$ AU and at dynamical time $l$ = 2,400 AU for simulations A--H.  The Alfv\'en and fast points are indicated by open triangles and squares, respectively.  The smooth profiles for simulations A--F are indicative of a steady state, whereas the ragged profiles for simulation H indicates no steady state has been reached.}
\end{figure*}

Certain quantities, such as the advance speed of the jet ($v_{\rm jet}$), reach an asymptotic limit early in the simulation (\eg\ after $\sim10$ yr), making comparisons among the simulations easy and straight-forward.  Other quantities such as the jet radius ($r_{\rm jet}$) continue to grow with time, and thus one must decide how such quantities are to be compared.  In particular, one could make comparisons at the same \emph{chronological time} or at the same \emph{dynamical time}, the latter defined as when a jet has reached a specified length.

Fig.\ \ref{fig:chrontim_v} shows the poloidal velocity, $v\pol$, and the fast magnetosonic surface (white contours) for all eight simulations at the same chronological time, $t\sim47$\,yr, when simulation A reaches the end of the domain.  This figure illustrates how the jet length and radius at a fixed time depend rather strongly upon $B\rmi$.

However, observationally, one can be more certain of comparing jets of the same length than of the same age, and thus Fig.\ \ref{fig:dyntim_v}, where the jets are shown at the same dynamical time $l=2$,400\,AU, is more practical.  In this case, there is little to distinguish the eight simulations geometrically.  At a given length, the radii of the sheath (indicated, for the most part, by the TD) and jet core increase slightly with decreasing $B\rmi$, but the dependence is so weak as to make these impractical observational comparators for inferring the magnetic field properties at the base of a jet.

\begin{figure}
  \begin{center}
   \includegraphics[width=0.5\columnwidth]{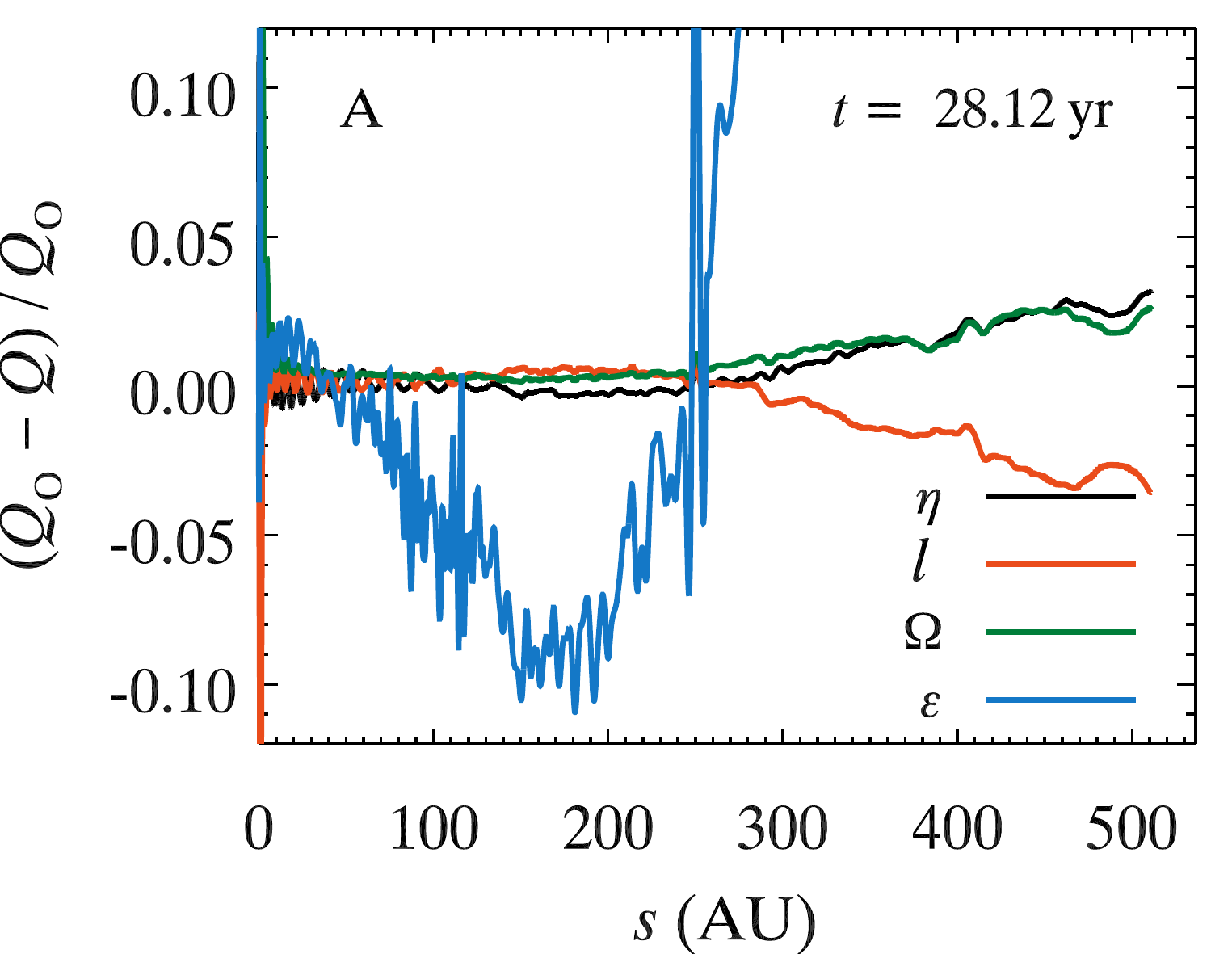}\includegraphics[width=0.5\columnwidth]{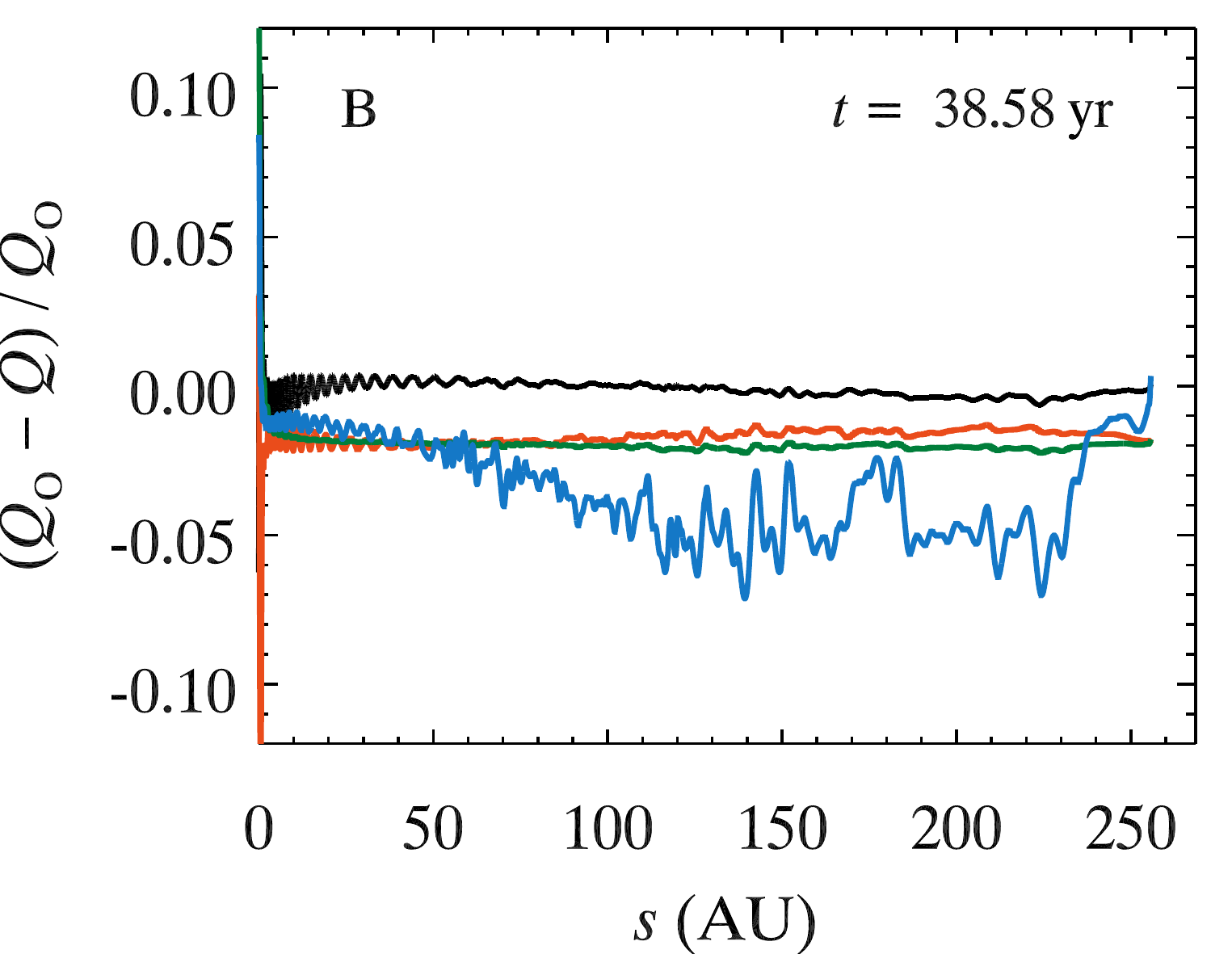}
   \includegraphics[width=0.5\columnwidth]{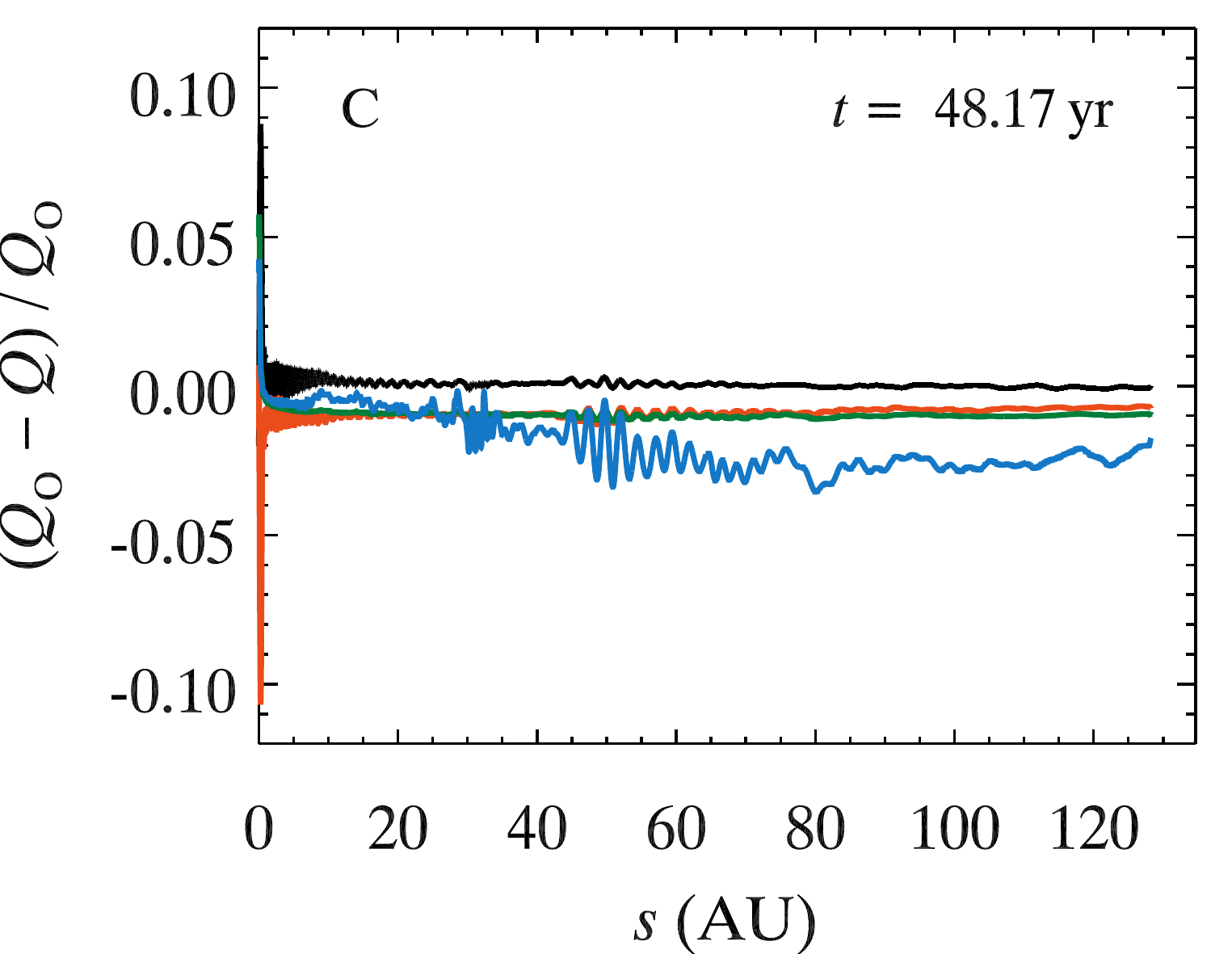}\includegraphics[width=0.5\columnwidth]{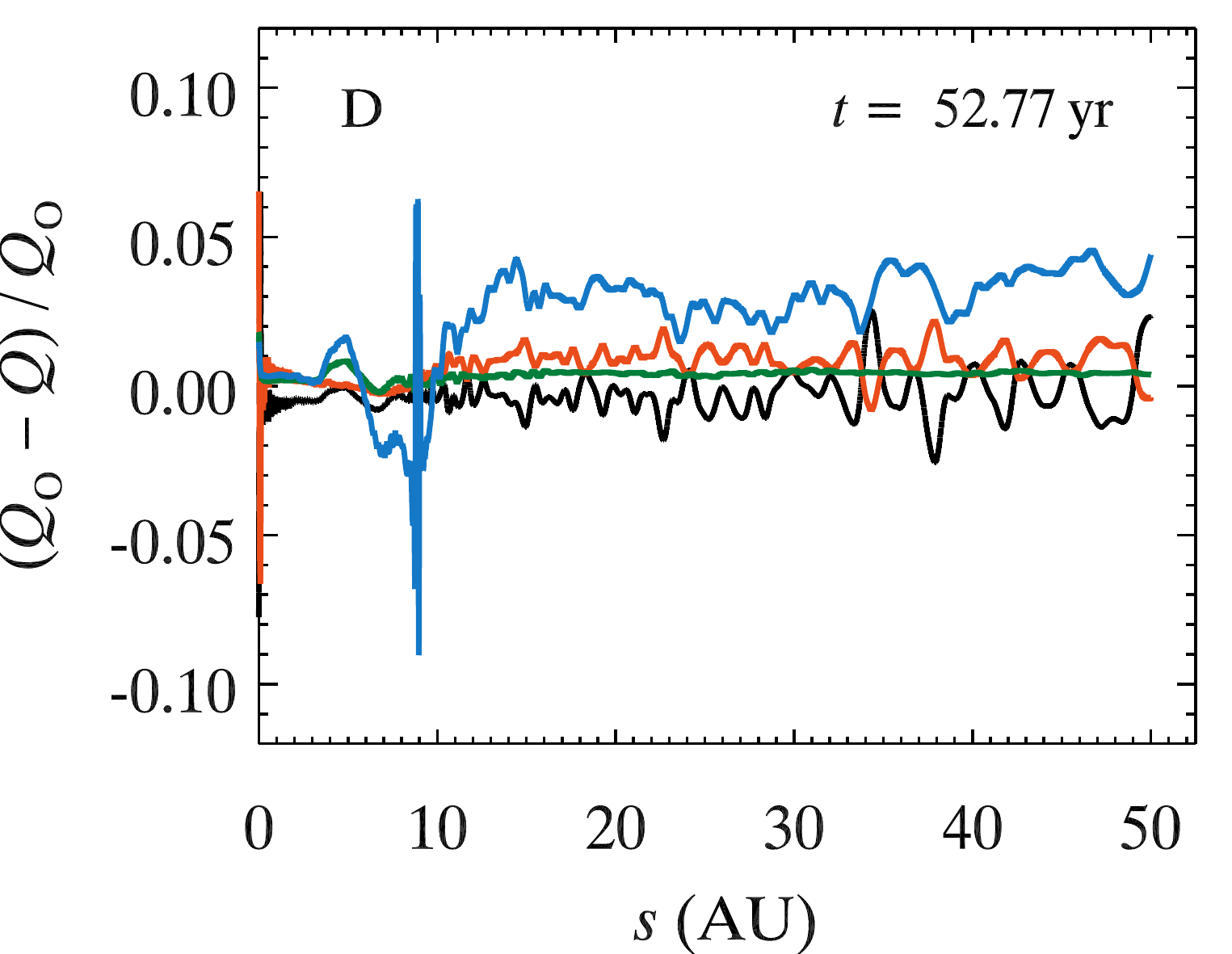}
   \includegraphics[width=0.5\columnwidth]{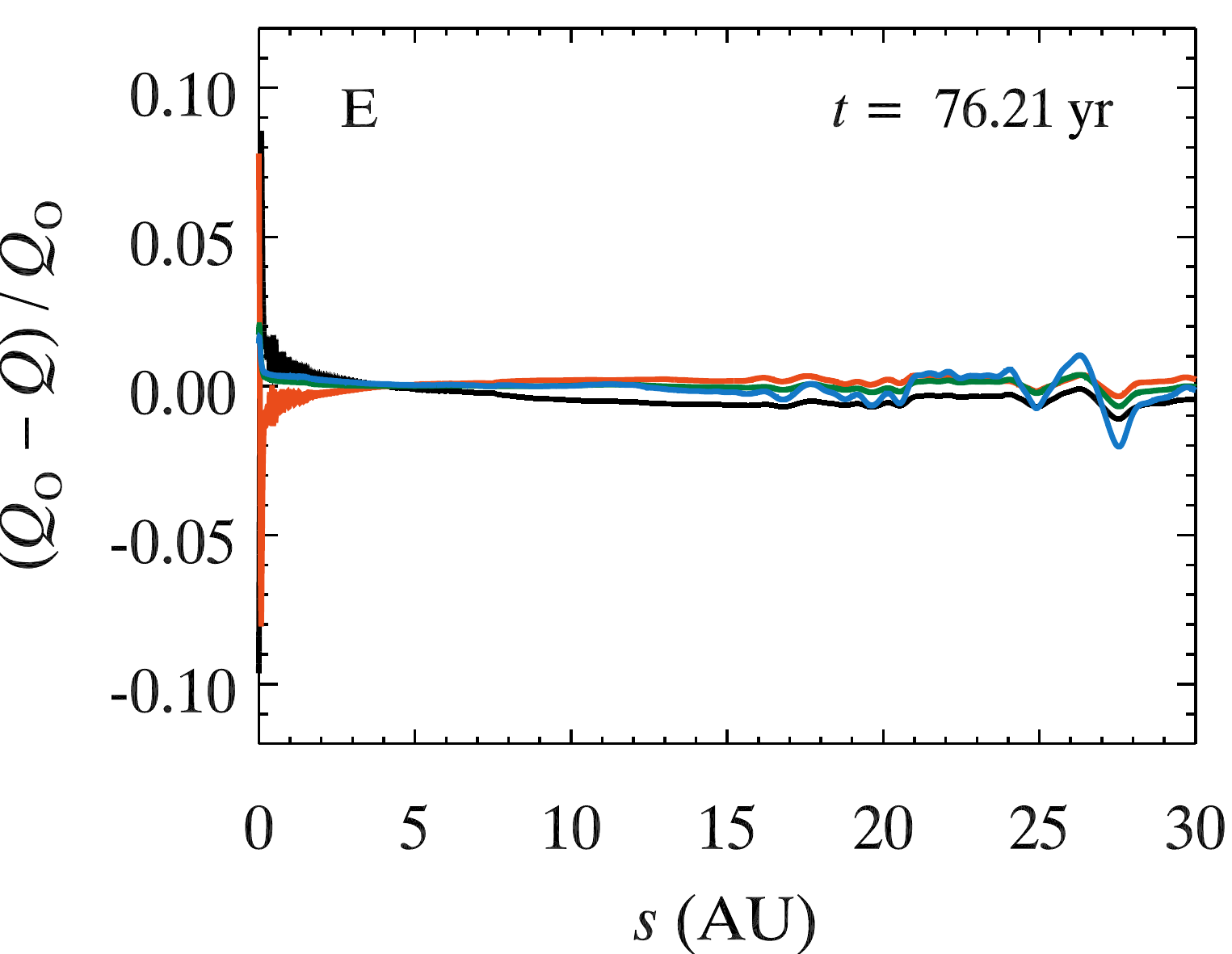}\includegraphics[width=0.5\columnwidth]{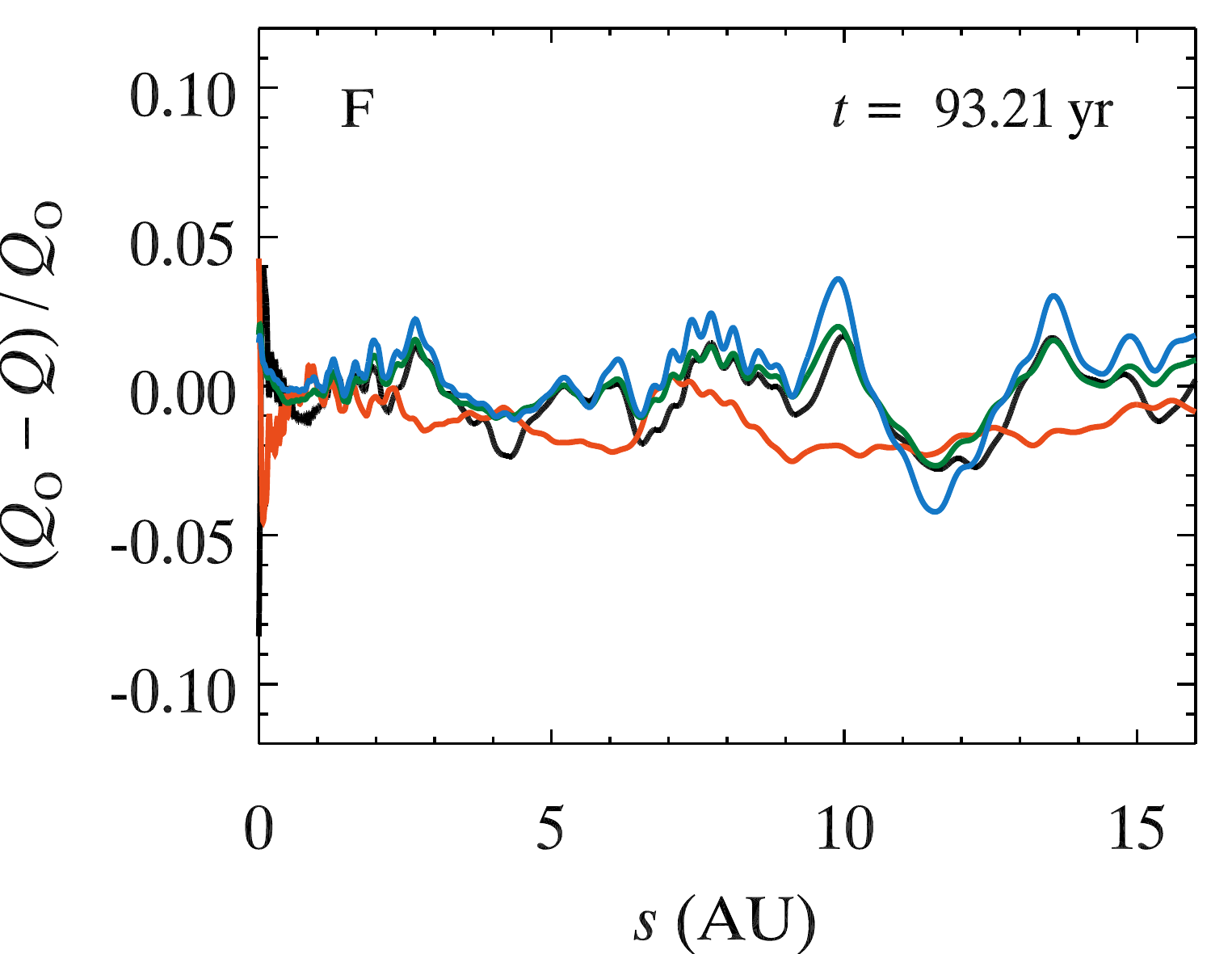}
  \end{center}
  \caption{\label{fig:steadystate} Fractional variation of the `WD constants' (eqs.\ \ref{eq:massload}--\ref{eq:energy}) with respect to expected values (eqs.\ \ref{eq:omegaE} and \ref{eq:etaL}) for simulations A--F along the field line anchored at 1 AU. Most variations remain within 3\%, with notable exceptions identified in the text.}
\end{figure}

\subsection{Weber-Davis constants revisited}
\label{sub:steadystate}

We now return to the WD constants defined in Sect.\ \ref{sec:steadystate} (Eqs.\ \ref{eq:massload}--\ref{eq:energy}), as well as Eq.\ (\ref{eq:fastpoint}) for the fast speed along a steady-state field line.  Since most of our global simulations do not reach a true steady state, these expressions will not be valid along all field lines.  However, as seen in Figs.\ \ref{fig:zoom0.1}, \ref{fig:zoom40} and, to a much lesser extent, Fig.\ \ref{fig:zoom160}, the first few hundred AU along field lines anchored in the disc at $0.5\lesssim r_0\lesssim 10$\,AU are smooth and appear to reach a quasi-steady state.  Within these regions the steady state constants can be examined, both to provide an \emph{in situ} check on our numerical methods, and for what they can tell us about the flow.

For convenience, we have chosen the field line anchored in the disc at $r_0=1$\,AU to perform this analysis as it exemplifies steady state behaviour for all but the weak-field simulations.  However, we emphasise that these results apply equally for portions of any field line anchored between 0.5 AU and 10 AU, with some simulations showing better steady state behaviour than others.  For the 1 AU field line, Fig.\ \ref{fig:velocitiesAH} shows plots of various speeds as a function of distance along the field line, $s$, including $v\pol$, $v_\varphi$, $a\pol$, $a_\varphi$, and $a_{\rm f}$, all defined in Sect.\ \ref{sec:steadystate}.  The dashed blue line shows $r\omega_0/2$ (where $\omega_0=v_{\rm K,0}/r_0$ is the angular speed at the anchor point of the field line), an important quantity in understanding the BWM (Sect.\ \ref{sub:driving}).

The Alfv\'en point ($v_{\rm p,A}$, where $v\pol$ and $a\pol$ intersect in Fig.\ \ref{fig:velocitiesAH}) and fast point ($v_{\rm p,f}$, where $v\pol$ and $a_{\rm f}$ intersect) are indicated in the figure with triangles and squares, respectively.  The quantities $v_{\rm p,A}$, $v_{\rm p,f}$, as well as the asymptotic poloidal speed attained along the 1 AU field line, $v_{\rm p,max}$, are included in Table \ref{tab:simsum2}.  For the 1 AU field line and most any other steady state field lines we examine, acceleration of jet material continues well beyond the fast point (via the MTM), less so for weaker $B\rmi$.

Figure \ref{fig:steadystate} shows profiles of the fractional differences of each WD constant (Eqs.\ \ref{eq:massload}--\ref{eq:energy}) relative to their expected values (Eqs.\ \ref{eq:omegaE} and \ref{eq:etaL}) in each of simulations A--F.  As can be seen, the WD constants remain---for the most part---constant to within 3\% along the 1 AU field line.  Notable exceptions include simulation F ($\pm5\%$), which is really a transitional simulation between those which exhibit strong steady state regions (A--E), and those that do not (G and H), and the specific energy constant, $\varepsilon$, for simulations A and B, which is dominated by the difference between two large and nearly equal numbers ($v_\varphi^2/2$ and $rv_\varphi\Omega$ in Eq.\ \ref{eq:energy}).  This dominance decreases with $B\rmi$ and, as such, $\varepsilon$ is constant to within 3\% for simulations D--F.  Simulation D also exhibits a transient feature in $\varepsilon$ at $\sim$9 AU from when the field line passes through one of the streaks mentioned in Sect.\ \ref{sub:strong-field}; this feature is not visible in the other steady state constants and is only barely visible in Fig.\ \ref{fig:velocitiesAH}. We note that the field lines embedded in quasi-steady state regions pass through several nested grids, and take this as evidence that our adaptive mesh and MHD algorithms maintain conserved quantities satisfactorily.

\begin{table}
  \begin{center}
  \caption{\label{tab:fastpoint} Comparison of the speed at the fast point as predicted by Eq.\ (\ref{eq:fastpoint}) and as measured directly along the 1 AU field line.}
  \begin{tabular}{c|ccccccc}
    \hline\hline
    $v_{\rm p,f}$ & A & B & C & D & E & F & G \\
    \hline
    Eq.\ (\ref{eq:fastpoint}) & 205.8 & 127.3 & 94.10 & 69.43 & 43.94 & 27.6 & 19.4 \\
    1 AU field line               & 203.5 & 128.8 & 94.64 & 69.45 & 43.86 & 27.9 & 18.5 \\
    \% difference                 & 1.1   & 1.2   & 0.57  & 0.03  & 0.2   & 1.1  & 4.5  \\
    \hline
  \end{tabular}
  \end{center}
\end{table}

Table \ref{tab:fastpoint} shows the expected values of $v_{\rm p,f}$ along the 1 AU field line (Eq.\ \ref{eq:fastpoint}) for simulations A--G compared to the values shown in Fig.\ \ref{fig:velocitiesAH}.  Notwithstanding simulation G, all values agree to within 1.2\%, once again indicating portions of the jets do attain a quasi-steady state, and that the code is able to recognise and maintain these regions.  In as much as there is a `numerical test' for these simulations, this would be it.

\begin{figure}
  \begin{center}
    \includegraphics[width=0.484\columnwidth]{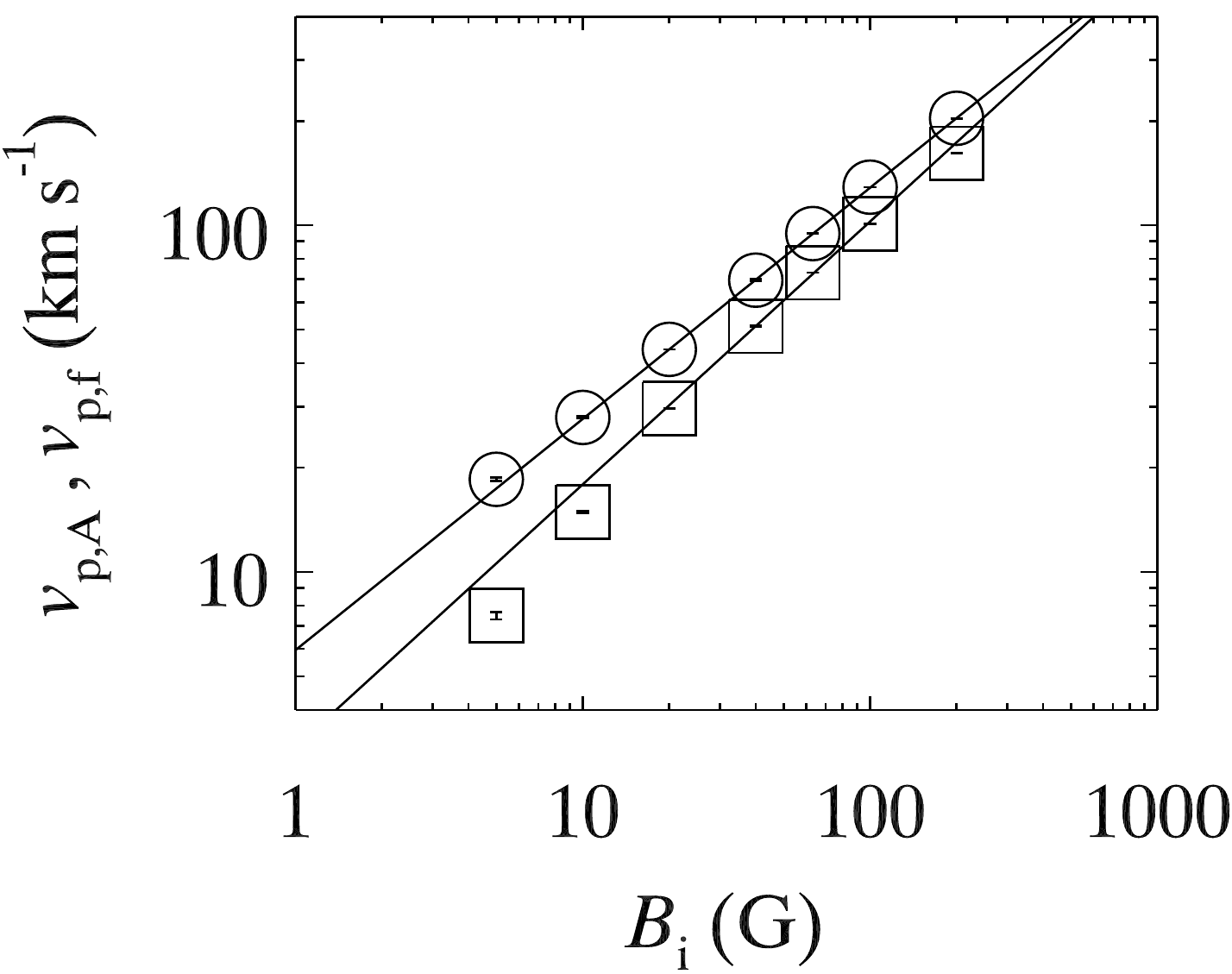} \includegraphics[width=0.504\columnwidth]{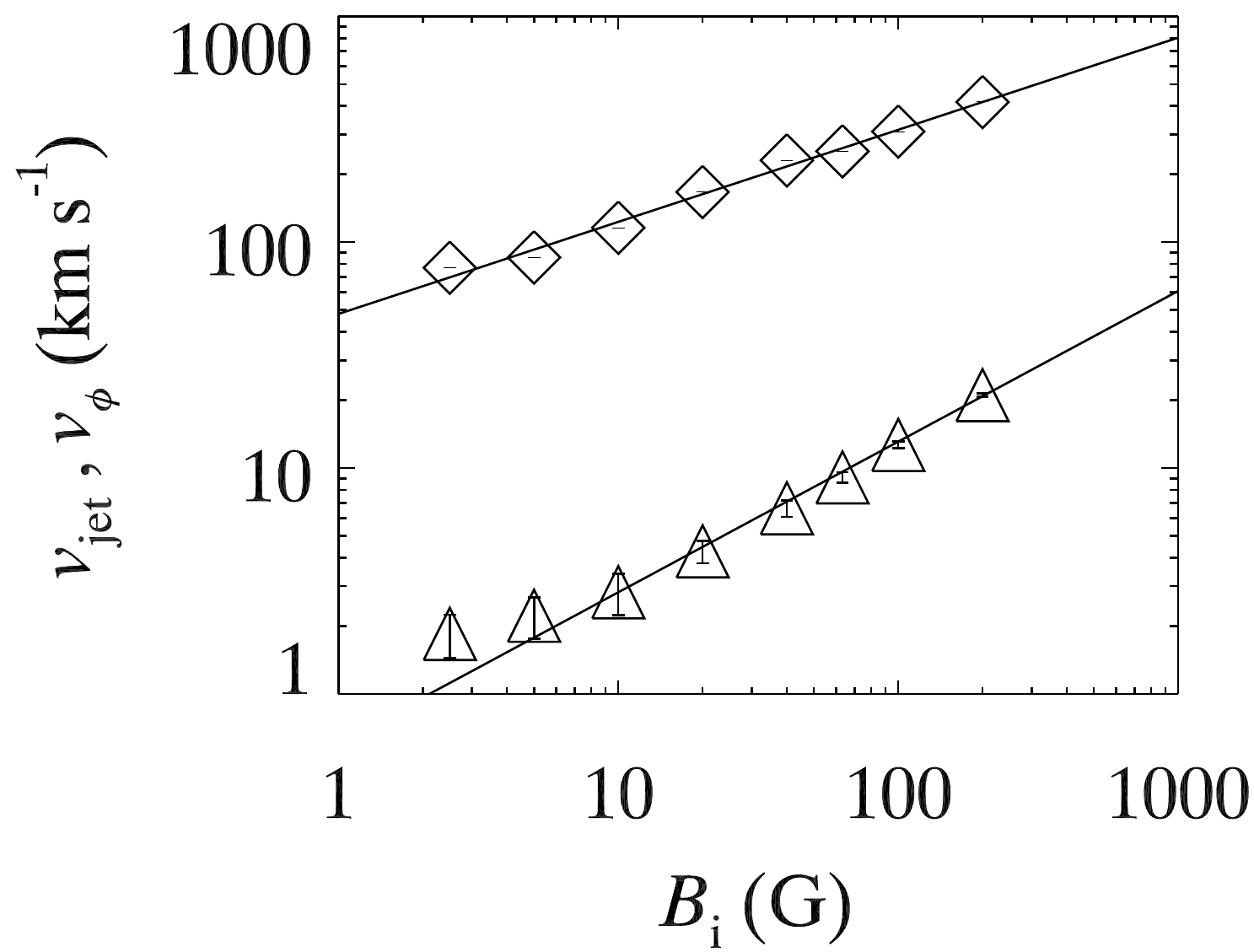}
  \end{center}
  \caption{\label{fig:powerlaws} \textit{Left:} Values of $v_{\rm p,f}$ (circles) and $v_{\rm p,A}$ (squares) on the 1 AU field line for simulations A--G (H is excluded since its 1 AU field line is not in steady state) with best-fit power laws: $v_{\rm p,f} \propto B\rmi^{0.67}$; $v_{\rm p,A} \propto B\rmi^{0.76}$ (uncertainties given in Table \ref{tab:simsum2}). \textit{Right:} Similar plot for $v_{\rm jet}$ (diamonds) and $\langle v_\varphi\rangle$ (triangles) with best-fit power laws: $v_\mathrm{jet} \propto B\rmi^{0.42}$; $v_{\phi} \propto B\rmi^{0.67}$.}
\end{figure}

As can be seen from Figure \ref{fig:powerlaws}, the poloidal speed at the fast point, $v_{\rm p,f}$, along the 1 AU field line follows a rather tight `2/3 power law' with $B\rmi$, contrary to the prediction made by equation (\ref{eq:vpfofBi}) and more in line with Eq.\ (\ref{eq:vpmax}).  Indeed, the fast point along all field lines passing through a quasi-steady state region show equally tight power laws $v_{\rm p,f}\sim B\rmi^\alpha$, with $\alpha=0.67\pm0.03$ (Table \ref{tab:simsum2}).  This is one of the most robust power-law relationships gleaned from the simulations, and we regard this result as firm.  

The discrepancy between the numerically determined power-law index and that predicted by Eq.\ (\ref{eq:vpfofBi}) can only be caused by the factor,
\begin{equation*}
f(z_{\rm f},r_{\rm f})\equiv\left[\frac1{4\pi\rho_{\rm f}}\left(\frac{r_{\rm f}^2}{r_0^2}+\frac{2r_0}{R_{\rm f}}-3\right)\right]^{1/4},
\end{equation*}
in Eq.\ (\ref{eq:fastpoint}), since it was assuming it to be independent of $B\rmi$ that led to the `1/2 power law' prediction in Eq.\ (\ref{eq:vpfofBi}).  Evidently, $f\sim B\rmi^{1/6}$, a weak but significant dependence that accounts for the difference between the measured and predicted power law indices for $v_{\rm p,f}$.  What was not anticipated in Sect.\ \ref{sec:steadystate} was the degree to which the fast point is pushed away from the disc for higher $B\rmi$ ($z_{\rm f}$ in Table \ref{tab:simsum2}).  Along the 1 AU field line, $z_{\rm f}\sim300\,$AU for simulation A, whereas $z_{\rm f}\sim7\,$AU for simulation F.  Since density is highly dependent upon $z$, this is where the dependence of $f$ upon $B\rmi$ arises.

We continue this discussion in the next subsection.

\subsection{Dependence of velocity on \texorpdfstring{$B\rmi$}{Bi}}
\label{sub:vofBi}

While both $v_{\rm p,f}$ and $v_{\rm p,A}$ follow a power law in $B\rmi$ along field lines in quasi-steady state, they do so with different indices, and the tightness of fit is somewhat stronger for $v_{\rm p,f}$ (Fig.\ \ref{fig:powerlaws}; Table \ref{tab:simsum2}).  This is true for other field lines in quasi-steady state as well, with similar power law indices measured.

In addition to speeds specific to the 1 AU field line, Table \ref{tab:simsum2} includes various observationally-accessible speeds for each jet, including the mass-weighted averages (exclusive of the second wind) of the rotational velocity, $\langle v_\phi\rangle$, the axial speed along the last 500 AU of the jet core, $\langle v_z\rangle$, and the advance speed of the jet tip into the quiescent atmosphere, $v_{\rm jet}$.  Each velocity varies with $B\rmi$ reasonably well as a power law (with power-law index $\alpha$ given in Table \ref{tab:simsum2}), indicating that the observable kinematics of the jet is highly and simply dependent upon the initial magnetic field strength.

For the toroidal velocity, we performed the mass-weighted average,
\begin{equation*}
\langle v_\varphi\rangle = \frac{\int_V \rho v_\varphi dV}{\int_V \rho dV},
\end{equation*}
where $V$ is the volume bounded by the TD.  As defined, this quantity should correspond to observationally-determinable rotation speeds of well-resolved jets such as that reported in \cite{woitasetal05}.  All integrations are performed from data taken at the same dynamical time, $l=2$,400 AU, which is comfortably beyond $l\sim1,\!000$\,AU where $\langle v_\varphi\rangle$ seems to reach its asymptotic and steady state value for all simulations.  As defined, the data in Table \ref{tab:simsum2} and plotted in Fig.\ \ref{fig:powerlaws} show that $\langle v_\varphi\rangle\sim B\rmi^{0.67\pm0.01}$.

Similarly, the quantity $\langle v_z\rangle$ is a mass-weighted average of the axial outflow speed given by: 
\begin{equation*}
\langle v_z\rangle = \frac{\int_V \rho v_z dV}{\int_V \rho dV},
\end{equation*}
where here, $V$ is the volume contained by the last 500 AU of the TD and within the jet core in which the outflow speed reaches its asymptotic and steady state value for all simulations.  Note that $\langle v_z\rangle$ should correspond to observed outflow speeds within jets.

Next, the advance speed of the jet tip into the ambient medium, $v_{\rm jet}$, is measured by fitting the asymptotic slope of the position of the tip of the jet as a function of time, and is plotted in Fig.\ \ref{fig:powerlaws}.  Such a velocity would be measured using multi-epoch observations of jet bow shocks.  In our simulations, we find that $v_{\rm jet}\sim \langle v_z\rangle$ (Table \ref{tab:simsum2}), and thus the jets are largely ballistic.  Given the greater mass density of the nose-cone relative to the ambient medium, and the confinement of the flow by strong toroidal fields, this is not a surprising result.  Both $\langle v_z\rangle$ and $v_{\rm jet}$ follow power-laws in $B\rmi$ with index $\alpha \sim0.43\pm0.01$ which, when combined with the $0.67$ power law for $\langle v_\varphi\rangle$, is consistent with the result,
\begin{equation}
\label{eq:vjvphi}
v_{\rm jet} \propto \langle v_\varphi\rangle^{2/3}.
\end{equation}
Finally, $v_{\rm entr}$ is a mass-weighted average of the forward speed of the entrained material (i.e.\ shocked, ambient material in the simulations identified as having a non-negligible forward velocity but very low $B_\varphi$) behind the forward 500 AU of the bow shock (the `second wind').  Such a velocity could be measured from CO molecular line observations, for example. It is worth noting, however, that the rotational velocity of the entrained material is negligible. In our simulations, $v_{\rm entr}$ ranges from 11 km\,s$^{-1}$ for simulation A to $1.1$ km\,s$^{-1}$ for simulation H, with a power law index of $\sim0.56$ (Table \ref{tab:simsum2}).  This combined with the $0.43$ power law index for $v_{\rm jet}$ is consistent with,
\begin{equation}
\label{eq:ventrvj}
v_{\rm entr} \propto v_{\rm jet}^{4/3}.
\end{equation}

Should either or both the `2/3 law' in Eq.\ (\ref{eq:vjvphi}) or the `4/3 law' in Eq.\ (\ref{eq:ventrvj}) be observed, this could be interpreted as indirect confirmation of the role played by $B\rmi$ in forming and driving protostellar jets.  Unfortunately, the proportionality constants in these power-law relationships may be and probably are different from jet to jet, which may make such measures challenging.

\subsection{The driving mechanism}
\label{sub:driving}

The jets are driven by the Lorentz force both directly, as a poloidal gradient in the toroidal magnetic pressure (the `magnetic tower mechanism'; MTM), and indirectly whereby the ability of $\mathbfit B$ to exert a substantial `normal force' when rotated allows it to be the `rigid wire' for the `bead-on-a-wire mechanism' (BWM).  Both mechanisms can be identified mathematically by a suitable analysis of the relevant forces.

\begin{figure}
  \begin{center}
    \includegraphics[width=0.65\columnwidth]{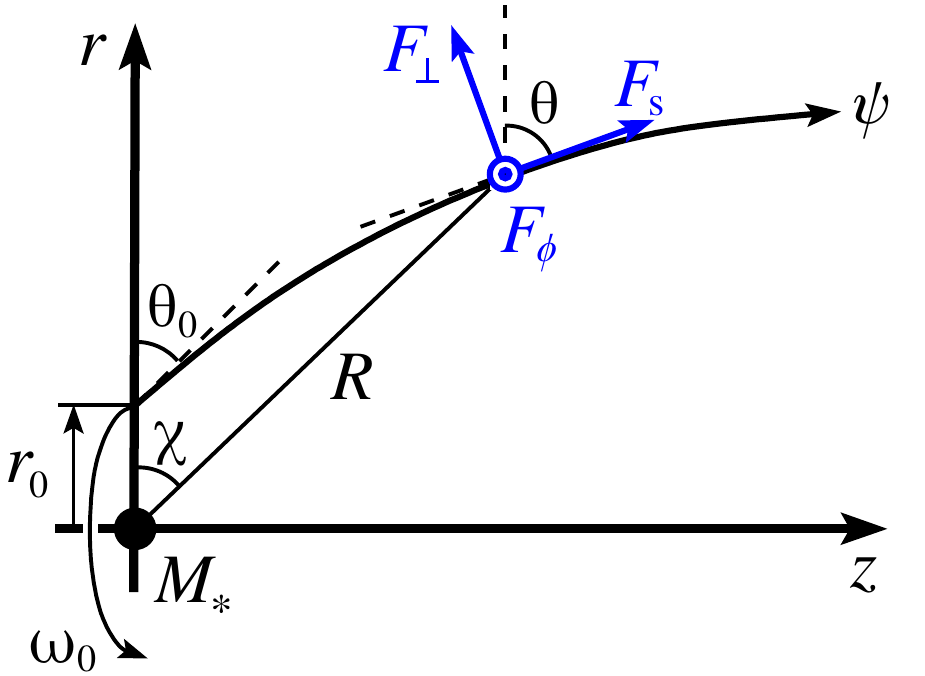}
  \end{center}
  \caption{\label{fig:beadonawire} Schematic of a single poloidal magnetic field line, $\psi$, that is anchored in the disc at a distance $r_0$ from the gravitational point mass, $M_*$, and rotating at the Keplerian angular speed $\omega_0$.  The components of the Lorentz force, $F_\perp$, $F_{\rm s}$, and $F_\varphi$ as described in the text are indicated for an arbitrary point along the field line.}
\end{figure}

In axisymmetric cylindrical coordinates, the Lorentz force is given by:
\begin{equation}
\label{eq:lorentz}
\mathbfit{F}_{\rm L}=\mathbfit{J}\times\mathbfit{B}=-\frac1{r^2}\nabla\pol\varpi_\varphi+J_\varphi\mathbfit{B}_\perp+\frac1r\mathbfit B\pol\cdot\nabla\pol b_\varphi\hat{\bmath{\varphi}},
\end{equation}
where $\nabla\pol=\hat{\mathbfit{r}}\partial_r+\hat{\mathbfit{z}}\partial_z$ is the poloidal gradient, $\varpi_\varphi=\frac1{8\pi}(rB_\varphi)^2$ is a radially weighted toroidal magnetic pressure, $J_\varphi=\frac1{4\pi}(\partial_zB_r-\partial_rB_z)$ is the $\varphi$-component of the current density, $\mathbfit B_\perp=-B_r\hat{\mathbfit{z}}+B_z\hat{\mathbfit{r}}$ is a vector perpendicular to and with the same magnitude as the poloidal magnetic field, $\mathbfit B\pol=B_z\hat{\mathbfit{z}}+B_r\hat{\mathbfit{r}}$, and $b_\varphi=\frac1{4\pi}rB_\varphi$.  The last two terms in Eq.\ (\ref{eq:lorentz}) are both `normal forces' exerted perpendicular to $\mathbfit B\pol$ ($F_\perp$ and $F_\varphi$ in Fig.\ \ref{fig:beadonawire}), whereas the first term, being the gradient of a function of $B_\varphi$ twisted out from the poloidal field, will lie along the general direction of $\mathbfit B\pol$ ($F_{\rm s}$ in Fig.\ \ref{fig:beadonawire}).  Note further that at $t=0$, $\mathbfit F_{\rm L}$ is identically zero since the initial `hour-glass' magnetic field configuration is force-free (Eqs.\ \ref{eq:aphi} and \ref{eq:binit}).  It is only after the disc begins to rotate and the magnetic field is distorted from its initial conditions that $\mathbfit F_{\rm L}\neq0$.

For convenience, consider the problem in the co-rotating reference frame of a point on the disc at distance $r_0$ from the origin (where the gravitating point mass $M_*$ is located; see Fig.\ \ref{fig:beadonawire}), whose angular speed is given by $\omega_0^2=GM_*/r_0^3$.  Next, consider the poloidal field line $\psi$ anchored at this point, emerging from the disc at an angle $\theta_0$.  If we think of the poloidal field line as the `rigid wire' for a `bead' of plasma on the surface of the disc then, as is widely known (\eg\ Fig.\ 1 in \citealt{bp82}), the `bead', when nudged, will accelerate out along the `wire' if $\theta_0<\theta_{\rm c}=60\degr$.

Now, unlike the classic `bead-on-a-wire' problem found in most sophomore mechanics texts, the poloidal field line is not truly `rigid', regardless of its strength.  When the disc starts to rotate, an Alfv\'en wave with speed $B\pol/\!\sqrt{4\pi\rho}$ is launched, twisting $\mathbfit B\pol$ in its wake.  The stronger $\mathbfit B\pol$ is, the faster the Alfv\'en wave propagates and the fewer number of turns per unit length suffered by the poloidal field.  Still, and regardless of its strength, $\mathbfit B\pol$ will be twisted by numerous full turns over a long enough distance, resulting in a restructuring of the field (rather than a perturbation) that ultimately shuts down and even reverses the effect of the BWM.

To see this, we must include the inertial (centrifugal and Coriolis) forces as observed in the co-rotating frame where the rotational speed of jet material is:
\begin{equation*}
v_\varphi^\prime=v_\varphi-\omega_0r,
\end{equation*}
and the inertial forces are given by:
\begin{equation}
\label{eq:inertial}
\mathbfit{F}_{\rm I}=-\rho\bmath{\omega}\times(\bmath{\omega}\times\mathbfit{r})-2\rho\bmath{\omega}\times\mathbfit{v}=\rho\omega_0(2v_\varphi-\omega_0r)\hat{\mathbfit{r}}-2\rho\omega_0 v_r\hat{\bmath{\varphi}},
\end{equation}
where $\bmath{\omega}=\omega_0\hat{\mathbfit{z}}$ and $\mathbfit{v}=v_z\hat{\mathbfit{z}}+v_r\hat{\mathbfit{r}}+v_\varphi^\prime\hat{\bmath{\varphi}}$.  

Combining Eqs.\ (\ref{eq:lorentz}) and (\ref{eq:inertial}), and resolving the radial inertial force into the two poloidal components, $\hat{\mathbfit{s}}$ and $\hat{\bmath{\perp}}$ as depicted in Fig.\ \ref{fig:beadonawire}, we get:
\begin{equation}
\label{eq:total}
\begin{aligned}
\mathbfit{F}=\mathbfit{F}_{\rm L}+\mathbfit{F}_{\rm I}&=\underbrace{-\frac1{r^2}\nabla\pol\varpi_\varphi}_{\textstyle \text{MTM}}+\underbrace{\rho\omega_0(2v_\varphi-\omega_0r)\cos\theta \hat{\mathbfit{s}}}_{\textstyle\text{BWM}}\\
&+\big(J_\varphi B\pol+\rho\omega_0(2v_\varphi-\omega_0r)\sin\theta\big)\hat{\bmath{\perp}}\\
&+\left(\frac1r\mathbfit{B}\pol\cdot\nabla\pol b_\varphi-2\rho\omega_0v_r\right)\hat{\bmath{\varphi}},
\end{aligned}
\end{equation}
where $\theta>\theta_0$ is the angle between the local field line and $\hat{\mathbfit{r}}$.  Other than $-\nabla p-\rho\nabla\phi$ which, in hydrostatic equilibrium, is zero, Eq.\ (\ref{eq:total}) represents all the forces acting on a `bead' of matter transported along a poloidal field line `wire', as observed in the co-rotating frame.  The omission of the pressure and gravitational gradients from Eq.\ (\ref{eq:total}) means any insight gained will be qualitative in nature; to do the full problem with all the forces properly accounted for is why we do the simulations.

The first two terms in Eq.\ (\ref{eq:total}) represent the MTM and BWM respectively, and are what drive the jet.  The third term is perpendicular to the poloidal magnetic field line within the poloidal plane and, in the steady state, should be zero.  If not, the poloidal field line would move, contrary to the assumption (for this analysis) of steady state.  The fourth term is a normal force in the symmetry direction, and is responsible for the field line acquiring a toroidal component.

From this it is evident that both the MTM and BWM contribute to the acceleration of the jet, regardless of poloidal field strength; it is simply a matter of which term dominates where.  In fact, while the MTM only requires the presence of an outwardly-pointing gradient in the toroidal magnetic pressure---a condition we find true for much of the jet length regardless of the strength of $\mathbfit B\rmi$---the conditions for the BWM are much more limiting.

Most importantly, Eq.\ (\ref{eq:total}) shows that the BWM term falls to \emph{zero} once $v_\varphi=\omega_0r/2$, where $\omega_0$ is the Keplerian angular speed at the anchor point.  As seen in Fig.\ \ref{fig:velocitiesAH}, this `cross-over' point---where the dashed blue and solid red lines cross---is located before the Alfv\'en point in all simulations, indicating that the BWM is only effective close to the disc.  Table \ref{tab:simsum2} includes the cross-over distance, $s_\times$, along the 1 AU field line where the BWM mechanism is shut down.  These range from $\sim\!20\,$AU for simulation A to $<1\,$AU for simulation G, with similar values for other field lines in quasi-steady state anchored in the disc between 0.5 and 10 AU.  Thus, while the BWM may be the dominant acceleration mechanism on and within short distances from the surface of the disc, the MTM is the dominant acceleration mechanism for most of the jet length in all our simulations.

Indeed, beyond $s_\times$, $2v_\varphi-\omega_0r<0$ and the BWM actually \emph{retards} outflow, pulling matter back toward the disc!  As we see in the simulations, however, there are two reasons why this doesn't happen.  First, poloidal field lines asymptote towards the $\hat{\mathbfit{z}}$-direction and $\cos\theta\rightarrow0$ in Eq.\ (\ref{eq:total}), minimising the BMW term.  Second, the MTM term is relentless, counteracting the negative but increasingly feeble BWM as one pulls away from $s_\times$.  Thus, along just about the entire length of the jet and regardless of $B\rmi$, the gradual, negative gradient in $B_\varphi$ gives rise to a net outward acceleration---however modest---well beyond the fast point.

As for the MTM, regardless of the strength of the poloidal field, the rotating disc twists the poloidal field into a dynamically active toroidal field with the passage of the Alfv\'en wave launched from the disc.  To see this, in a time $\Delta t$, the Alfv\'en wave travels a distance $l=B\pol \Delta t/\!\sqrt{4\pi\rho}$, during which time the disc has wound up the field into $n=v_{\rm K}(r)\Delta t/(2\pi r)$ coils of radius $r$.  Thus,
\begin{equation*}
\frac{B_\varphi}{B\pol}\sim\frac{2\pi rn}l=v_{\rm K}\frac{\!\sqrt{4\pi\rho}}{B\pol}~~\Rightarrow~~ B_\varphi\sim v_{\rm K}\!\sqrt{4\pi\rho} \sim\!\sqrt{\frac{4\pi\gamma p}{\gamma -1}},
\end{equation*}
since, for quasi-hydrostatic equilibrium, $v_{\rm K}\sim c_{\rm s}/\!\sqrt{\gamma-1} $ (Eq.\ \ref{eq:csvk}).  Thus,
\begin{equation*}
\beta_\varphi=\frac{8\pi p}{B_\varphi^2}\sim\frac{2(\gamma-1)}{\gamma}\sim1,
\end{equation*}
and $-\nabla\pol \varpi_\varphi/r^2$ is a dynamically important outward-pointing force, regardless of the initial poloidal field.  This means that in principle, even a \emph{trace} poloidal magnetic field at the disc surface is sufficient to launch an outflow into the ambient medium, however slow that outflow may be, an observation borne out by our simulation H.

We remind the reader, however, that this analysis is strictly for 2-D axisymmetry which, as has been pointed out, provides the ideal environment for the MTM.  In 3-D, the rotation of the fluid which encourages $B_\varphi$ is accompanied by numerous modes of instability which discourages its development.  We therefore anticipate a much more complicated picture in 3-D, particularly for weaker values of $B\rmi$.

\subsection{The knot generator}
\label{sub:knotgen}

One of the most striking features of simulation F is the regularity with which `knots' or `towers' (rings or discs in 3-D) are launched from the inner disc, roughly in the region $r\rmi<r<2r\rmi$ (lowermost panel of Fig.\ \ref{fig:zoom40}).  As observed in Sect.\ \ref{sub:medium-field}, the knots take a little while to establish themselves as various early transients occur in the simulation but, after $\sim9$\,yr, knot production in simulation F remains steady for the remainder of the run.

The only other simulation in which knots of any significance are observed is simulation E, and then only sporadically.  Thus, we consider the production of knots in simulation F as `transitionary' between runs with stronger $B\rmi$ characterised by more steady, even laminar flow (\eg, Fig.\ \ref{fig:zoom0.1}), and runs with weaker $B\rmi$ characterised by much more chaotic flow within the sheath and jet core (\eg, Fig.\ \ref{fig:zoom160}).  We note that in the region $0.1<r<0.2\,$AU in the lower panel of Fig.\ \ref{fig:zoom160}, several distinct `knot bases' are present where knots might have formed were the confinement by a poloidal magnetic field more effective.

When produced, the temperature and density within the knots is commensurate with $\rho$ and $T$ near the centre of the gravitational potential well, and these values are maintained as they venture away from the disc.  Thus, by the time they reach a distance of $\sim50$ AU from the disc, their temperature and density can be $10^3$ and ten times higher, respectively, than that of the surrounding medium (\eg\ Fig.\ \ref{fig:zslice40}).  Such knots, therefore, have a \emph{thermal} pressure excess of \emph{four} orders of magnitude compared to their immediate surroundings, and it is only the enshrouding poloidal magnetic flux loops that maintain equilibrium. These knots, therefore, may be considered plasmoids (\citealp{bostick56}) confined by naturally occurring magnetic bottles reminiscent of hydrogen pellet confinement in a thermonuclear reactor.

The mechanism by which knots are launched is very simple and quite unlike that described by \cite{op97b}.  In our simulations, knots are generated directly from the disc, and from poloidal field lines emerging from the disc at or very near the critical angle for the BWM ($\theta_{\rm c}=60\dgr$; Sects.\ \ref{sub:overview} and \ref{sub:driving}).  Although the knots form from warm material, they occur at a transition to a cold medium (bottom panel of Fig.\ \ref{fig:zoom40}), and we indeed observe that the magnetic field angle at this location is much closer to the dynamically cold critical value of $60^\circ$ \citep{bp82} than a dynamically warm value of $70^\circ$ \citep{pp92}.  For strong $B\rmi$, field lines are rigid at the disc surface, and plasma is guided subserviently by the BWM.  For weak $B\rmi$, field lines are completely at the mercy of the inertia of the plasma, and it is up to the MTM to organise a sufficiently strong toroidal field to launch the jet.  However, for simulation F where neither the rigidity of the field lines nor the inertia of the plasma dominate the dynamics, there is much `give and take' at the critical angle.

For field lines of marginal strength, perturbations in the outflow cause the critical field lines to wiggle back and forth across $\theta_{\rm c}$.  When $\theta_0$ is slightly greater than $\theta_{\rm c}$, hot, dense material near the depth of the gravitational potential accumulates at the base of the field line without moving outward.  As the field line is `loaded up', the centrifugal inertia of the growing plasmoid bends the field line outward so that $\theta_0$ falls below $\theta_{\rm c}$, and the plasmoid is launched.  With the field line now relieved of its mass load, it bends back to something greater than $\theta_{\rm c}$, outflow is quashed, and the cycle repeats.  As in any oscillator, the system `overshoots' equilibrium, establishing the regular periodicity observed.

\begin{figure}
  \begin{center}
    \includegraphics[width=0.56\columnwidth]{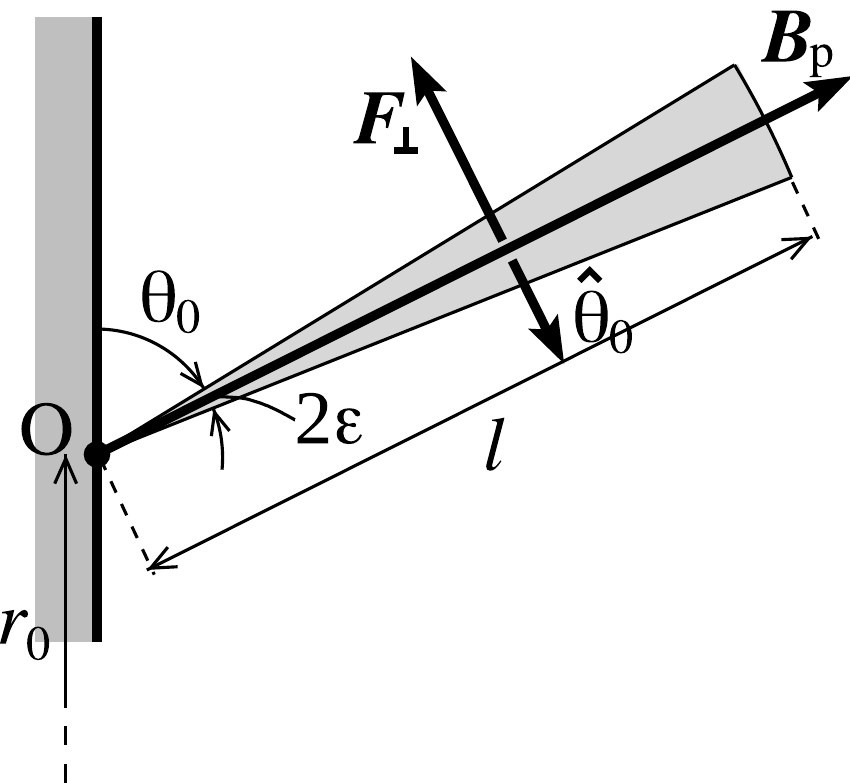}
  \end{center}
  \caption{\label{fig:knotgen} A close-up of the poloidal field line depicted in Fig.\ \ref{fig:beadonawire} illustrating the simple harmonic oscillator established in simulation F to generate the knots.  The shaded wedge of length $l$ is the cross section through a `truncated cone' of mass loaded on the field line.}
\end{figure}

The regularity of knot spacing is indicative of a simple harmonic oscillator whose dependence on $B\rmi$ can be understood as follows.  If a force-free magnetic field, $(B_{z_0},B_{r_0})$, is perturbed by changing the angle at which it emerges from the disc, $\theta_0$, by a small angle $\veps\ll\theta_0$ (Fig.\ \ref{fig:knotgen}), then the components of the perturbed field become:
\begin{equation*}
B_z=B_{z_0}+B_{r_0}\veps;\qquad B_r=B_{r_0}-B_{z_0}\veps,
\end{equation*}
giving rise to a non-zero toroidal current density,
\begin{equation*}
J_\varphi=-\frac1{4\pi}(\partial_zB_{z_0}+\partial_rB_{r_0})\veps=\frac{B_{r_0}}{4\pi r}\veps,
\end{equation*}
using $\nabla\cdot\mathbfit B=0$.  Thus, the restoring force, $\mathbfit F_\perp$ (Eq.\ \ref{eq:lorentz}), is given by:
\begin{equation}
\label{eq:rest_force}
\mathbfit{F}_\perp=J_\varphi\mathbfit{B}_\perp=\frac{B_{r_0}}{4\pi r}\veps\,B\pol(-\hat{\bmath{\theta}}_0)=-\frac{B_p^2}{8\pi r}\veps\,\hat{\bmath{\theta}}_0,
\end{equation}
since $\hat{\bmath{\theta}}_0$, a unit vector in the direction of increasing $\theta_0$, is antiparallel to $\mathbfit F_\perp$, and $B_{r_0}=B\pol\cos\theta_0=B\pol/2$ for $\theta_0=\theta_{\rm c}=60\degr$.

To find the equation of motion for this simple harmonic oscillator, consider the torque density, $\tau$, generated by $\mathbfit F_\perp$ about point O on the mass within the shaded wedge subtended by angle $2\veps$ in Fig.\ \ref{fig:knotgen} which, in axisymmetry, is a hollow truncated cone of length $l$ (a portion of this mass is what gets launched to form a knot).  For $l\ll r_0$, the moment of inertia per unit volume of this wedge is $I\sim\rho l^2/2$, and we have using Eq.\ (\ref{eq:rest_force}):
\begin{equation*}
\left.
\begin{aligned}
\tau&= I\ddot\veps\sim \frac{\rho l^2}2\ddot\veps\\
&\propto F_\perp l=-\frac{B\pol^2l}{8\pi r_0}\veps,
\end{aligned}
\quad\right\}\quad\Rightarrow\quad\ddot\veps\propto-\frac1{lr_0}\frac{B\pol^2}{4\pi\rho}\veps,
\end{equation*}
which has the classic form of a simple harmonic oscillator in $\veps$, with a frequency of oscillation given by $\omega\propto a\pol/\!\sqrt{lr_0}\sim B\rmi$.

Since no simulation other than F generated a steady stream of knots, we are unable to test the predicted dependency of $\omega$.  Thus, the analysis is included here only as an attempt to identify the physics responsible for the simple harmonic behaviour of the knots in simulation F. We also remark that the generation of knots, although occurring well inside the simulation domain, is a physical result of our prescribed boundary conditions; were the disc self-consistently included in these simulations, it is far from guaranteed that the knots would persist in their observed form.

\subsection{Fluxes of mass, momentum, and energy}
\label{sub:fluxes}

We now turn to the transport of mass, linear and angular momentum, and kinetic energy within the jet and consider these separately from that transported by material entrained by the bow shock (\ie, the `second wind').  In addition to the velocities discussed above, fluxes are among the few concrete physical values that can be determined from observations of jets.  All fluxes reported here are axial fluxes measured at a height of $z = 1,\!000$ AU above the disc using:
\begin{align*}
  \dot{M} &= 2\pi \dot{M}\rmi \int \rho v_z r\, {\rm d}r;\\
  \dot{p} &= 2\pi \dot{p}\rmi \int \rho v_z^2 r\, {\rm d}r;\\
  \dot{L} &= 2\pi \dot{L}\rmi \int \rho v_z v_\varphi r^2\, {\rm d}r;\\
  \dot{K} &= 2\pi \dot{K}\rmi \int \rho v_z^3 r\, {\rm d}r,
\end{align*}
where the integral is performed as a sum over grid points in the $r$-direction.  For fluxes transported by the jet core and sheath, we mask the data by requiring $M>5$ (similar to the masking used to generate Fig.\ \ref{fig:avgbeta}), which effectively identifies all material between the jet axis and the TD.  For fluxes transported between the TD and bow shock (the second wind), the data are masked by requiring $M<5$.  While this includes all points beyond the bow shock as well, there the velocities are zero and thus do not contribute to the fluxes.

Applying the scaling relations in Sect.\ \ref{sub:scaling}, one can convert the fluxes from code to physical units via:
\begin{align*}
  \dot{M}\rmi &= \left(3.1\!\times\! 10^{-7}\!M_\odot\, {\rm yr}^{-1}\!\right)\!\left(\frac{\beta\rmi}{40}\right)\!\left(\frac{B\rmi}{10\,{\rm G}}\right)^2\!\!\left(\frac{r\rmi}{0.05\,{\rm AU}}\right)^{5/2}\!\!\left(\frac{0.5 M_\odot}{M_*}\right)^{1/2}\!\!\!\!\!\!;\\
  \dot{p}\rmi &= \left(2.4\!\times\! 10^{-5}M_\odot\,{\rm yr^{-1}\, km\, s^{-1}}\right)\!\left(\frac{\beta\rmi}{40}\right)\!\left(\frac{B\rmi}{10\,{\rm G}}\right)^2\!\!\left(\frac{r\rmi}{0.05\,{\rm AU}}\right)^2\!\!;\\
  \dot{L}\rmi &= \left(1.2\!\times\! 10^{-6} M_\odot\,{\rm yr^{-1} AU\, km\, s^{-1}}\right)\!\left(\frac{\beta\rmi}{40}\right)\!\left(\frac{B\rmi}{10\,{\rm G}}\right)^2\!\!\left(\frac{r\rmi}{0.05\,{\rm AU}}\right)^{3}\!\!;\\
  \dot{K}\rmi &= \left(1.1\!\times\! 10^{33}\,{\rm erg\, s}^{-1}\!\right)\!\left(\frac{\beta\rmi}{40}\right)\!\left(\frac{B\rmi}{10\,{\rm G}}\right)^2\!\!\left(\frac{r\rmi}{0.05\,{\rm AU}}\right)^{3/2}\!\!\left(\frac{M_*}{0.5 M_\odot}\right)^{1/2}\!\!\!\!\!\!.
\end{align*}

\begin{figure}
  \begin{center}
    \includegraphics[width=0.499\columnwidth,clip,trim=0 35 0 0]{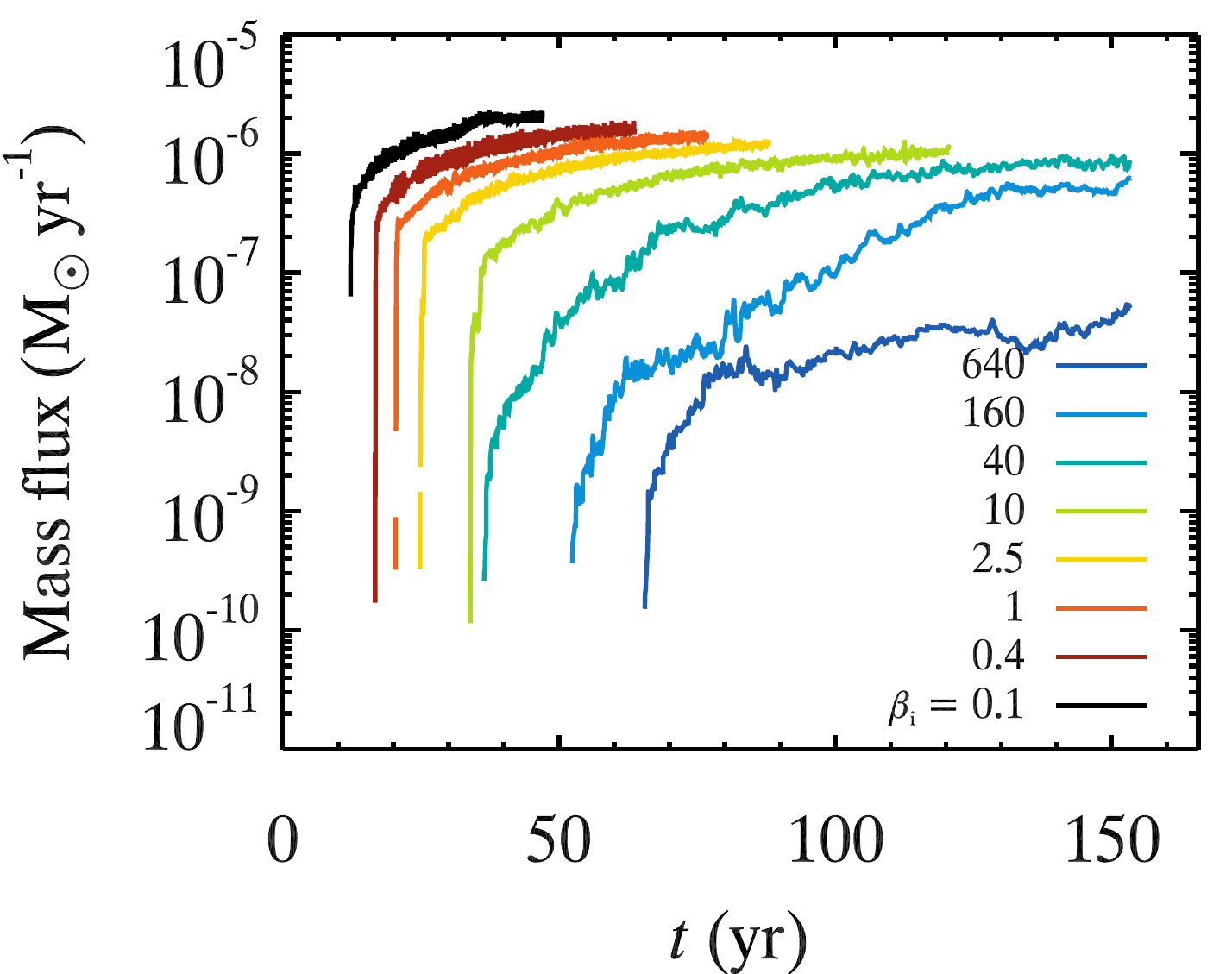}\includegraphics[width=0.499\columnwidth,clip,trim=0 35 0 0]{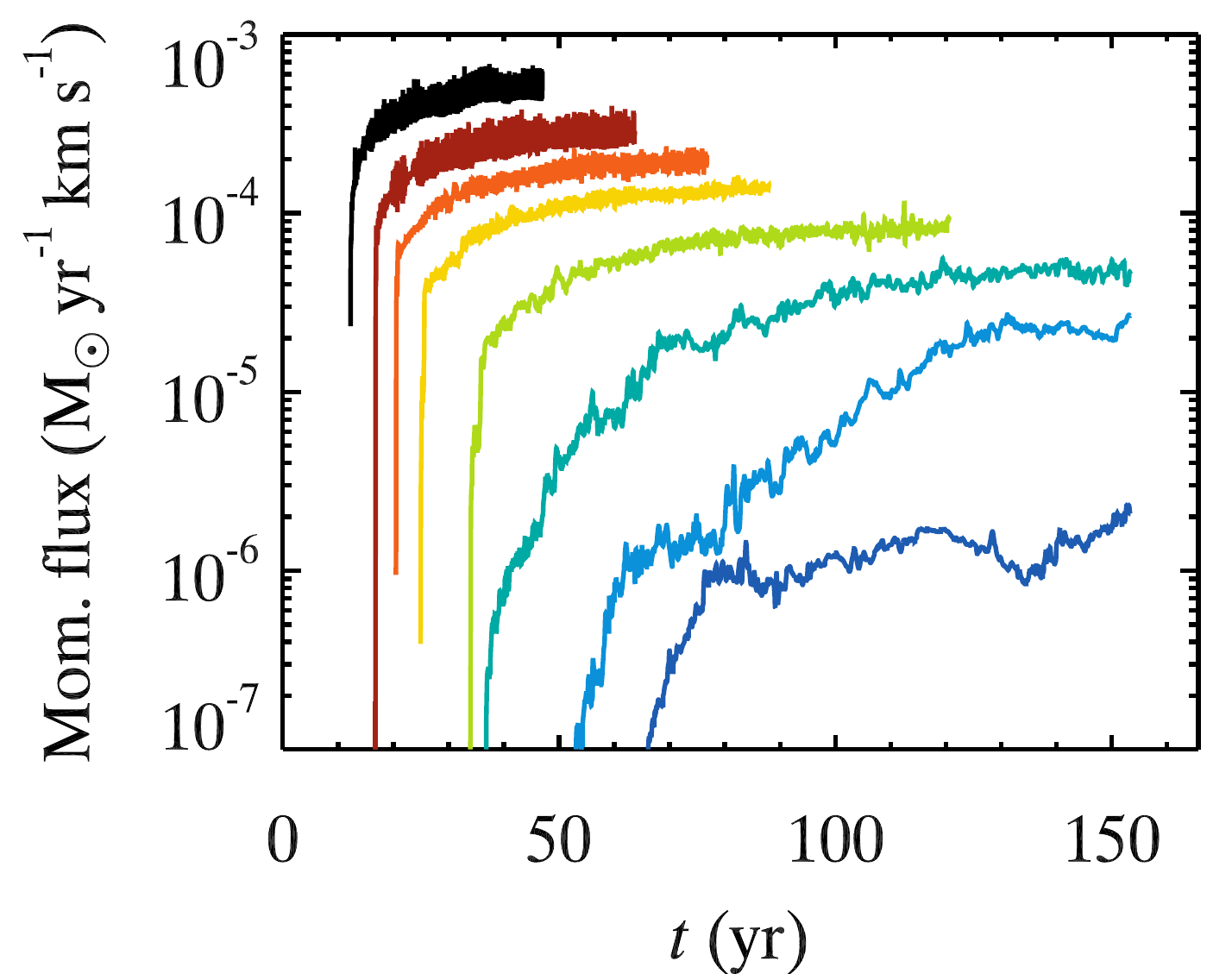}\\[0.5ex]
    \includegraphics[width=0.499\columnwidth]{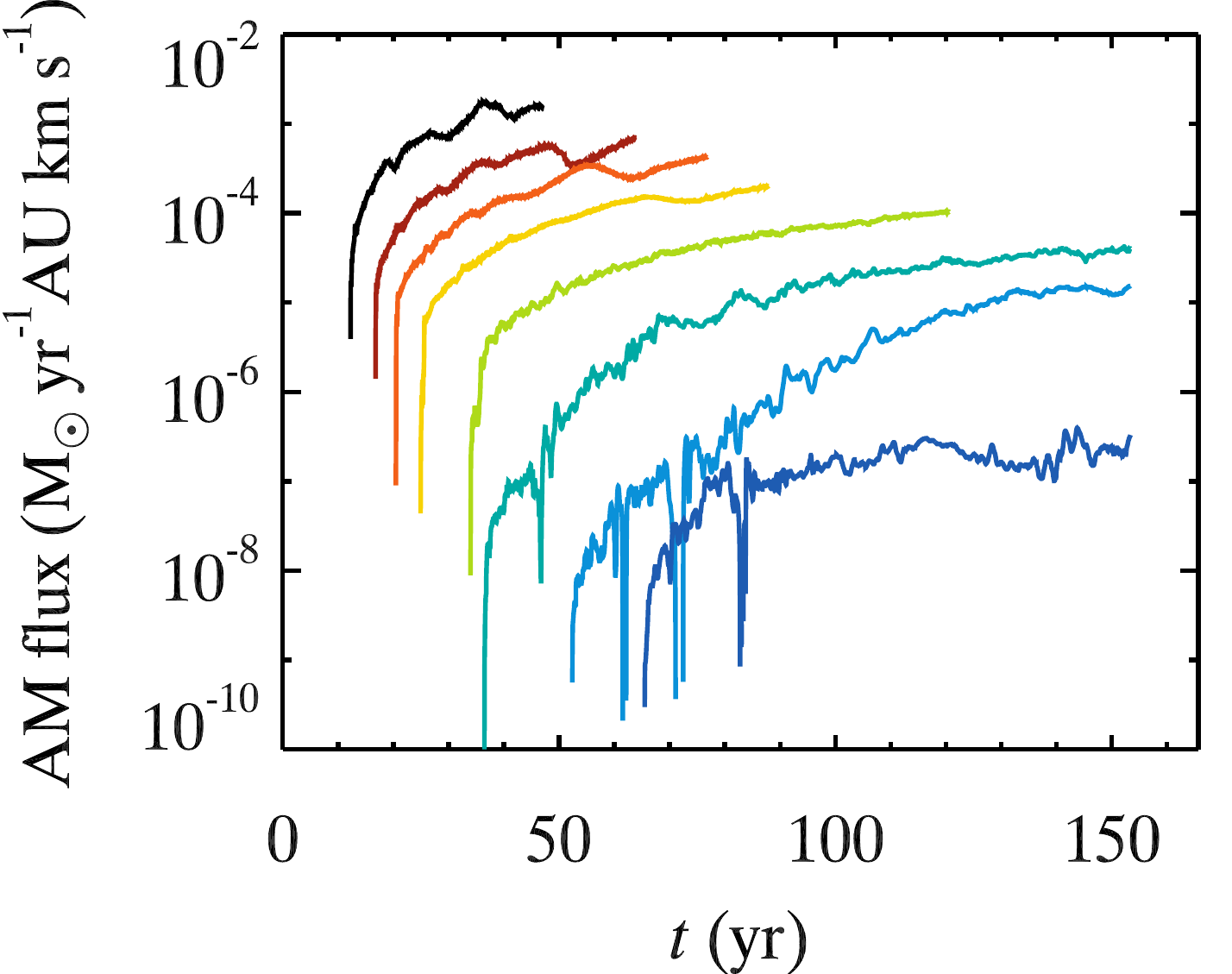}\includegraphics[width=0.499\columnwidth]{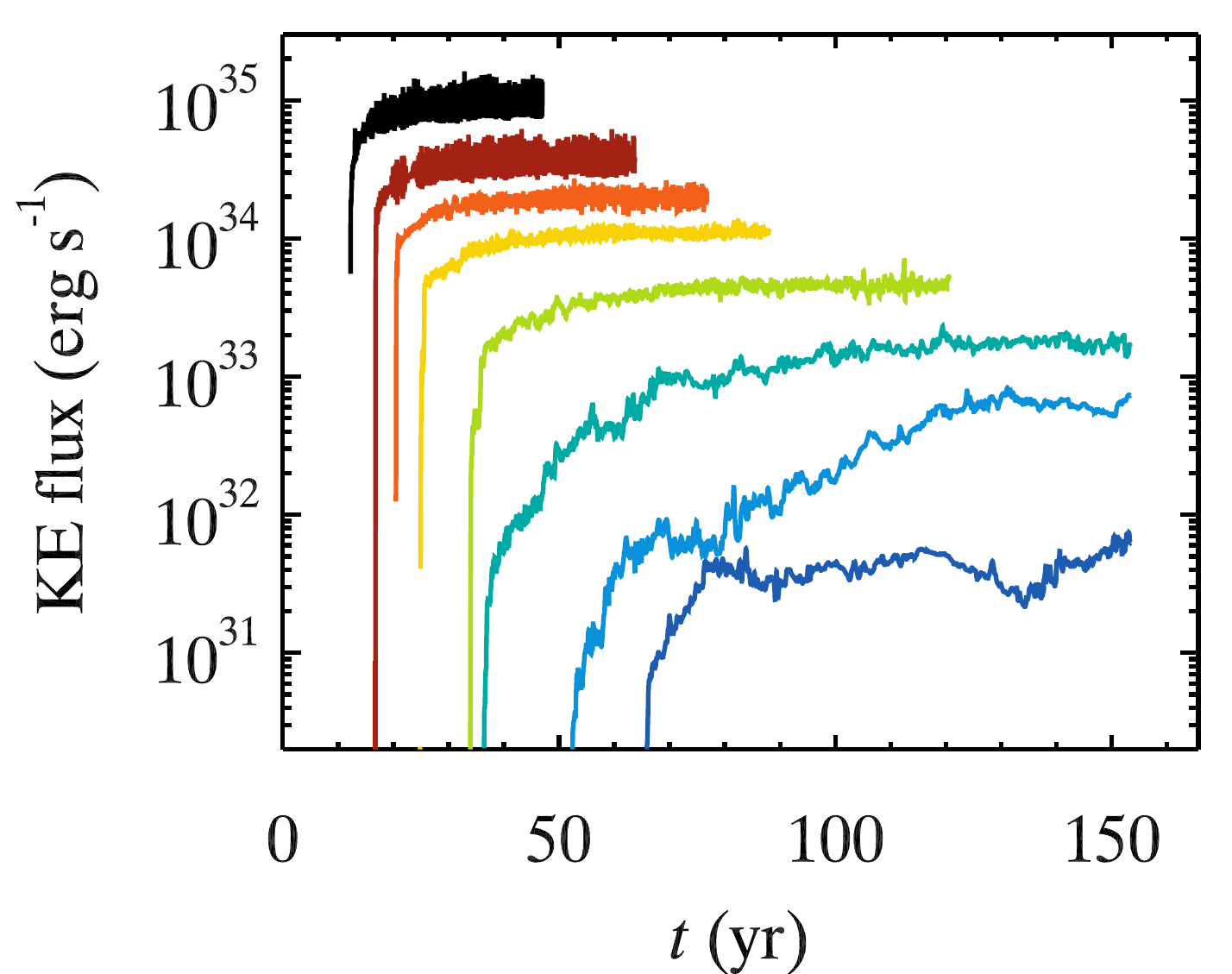}
  \end{center}
  \caption{\label{fig:fluxes} Fluxes of mass, linear momentum,  angular momentum (AM), and kinetic energy (KE) at a height of $z = 1,\!000$ AU above the disc as a function of time for simulations A (black) through H (dark blue).  The data are masked to include only regions where $M \geq 5$, which selects all outflow between the jet axis and the TD.}
\end{figure}

Figure \ref{fig:fluxes} shows the fluxes calculated from each simulation as a function of time.  Notably, each flux reaches (or nearly reaches) an asymptotic value by the end of each simulation, including, to some extent, simulation H.  This is confirmed by calculating the fluxes at heights $z = 750$ and $1,\!500$ AU which give the same results, short of an appropriate offset in time.  We also note that similar loci for fluxes within the second wind (not shown) also converge to asymptotic values, although these data are rather noisier.  Typically, second wind fluxes are $\sim\!1$\% of the corresponding jet fluxes.

The asymptotic jet fluxes (averaged over the last ten years of each simulation) are included in Table \ref{tab:simsum2}.  Because the fluxes depend upon the velocities, they inherit the dependence the velocities have on $B\rmi$ (Sect.\ \ref{sub:vofBi}), and thus the power law relationships indicated in the Table.

In Table \ref{tab:fluxes}, we summarise ranges of velocities ($v_{\rm jet}$, $\langle v_\varphi\rangle \sim\langle v_{\rm rot}\rangle$, and $v_{\rm entr}$) from Table \ref{tab:simsum2} along with ranges for the fluxes measured from the last 10 yr of each simulation. The upper (lower) half of the table includes jet (second wind) fluxes where $M\geq5$ ($M<5$). On comparing Tables \ref{tab:obsparms} and \ref{tab:fluxes}, it is evident that these aspects of our simulations agree comfortably with the observations, including both `jet' and `entrained' fluxes.  The overlap is not perfect, of course.  Our mass fluxes fall within the observational range while the observed angular momentum fluxes fall within the range of our simulations.  As extensive as these simulations are, they are still limited by initial conditions, geometry, evolution time, and even some physics. With the relaxation of any of these constraints, specific comparisons may well improve.  Furthermore, the results presented in Table \ref{tab:fluxes} depend somewhat on the values adopted in the scaling relations (Eqs.\ \ref{eq:pscale}--\ref{eq:tscale}) for the stellar mass ($M_*$), inner disc radius ($r\rmi$), plus initial magnetic field and plasma-$\beta$ values at $r\rmi$ ($B\rmi$, $\beta\rmi$), and should be adjusted appropriately for specific comparisons.  Nevertheless, we take the agreement between Tables \ref{tab:obsparms} and \ref{tab:fluxes} as encouragement that these global simulations have captured the essence of both the production and propagation of protostellar outflows, and are among the first to do so.

\begin{table}
\begin{center}
\caption{\label{tab:fluxes} Estimates made from our global simulations for the observational parameters listed in Table \ref{tab:obsparms}.  Ranges for $\dot M$, $\dot p$, $\dot L$, and $\dot K$ measured from the last 10 yr of data for each simulation at a height of $z = 1,\!000$ AU above the disc. `Jet' values are defined as regions interior to the TD (masked using $M \geq 5$), while `entrained' material is defined as the shocked ambient just beyond the TD (masked using $M < 5$).}
\begin{tabular}{r|l}
\hline
 & ~Jet\\
$v_{\rm jet}$ & ~74 -- 420 km\,s$^{-1}$ \\
$\langle v_\varphi\rangle$ & 1.8 -- 21 km\,s$^{-1}$ \\
$\dot M$ & ~$2.9\times 10^{-8}$ -- $2.3\times 10^{-6}$ $M_{\odot}$\,yr$^{-1}$ \\
$\dot p$ & ~$1.3\times 10^{-6}$ -- $6.8\times 10^{-4}$ $M_{\odot}$\,yr$^{-1}$\,km\,s$^{-1}$ \\
$\dot L$ & ~$1.5\times 10^{-7}$ -- $1.7\times 10^{-3}$ $M_{\odot}$\,yr$^{-1}$\,AU\,km\,s$^{-1}$ \\
$\dot K$ & ~$3.9\times 10^{31}$ -- $1.6\times 10^{35}$ erg\,s$^{-1}$ \\
\hline
& ~Entrained\\
$v_{\rm entr}$ & ~1.1 -- 11.5 km\,s$^{-1}$ \\
$\langle v_\varphi\rangle$ & ~--- \\
$\dot M$ & ~$4.8\times 10^{-9}$ -- $2.3\times 10^{-7}$ $M_{\odot}$\,yr$^{-1}$ \\
$\dot p$ & ~$1.6\times 10^{-8}$ -- $1.9\times 10^{-5}$ $M_{\odot}$\,yr$^{-1}$\,km\,s$^{-1}$ \\
$\dot L$ & ~$7.7\times 10^{-10}$ -- $1.4\times 10^{-4}$ $M_{\odot}$\,yr$^{-1}$\,AU\,km\,s$^{-1}$ \\
$\dot K$ & ~$4.2\times 10^{28}$ -- $1.0\times 10^{33}$ erg\,s$^{-1}$\\
\hline
\end{tabular}
\end{center}
\end{table}

We note that observational measurements of the linear momentum flux in jets are rare.  Still, that the few values of which we are aware (see \citealt{podioetal06}; \citealt{hartiganetal94}; $10^{-6}$ -- $1.4\times 10^{-4}\, M_\odot\, {\rm yr}^{-1}\, {\rm km}\, {\rm s}^{-1}$) fall within the range of our calculated fluxes is encouraging.  Meanwhile, to the best of our knowledge, there are no observational estimates of kinetic energy fluxes in protostellar outflows, so our values in Table \ref{tab:fluxes} serve as a prediction should such measurements ever be made.

\section{Summary and Conclusions}
\label{sec:conclusions}

We present the first simulations to resolve the inner launching region of a protostellar (non-relativistic) jet, and then follow the jet to observational length scales.  The simulations were performed with our adaptive grid MHD code, \azeus\ \citep{rcm12}, with an effective dynamic range in resolution of $\sim 6.5\times 10^5$.

With a putative protostellar mass of $0.5\,M_\odot$, results from eight axisymmetric simulations are discussed, each identical except for the value of the plasma-beta at the inner radius of the accretion disc, $\beta\rmi$, ranging from $0.1$ (simulation A) to 640 (simulation H).  In each simulation, a jet is launched from the inner portion of a Keplerian accretion disc maintained as boundary conditions by magneto-rotational dynamics on the grid, and outflow is followed until the leading tip of the bow shock excited in the primordial atmosphere reaches the end of the coarsest grid.  One of the primary features of these simulations that sets them apart from others is \emph{at no time does any part of the outflow leave the computational domain.}

While our jets are still extremely `young' (lengths range from $\sim\!2,\!300$ to $\sim\!4,\!100$\,AU; ages range from $\sim\!50$ to $\sim\!150$\,yr), each has developed sufficiently to be directly comparable to observed jets, and to have established certain key observational properties such as asymptotic speeds, mass, momentum, and energy fluxes, rotation rates, \etc

Based on their characteristics, we have divided our eight simulations (Table \ref{tab:simsum}) into three sub-groups: `strong-field' (simulations A--D), `medium-field' (simulations E, F), and `weak-field' (simulations G, H).  With this distinction in mind, we draw the following conclusions:

\begin{enumerate}[label={\arabic*}.,leftmargin=*]

\item All eight simulations generate an organised, supersonic, magnetically confined outflow.

\vspace{6pt}

\item In regions of quasi-steady state, the `Weber-Davis constants' (equations 
\ref{eq:massload}--\ref{eq:energy}) remain constant for the most part to within 3\%, attesting to the numerical integrity of our AMR-MHD scheme.

\vspace{6pt}

\item On the observational scale ($>\!1,\!000\,$AU), each jet resembles the MHD simulations of \citet{clarkenormanburns86}.  Thus, they are confined by an internal toroidal magnetic field twisted out of the poloidal magnetic field in the atmosphere, and develop: (i) a hot, dense, super-fast `jet core' we identify with the observed jet; (ii) a rarefied, cold, magnetised, trans-fast `sheath' which we suggest may show up as an `emission cavity' surrounding the jet; and (iii) a narrow, laminar, magnetically confined `nose-cone' leading the jet.  Surrounding the jet is a \emph{tangential discontinuity} (TD) that separates jet material from the warm, trans-slow, shocked ambient medium entrained by the bow shock excited by the passage of the jet (`second wind'; top panel of Fig.\ \ref{fig:zoom0.1}). Due to the very large dynamic range and adaptive resolution, to the best of our knowledge, this is the first study of its kind that self-consistently includes and clearly differentiates between these different outflow components.

\vspace{6pt}

\item In agreement with local studies of outflow launching, strong-field jets (e.g.\ Fig.\ \ref{fig:zoom0.1}) are initially launched by the `bead-on-a-wire mechanism' (BWM; \citealt{bp82}) but is shut down even before the Alfv\'en point.  Beyond that, the `magnetic tower mechanism' (MTM; \citealt{lyndenbell96}) takes over, accelerating outflow well beyond the fast point.  They are characterised by quasi-steady state flow even near the jet launching region.

\vspace{6pt}

\item Also in agreement with local studies, weak-field jets (e.g.\ Fig.\ \ref{fig:zoom160}) are launched and accelerated by the MTM, in which relentless twisting of the weak poloidal field builds up sufficient toroidal magnetic pressure to drive the outflow.  These are characterised by highly turbulent flow near the launching region, with turbulence encroaching the TD as $\beta\rmi$ increases. For very high $\beta\rmi$, we speculate that the encroachment of turbulence on the TD will cause enough mixing between the shocked jet and ambient media to disrupt the jet in 3-D (\citealt{hardeeclarkerosen97}).

\vspace{6pt}

\item Medium-field jets (e.g.\ Fig.\ \ref{fig:zoom40}) are transitional between the strong and weak field cases.  They are launched by the BWM but accelerated soon thereafter by the MTM.  Neither laminar nor turbulent in the jet launching region, medium-field jets can exhibit `knot-like' structures which are generated periodically from the inner disc and propagate with the flow.

\vspace{6pt}

\item With the exception of simulation H (our weakest-field simulation), this study confirms that the interior dynamics of jets are dominated by a strong toroidal magnetic field.  Indeed, we find that plasma-$\beta$ averaged over rotating regions where $M>5$, $\langle\beta_{\rm tot}\rangle$, asymptotes to 0.2--0.4 regardless of $\beta\rmi$.  Simulation H struggles to maintain outflow, but even still $\langle\beta_{\rm tot}\rangle\rightarrow1$. Thus, this study implies that measures of magnetic field strength within the jet may not reveal much about the magnetic environment near the protostar.

\vspace{6pt}

\item The global nature of these simulations reveal that jets at similar `dynamical times' (\eg, of equal length) have similar radii and bow shock shape.  Thus, these cannot be used as a measure of magnetic field strength near the protostar.

\vspace{6pt}

\item Since no part of the outflow leaves the domain, we are able to determine that the advance speed of the jet (as would be measured by time-lapse images), $v_{\rm jet}$, and the average flow speed within the leading portion of the jet (as could be measured from line emission observations), $\langle v_z\rangle$, both vary as $B\rmi^{\sim4/9}$, where $B\rmi$ is the initial magnetic field strength at the inner radius of the disc. The fact that $v_{\rm jet}\sim\langle v_z\rangle$ indicate the jets are essentially ballistic.

\vspace{6pt}

\item The average rotation speed along the latter portion of the jet varies as $\langle v_\varphi\rangle\sim B\rmi^{2/3}$.  This, along with our other results, leads to possible observable evidence for the magnetic character of protostellar jets: $v_{\rm jet}\sim\langle v_\varphi\rangle^{2/3}$ (Eq.\ \ref{eq:vjvphi}).

\vspace{6pt}

\item The jet advance speed, $v_{\rm jet}$, and the advance speed of material entrained by the bow shock, $v_{\rm entr}$, all fall nicely within the realm of observational constraints (Table \ref{tab:obsparms}). Together, they indicate $v_{\rm entr}\sim v_{\rm jet}^{4/3}$ (Eq.\ \ref{eq:ventrvj}).

\vspace{6pt}

\item Because no jet leaves the computational domain, we have been able to make estimates of fluxes for mass, momentum (linear and angular), and kinetic energy in our jets.  Based in part on the consistency of these fluxes and velocities with those measured from observational data (Tables \ref{tab:obsparms} and \ref{tab:fluxes}), we conclude that our \emph{global} simulations are able to make the link to observations where \emph{local} simulations cannot.  In particular, we have shown that numerical models based solely on gravito-magneto-rotational fluid dynamics are capable of launching and driving jets that are consistent morphologically and quantitatively with the observations.

\end{enumerate}

\section*{Acknowledgements}
We thank ACEnet technicians Phil Romkey and Sergiy Khan for their assistance in optimising \azeus for the ACEnet facilities.  This work was supported, in part, by an NSERC Discovery Grant to DAC.  The Centre for Star and Planet Formation is funded by the Danish National Research Foundation (DNRF97). JPR was supported, in part, by the Virginia Initiative on Cosmic Origins (VICO).  Computing resources were provided, in part, by ComputeCanada via ACEnet and Calcul Queb\' ec. Additional computing facilities were provided by the University of Copenhagen HPC centre, funded in part by Villum Fonden (VKR023406), and the University of Copenhagen Electronic Research Data Archive (ERDA). This work made use of the SAO/NASA Astrophysics Data System and the MPFIT least-squares fitting package \citep{markwardt2009_mpfit}. We also thank the anonymous referee for a timely and constructive report.

\bibliographystyle{mnras}
\bibliography{./rcIII}
\bsp
\label{lastpage}
\end{document}